\documentclass[a4paper,twocolumn,showpacs,notitlepage,floatfix,nofootinbib,superscriptaddress,prc]{revtex4-1}
\usepackage{graphicx,color}
\usepackage{amsfonts}
\usepackage{amsmath}
\usepackage{hyperref}
\usepackage{float}
\usepackage{amssymb}
\usepackage{epstopdf}

\begin{document}

\title{Closing the equations of motion of anisotropic fluid dynamics by a judicious choice of moment of the Boltzmann equation}

\author{E.\ Moln\'ar}
\affiliation{Institut f\"ur Theoretische Physik, Johann Wolfgang Goethe-Universit\"at,
Max-von-Laue-Str.\ 1, D-60438 Frankfurt am Main, Germany}

\author{H.\ Niemi}
\affiliation{Institut f\"ur Theoretische Physik, Johann Wolfgang Goethe-Universit\"at,
Max-von-Laue-Str.\ 1, D-60438 Frankfurt am Main, Germany}

\author{D.\ H.\ Rischke}
\affiliation{Institut f\"ur Theoretische Physik, Johann Wolfgang Goethe-Universit\"at,
Max-von-Laue-Str.\ 1, D-60438 Frankfurt am Main, Germany}


\begin{abstract}
In Moln\'ar \textit{et al.} [Phys.\ Rev.\ D \textbf{93}, 114025 (2016)] the equations of anisotropic
dissipative fluid dynamics were obtained from the moments of the Boltzmann equation 
based on an expansion around an arbitrary anisotropic single-particle distribution function. 
In this paper we make a particular choice for this distribution function
and consider the boost-invariant expansion of a fluid in one dimension.
In order to close the conservation equations, we need to choose an additional moment of
the Boltzmann equation. 
We discuss the influence of the choice of this moment on the time evolution of fluid-dynamical 
variables and identify the moment that provides the best match of anisotropic fluid dynamics to 
the solution of the Boltzmann equation in the relaxation-time approximation.
\end{abstract}

%\date{\today }
%Initial submission: version 28 June 2016
%Corrections: version 12 November 2016
\maketitle

\section{Introduction}
\label{Introduction}

Relativistic fluid dynamics has been successfully applied to
understand a wide variety of phenomena in the fields of astrophysics,
cosmology, cold-atoms, and heavy-ion collisions 
\cite{Csernai_book,Rezzolla_book,Schaefer:2014awa,Braun-Munzinger:2015hba}. 
In particular, relativistic dissipative fluid dynamics has become one of 
the main tools in understanding the dynamics and properties of strongly 
interacting matter formed in ultrarelativistic heavy-ion collisions at 
BNL's Relativistic Heavy Ion Collider (RHIC) and at CERN's Large Hadron Collider (LHC). 
Such investigations have led to tremendous progress in our understanding of the properties 
of such matter, e.g.\ its equation of state and transport coefficients 
\cite{Niemi:2011ix,Heinz:2013th,Gale:2013da,Huovinen:2013wma,Ryu:2015vwa,Niemi:2015qia}.
A necessary prerequisite for these investigations is, however, to know the regime of applicability and 
limitations of relativistic fluid dynamics.

The applicability of traditional dissipative fluid-dynamical theories is
restricted to the vicinity of local thermodynamical equilibrium. This implies that the
deviations of the single-particle distribution function from its form in
local thermodynamical equilibrium are small.
However, the system formed in relativistic
heavy-ion collisions is not macroscopically large and it undergoes rapid expansion. 
Such conditions are challenging for fluid dynamics, as this can create situations
where momentum-space anisotropies are of major significance. In particular, this is true
in the very early stages of heavy-ion collisions.

To overcome such limitations of fluid-dynamical theories, in 
the late 1980's Barz, K\"ampfer, Luk\'acs, Martin\'as, and Wolf \cite{Barz:1987pq} proposed
an energy-momentum tensor which incorporated the momentum anisotropy in terms of a space-like four-vector 
$l^\mu$. Quite recently, two research groups aimed at including a large momentum-space anisotropy into the 
fluid-dynamical framework for ultrarelativistic heavy-ion collisions. 
Florkowski and Ryblewski \cite{Florkowski:2008ag,Florkowski:2010cf,Ryblewski:2010bs,
Ryblewski:2011aq,Ryblewski:2012rr} and Martinez and Strickland 
\cite{Martinez:2009ry,Martinez:2010sc,Martinez:2010sd} effectively rediscovered
anisotropic fluid dynamics and initiated a new line of research, 
cf.\ Refs.\ \cite{Denicol:2014mca,Bazow:2013ifa,Bazow:2015zca,Nopoush:2015yga}.

Their approach is based on a 
single-particle distribution function in momentum space, termed $\hat{f}_{0 \mathbf{k}}$ in the following, 
which is a deformed ellipsoid in the local rest (LR) frame of matter \cite{Romatschke:2003ms}. 
The momentum-space anisotropy is controlled by a single parameter $\xi$,
such that $\lim_{\xi \rightarrow 0} \hat{f}_{0 \mathbf{k}} = f_{0 \mathbf{k}}$, where
$f_{0 \mathbf{k}}$ is the single-particle distribution function in local thermodynamical equilibrium,
which is isotropic in the LR frame. In anisotropic fluid dynamics the momentum-space anisotropy can in 
principle be arbitrarily large, which is in contrast to conventional dissipative fluid dynamics, which is
based on the assumption of small deviations from local equilibrium.

The particle number four-current $\hat{N}^\mu$ and energy-momentum tensor $\hat{T}^{\mu \nu}$ 
are given by the first and second moments of $\hat{f}_{0 \mathbf{k}}$. The difference compared to an ideal fluid,
where $f_{0 \mathbf{k}}$ is the single-particle distribution function, is
that now the conserved quantities are functions of the parameters $\xi$ and $l^\mu$,
in addition to temperature $T$, chemical potential $\mu$, and fluid four-velocity $u^\mu$, which specify
$f_{0 \mathbf{k}}$. Therefore, the conservation equations and the equation of state do no longer
form a closed set of equations, and additional equations determining $\xi$ and $l^\mu$ are needed.
Usually, $l^\mu$ is fixed 
by the requirement that it is orthogonal to $u^\mu$, $u_\mu l^\mu =0$, and normalized,
$l_\mu l^\mu = -1$. For the sake of simplicity it may be chosen to have no components in the plane transverse 
to the beam ($z-$)direction, such that $l^\mu = \gamma _{z}\left(v_{z},0,0,1\right) $, 
where $\gamma_z = (1- v_z^2)^{-1/2}$. Thus, only one additional
equation is needed, which determines the time evolution of $\xi$.

If the particle number (or, in the relativistic context, net-charge) is conserved, 
the zeroth moment of the Boltzmann equation provides the corresponding conservation equation.
However, in situations where particle number (or net-charge) is not conserved 
(for instance, when particles are produced or annihilated), the
zeroth moment of the collision term does not vanish. The zeroth moment of the
Boltzmann equation can then be used to determine the momentum-space anisotropy.
However, due to fact that there is an infinite hierarchy of moment equations, also higher moments of
the Boltzmann equation could be used to provide closure of the equations of motion. This strategy
was employed in Refs.\ \cite{Nopoush:2014pfa,Alqahtani:2015qja,Florkowski:2015cba,tinti}, where specific
projections of the second moment of the Boltzmann equation were used.

In principle, the ambiguity in the choice of moment can be resolved by
comparing the fluid-dynamical solution to that of the Boltzmann equation.
This is the purpose of the present paper. We study several possible choices for
the moment that closes the equations of motion, both in the case with and 
without particle-number conservation. 

The paper is organized as follows. In Sec.\ \ref{Equations_of_motion}
we recall the tensor decomposition and the equations of motion
in the case of an arbitrary anisotropic distribution function from Ref.\ \cite{Molnar:2016vvu}. In Sec.\
\ref{applications} we apply this formalism to a specific example, the so-called
Romatschke-Strickland distribution function, and provide the Landau matching
conditions to calculate temperature and, in the case of particle-number conservation, chemical potential.
Assuming 0+1 dimensional Bjorken flow \cite{Bjorken:1982qr} and the relaxation-time approximation (RTA) for
the collision term \cite{Bhatnagar:1954zz,Anderson_Witting}, we present the conservation
equations and the various choices for the moment equation which is used to provide
closure. In Sec.\ \ref{results} we systematically study these choices 
and compare them to the solution of the Boltzmann equation.
We conclude this work in Sec.\ \ref{Conclusions} with a summary and an outlook. 
Technical details are relegated to the Appendices.

We adopt natural units, $\hbar =c=k_{B}=1$, throughout this work. 
The elementary projection operator orthogonal to $u^{\mu }$ is denoted by 
$\Delta ^{\mu \nu }=g^{\mu \nu }-u^{\mu }u^{\nu }$, where $g^{\mu \nu}
=g_{\nu \mu }=\text{diag}(1,-1,-1,-1)$ is the Minkowski metric tensor of
flat space-time. The four-momentum of particles, $k^{\mu}=(k_{0},k_{x},k_{y},k_{z})$ 
is normalized to the rest mass $m_{0}$ of the
particles, $k^{\mu }k_{\mu }=m_{0}^{2}$, and can be decomposed into two parts, 
$k^{\mu }=E_{\mathbf{k}u}u^{\mu }+k^{\left\langle \mu \right\rangle }$, where 
$E_{\mathbf{k}u}=k^{\mu }u_{\mu }$ is the (relativistic on-shell) energy,
while $k^{\left\langle \mu \right\rangle }=\Delta ^{\mu \nu }k_{\nu }$ is
the particle momentum orthogonal to the flow velocity. For an arbitrary
anisotropy the projection tensor orthogonal to both $u^{\mu }$ and $l^{\mu }$
will be denoted by $\Xi ^{\mu \nu }\equiv g^{\mu \nu }-u^{\mu }u^{\nu
}+l^{\mu }l^{\nu }=\Delta ^{\mu \nu }+l^{\mu }l^{\nu }$ \cite{Gedalin_1991,Gedalin_1995,Huang:2009ue,Huang:2011dc}. 
Thus, the four-momentum of particles can be decomposed as $k^{\mu }
=E_{\mathbf{k}u}u^{\mu }+E_{\mathbf{k}l}l^{\mu }+k^{\left\{ \mu \right\} }$, where 
$E_{\mathbf{k}l}=-k^{\mu }l_{\mu }$ is the particle momentum in the direction of
the anisotropy and $k^{\left\{ \mu \right\} }=\Xi ^{\mu \nu }k_{\nu }$ are
the components of the momentum orthogonal to both $u^{\mu }$ and $l^{\mu }$.

\section{The general equations of motion of anisotropic fluids}
\label{Equations_of_motion}

The starting point of relativistic kinetic theory is the
Boltzmann equation \cite{deGroot,Cercignani_book},
\begin{equation}
k^{\mu }\partial _{\mu }f_{\mathbf{k}}=C\left[ f\right] \; ,
\label{Boltzmann_eq}
\end{equation}
where $f_{\mathbf{k}}=f\left( x^{\mu },k^{\mu }\right) $ is the single-particle distribution function 
at space-time coordinate $x^{\mu }$, while $\partial _{\mu }\equiv \partial /\partial x^{\mu }$ is the space-time
derivative. The collision integral (for binary collisions only) is
\begin{eqnarray}
C\left[ f\right] & =& \frac{1}{2}\int dK^{\prime }dPdP^{\prime }\,
W_{\mathbf{kk}\prime \rightarrow \mathbf{pp}\prime } \nonumber \\
&  & \hspace*{0.5cm} \times  \left( f_{\mathbf{p}}f_{\mathbf{p}^{\prime }}\tilde{f}_{\mathbf{k}}\tilde{f}_{\mathbf{k}^{\prime}}
-f_{\mathbf{k}}f_{\mathbf{k}^{\prime }}\tilde{f}_{\mathbf{p}}\tilde{f}_{\mathbf{p}^{\prime }}\right) \,.
\label{COLL_INT}
\end{eqnarray}
Here, $\tilde{f}_{\mathbf{k}}= 1-af_{\mathbf{k}} $, where $a=\pm 1$ for fermions/bosons, while 
$a = 0$ corresponds to classical, indistinguishable particles. The
invariant momentum-space volume is $dK=gd^{3}\mathbf{k/}\left[ (2\pi )^{3}k^{0}\right] $, where $g$ 
denotes the number of internal degrees of
freedom. Furthermore, $W_{\mathbf{kk}\prime \rightarrow \mathbf{pp}\prime }$
is the invariant transition rate. 

Following Ref.\ \cite{Molnar:2016vvu} we denote the anisotropic distribution function as
$\hat{f}_{0\mathbf{k}}\left( \hat{\alpha},\hat{\beta}_{u} E_{\mathbf{k}u},\hat{\beta}_{l} E_{\mathbf{k}l}\right)$, 
which characterizes an anisotropic state as a function of three scalar parameters, $\hat{\alpha}$, 
$\hat{\beta}_{u}$, and $\hat{\beta}_{l}$, as
well as the on-shell energy $E_{\mathbf{k}u}$ and the momentum component $E_{\mathbf{k}l}$
in the direction of the anisotropy. We also demand that
\begin{equation}
\lim_{\hat{\beta}_{l}\rightarrow 0}\hat{f}_{0\mathbf{k}}\left( \hat{\alpha},\hat{\beta}_{u}E_{\mathbf{k}u},
\hat{\beta}_{l}E_{\mathbf{k}l}\right) =f_{0\mathbf{k}}\left( \hat{\alpha},\hat{\beta}_{u}E_{\mathbf{k}u}\right)\; ,
\label{hat_f->f_0}
\end{equation}
i.e., in the limit of vanishing anisotropy parameter $\hat{\beta}_{l}$ the
anisotropic distribution converges to the distribution function in local thermodynamical equilibrium.
This is the so-called J\"{u}ttner
distribution function \cite{Juttner,Juttner_quantum},
\begin{equation}
f_{0\mathbf{k}}\left( \alpha _{0},\beta _{0}E_{\mathbf{k}u}\right) =\left[
\exp (-\alpha _{0}+\beta _{0}E_{\mathbf{k}u})+a\right] ^{-1}\;,
\label{Equilibrium_dist}
\end{equation}
where $\beta _{0}=1/T$ and $\alpha _{0}=\mu \beta _{0}$.

The moments of tensor
rank $n$ of the anisotropic distribution function $\hat{f}_{0\mathbf{k}}$ are defined as
\begin{equation}
\hat{\mathcal{I}}_{ij}^{\mu _{1}\cdots \mu _{n}}=\left\langle E_{\mathbf{k}u}^{i}E_{\mathbf{k}l}^{j}
k^{\mu _{1}}\cdots k^{\mu _{n}}\right\rangle _{\hat{0}}\;,  \label{I_ij_tens}
\end{equation}
where $\left\langle \cdots \right\rangle _{\hat{0}}=\int dK\left( \cdots
\right) \hat{f}_{0\mathbf{k}}$.
These moments are expanded as
\begin{align}
& \!\!\hat{\mathcal{I}}_{ij}^{\mu _{1}\cdots \mu _{n}}=\sum_{q=0}^{\left[ n/2\right] }
\sum_{r=0}^{n-2q}\left( -1\right) ^{q}b_{nrq}\hat{I}_{i+j+n,j+r,q} 
\notag \\
& \!\!\times \Xi ^{\left( \mu _{1}\mu _{2}\right. }\cdots \Xi ^{\mu
_{2q-1}\mu _{2q}}l^{\mu _{2q+1}}\cdots l^{\mu _{2q+r}}u^{\mu
_{2q+r+1}}\cdots u^{\left. \mu _{n}\right) }\;,  \label{I_ij_tens_expanded}
\end{align}
where $n$, $r$, and $q$ are natural numbers and $\left[ n/2\right] $ denotes the integer
part of $n/2$. The number of permutations of indices that lead to distinct 
tensors $\Xi ^{\left( {}\right. }\cdots
l\cdots u^{\left. {}\right) }$ is $b_{nrq}=n!\left( 2q-1\right)
!!/[\left( 2q\right)! r ! \left( n-2q-r\right) !]$. The double factorials of even and
odd numbers are defined as $\left( 2q\right) !!=2^{q} q !$ and 
$\left( 2q-1\right) !!=\left( 2q\right) !/\left( 2^{q} q !\right)$, respectively. 
Finally, the generalized thermodynamic integrals are defined as
\begin{equation}
\hat{I}_{nrq}=\frac{\left( -1\right) ^{q}}{\left( 2q\right) !!}\left\langle
E_{\mathbf{k}u}^{n-r-2q}E_{\mathbf{k}l}^{r}\left( \Xi ^{\mu \nu }k_{\mu
}k_{\nu }\right) ^{q}\right\rangle _{\hat{0}}\;.  \label{I_nrq}
\end{equation}
Note that in analogy to Eq.\ (\ref{I_ij_tens}) we define the
generalized moments of $f_{0\mathbf{k}}$ as
\begin{equation}
\lim_{\hat{\beta}_{l}\rightarrow 0}\hat{\mathcal{I}}_{ij}^{\mu _{1}\cdots
\mu _{n}}\equiv \mathcal{I}_{ij}^{\mu _{1}\cdots \mu _{n}}
=\left\langle E_{\mathbf{k}u}^{i}E_{\mathbf{k}l}^{j}k^{\mu _{1}}\cdots k^{\mu
_{n}}\right\rangle _{0}\;,  \label{I_ij_eq_tens}
\end{equation}
where $\left\langle \cdots \right\rangle _{0}=\int dK\left( \cdots \right)
f_{0\mathbf{k}}$. The thermodynamic integrals in equilibrium are thus given by
\begin{equation}
I_{nrq}=\lim_{\hat{\beta}_{l}\rightarrow 0}\hat{I}_{nrq} \; ,
\label{I_nrq_eq}
\end{equation}
i.e., they are given by Eq.\ (\ref{I_nrq}) upon replacing $\left\langle \cdots \right\rangle _{\hat{0}}\rightarrow
\left\langle \cdots \right\rangle _{0}$.

Using the expansion (\ref{I_ij_tens_expanded}), we readily
obtain the conserved quantities $\hat{N}^{\mu }\equiv \hat{\mathcal{I}}_{00}^{\mu }$ 
and $\hat{T}^{\mu \nu } \equiv \hat{\mathcal{I}}_{00}^{\mu \nu}$ decomposed 
with respect to $u^{\mu }$, $l^{\nu }$, and $\Xi^{\mu \nu }$, 
\begin{align}
& \hat{N}^{\mu }\equiv \langle k^{\mu} \rangle_{\hat{0}}=\hat{n}u^{\mu }+
\hat{n}_{l}l^{\mu }\;,  \label{N_mu_aniso} \\
& \hat{T}^{\mu \nu }\equiv\langle k^{\mu} k^{\nu} \rangle_{\hat{0}} =\hat{e}
u^{\mu }u^{\nu }+2\hat{M}u^{\left( \mu \right. }l^{\left. \nu \right) }+\hat{P}_{l}l^{\mu }l^{\nu }
-\hat{P}_{\perp }\Xi ^{\mu \nu }\;. \quad
\label{T_munu_aniso}
\end{align}
The coefficients of the various tensor structures can be expressed in terms of
generalized thermodynamic integrals or, equivalently, by different projections of the
tensor moments (\ref{I_ij_tens}),
\begin{align}
\hat{n}& \equiv \hat{N}^{\mu }u_{\mu }=\hat{I}_{100}=\hat{\mathcal{I}}_{10}\;,
\label{n_hat} \\
\hat{n}_{l}& \equiv -\hat{N}^{\mu }l_{\mu }=\hat{I}_{110}=\hat{\mathcal{I}}_{01}\;,  \label{n_l_hat} \\
\hat{e}& \equiv \hat{T}^{\mu \nu }u_{\mu }u_{\nu }=\hat{I}_{200}=\hat{\mathcal{I}}_{20}\;,  \label{e_hat} \\
\hat{M}& \equiv -\hat{T}^{\mu \nu }u_{\mu }l_{\nu }=\hat{I}_{210}=\hat{\mathcal{I}}_{11}\;,  \label{M_hat} \\
\hat{P}_{l}& \equiv \hat{T}^{\mu \nu }l_{\mu }l_{\nu }=\hat{I}_{220}=\hat{\mathcal{I}}_{02}\;,   \label{P_l_hat} \\
\hat{P}_{\perp }& \equiv -\frac{1}{2}\hat{T}^{\mu \nu }\Xi _{\mu \nu }
=\hat{I}_{201}=-\frac{1}{2}\left( m_{0}^{2}\hat{\mathcal{I}}_{00}-\hat{\mathcal{I}}_{20}
+\hat{\mathcal{I}}_{02}\right)\; .  \label{P_t_hat}
\end{align}
The particle density is $\hat{n}$, and $\hat{n}_{l}$ is the part of
the particle diffusion current that points into the $l^{\mu}-$direction. The energy density is 
$\hat{e}$, while $\hat{M}$ is the part of the energy diffusion current along the $l^{\mu }-$direction. 
The pressure component in the direction of the momentum anisotropy is $\hat{P}_{l}$, while the pressure in
the direction transverse to $l^{\mu }$ is $\hat{P}_{\perp} $. The isotropic pressure is defined as 
\begin{equation}
\hat{P}\equiv -\frac{1}{3}\hat{T}^{\mu \nu }\Delta _{\mu \nu }=\frac{1}{3}%
\left( \hat{P}_{l}+2\hat{P}_{\perp }\right)\; .  \label{P_iso_relation}
\end{equation}

Therefore, the particle four-current and energy-momentum tensor defined in
Eqs.\ (\ref{N_mu_aniso}) and (\ref{T_munu_aniso}) contain nine unknowns: the four-vector $u^\mu$ 
with three independent components and six scalars, $\hat{n}$, $\hat{e}$, $\hat{n}_{l}$, $\hat{M}$, 
$\hat{P}_{l}$, and $\hat{P}_{\perp }$. (We assume that $l^\mu$ is already fixed
as described in Sec.\ \ref{Introduction}.) However, since these latter quantities are
functions of three independent scalars $\hat{\alpha}$, $\hat{\beta}_{u}$, and 
$\hat{\beta}_{l}$, only three of the above six scalar variables are independent.

One still needs to assign a physical meaning to the fluid four-velocity, i.e., 
one needs to determine which 
physical quantity is actually at rest in the LR frame.
Eckart's choice \cite{Eckart:1940te} is the flow of particles,
\begin{equation}
u^{\mu }\equiv \frac{\hat{N}^{\mu }}{\sqrt{\hat{N}^{\nu }\hat{N}_{\nu }}}\;.
\label{LR_Eckart_1}
\end{equation}
This implies that there is no particle diffusion,
i.e., $\hat{n}_{l}=0$. Landau and Lifshitz \cite{Landau_book} choose to
define the LR frame in terms of the flow of energy, 
\begin{equation}
u^{\mu }\equiv \frac{\hat{T}^{\mu \nu }u_{\nu }}{\sqrt{u^{\lambda }
\hat{T}_{\alpha \lambda }\hat{T}^{\alpha \beta }u_{\beta }}}\;,  \label{LR_Landau_1}
\end{equation}
which leads to a vanishing energy diffusion current, i.e., $\hat{M}=0$.

However, neither of these choices removes one of the six unknowns
$u^\mu$, $\hat{\alpha}$, $\hat{\beta}_{u}$, and $\hat{\beta}_{l}$.
The conservation equations, $\partial _{\mu }\hat{N}^{\mu }=0$ and 
$\partial _{\mu }\hat{T}^{\mu \nu }=0$, provide only five constraints for these six
independent variables, hence we need an additional equation for the
remaining variable. Naturally, in kinetic theory
this can be provided by choosing an equation from the infinite hierarchy of
moment equations of the Boltzmann equation. For an anisotropic
distribution function these equations have the following form
\begin{equation}
\partial _{\lambda }\hat{\mathcal{I}}_{00}^{\mu _{1}\cdots \mu _{n}\lambda }=
\hat{\mathcal{C}}_{00}^{\mu _{1}\cdots \mu _{n}}\;,  \label{Hierarchy_moments}
\end{equation}
where the collision integral is defined as 
\begin{equation}
\hat{\mathcal{C}}_{ij}^{ \mu _{1}\cdots \mu _{n} }=\int dKE_{\mathbf{k}u}^{i}E_{\mathbf{k}l}^{j}
k^{\mu _{1}}\cdots k^{\mu _{n}} C[\hat{f}_{0\mathbf{k}}] \;.  \label{hat_coll_int}
\end{equation}
Contracting Eq.\ (\ref{Hierarchy_moments}) with projection tensors
built from $u^\mu$, $l^\nu$, and $\Xi^{\mu \nu}$ leads to the
following tensor equations,
\begin{align}
& u_{\mu _{1}}\cdots u_{\mu _{i}}l_{\mu _{i+1}}\cdots l_{\mu _{i+j}}\Xi
_{\mu _{i+j+1}\cdots \mu _{n}}^{\alpha _{i+j+1}\cdots \alpha _{n}}\partial
_{\lambda }\hat{\mathcal{I}}_{00}^{\mu_{1}\cdots \mu_{n}\lambda }  \notag
\\
& =u_{\mu _{1}}\cdots u_{\mu_{i}}l_{\mu_{i+1}}\cdots l_{\mu_{i+j}}\Xi
_{\mu_{i+j+1}\cdots \mu _{n}}^{\alpha_{i+j+1}\cdots \alpha_{n}}
\hat{\mathcal{C}}_{00}^{\mu_{1}\cdots \mu_{n}}\;,  \label{Hierarchy_projections}
\end{align}
where $\Xi _{\nu _{1}\cdots \nu _{n}}^{\mu _{1}\cdots \mu _{n}}$ are
irreducible projection operators constructed from the $\Xi^{\mu\nu}$'s, 
such that for any $n\geq 2$ they are
symmetric, traceless, and orthogonal to both $u^{\mu}$ and $l^{\nu}$.

In Ref.\ \cite{Molnar:2016vvu} we derived the equations of motion for the irreducible
moments $\hat{\rho}_{ij}^{\mu_1 \cdots \mu_\ell}$ of $\delta \hat{f}_{\mathbf{k}} \equiv f_{\mathbf{k}} - 
\hat{f}_{0\mathbf{k}}$. Although we only wrote them down explicitly up to tensor rank $n=2$,
they follow from a tensor equation similar to Eq.\ (\ref{Hierarchy_projections}). The latter is then simply
the special case of that tensor equation obtained by setting $\hat{\rho}_{ij}^{\mu_1 \cdots \mu_\ell}\equiv 0$.
Ultimately, we can take the equation for the scalar moment, Eq.\ (110) of Ref.\ \cite{Molnar:2016vvu}, and put all
irreducible moments $\hat{\rho}_{ij}^{\mu_1 \cdots \mu_\ell}\equiv 0$ to obtain
\begin{align}
& \hat{\mathcal{C}}_{i-1,j}=D\hat{\mathcal{I}}_{ij}-\left( \hat{i\mathcal{I}}_{i-1,j+1}
+j\hat{\mathcal{I}}_{i+1,j-1}\right) l_{\alpha }Du^{\alpha } 
\notag \\
& -D_{l}\hat{\mathcal{I}}_{i-1,j+1}+\left[ \left( i-1\right) \hat{\mathcal{I}}_{i-2,j+2}
+\left( j+1\right) \hat{\mathcal{I}}_{ij}\right] l_{\alpha}D_{l}u^{\alpha }  \notag \\
& -\frac{1}{2}\left[ m_{0}^{2}\left( i-1\right) \hat{\mathcal{I}}_{i-2,j}-\left( i+1\right) \hat{\mathcal{I}}_{ij}
+\left( i-1\right) \hat{\mathcal{I}}_{i-2,j+2}\right] \tilde{\theta}  \notag \\
& +\frac{1}{2}\left[ m_{0}^{2}j\hat{\mathcal{I}}_{i-1,j-1}-j\hat{\mathcal{I}}_{i+1,j-1}
+\left( j+2\right) \hat{\mathcal{I}}_{i-1,j+1}\right] \tilde{\theta}_{l}\;.  \label{General_aniso_scalar}
\end{align}
Similarly, taking Eq.\ (111) of Ref.\ \cite{Molnar:2016vvu} and setting all
irreducible moments $\hat{\rho}_{ij}^{\mu_1 \cdots \mu_\ell}\equiv 0$ we
obtain the equation for the vector moment,
\begin{align}
&2\, \hat{\mathcal{C}}_{i-1,j}^{\left\{ \mu \right\} }=\tilde{\nabla}^{\mu }\left( m_{0}^{2}
\hat{\mathcal{I}}_{i-1,j}-\hat{\mathcal{I}}_{i+1,j}+
\hat{\mathcal{I}}_{i-1,j+2}\right)  \notag \\
& -\left[ m_{0}^{2}i\hat{\mathcal{I}}_{i-1,j}-\left( i+2\right) 
\hat{\mathcal{I}}_{i+1,j}+i\hat{\mathcal{I}}_{i-1,j+2}\right] \Xi _{\alpha}^{\mu }Du^{\alpha }  \notag \\
& +\left[ m_{0}^{2}j\hat{\mathcal{I}}_{i,j-1}-j\hat{\mathcal{I}}_{i+2,j-1}
+\left( j+2\right) \hat{\mathcal{I}}_{i,j+1}\right] \Xi _{\alpha}^{\mu }Dl^{\alpha }  \notag \\
& +\left[ m_{0}^{2}\left( i-1\right) \hat{\mathcal{I}}_{i-2,j+1}
-\left( i+1\right) \hat{\mathcal{I}}_{i,j+1}\right] \Xi _{\alpha}^{\mu }D_{l}u^{\alpha }  \notag \\
& -\left[ m_{0}^{2}\left( j+1\right) \hat{\mathcal{I}}_{i-1,j}
-\left( j+1\right) \hat{\mathcal{I}}_{i+1,j}\right] \Xi _{\alpha}^{\mu }D_{l}l^{\alpha }  \notag \\
& +\left( i-1\right) \hat{\mathcal{I}}_{i-2,j+3}\Xi _{\alpha}^{\mu }D_{l}u^{\alpha }
-\left( j+3\right)\hat{\mathcal{I}}_{i-1,j+2}
\Xi _{\alpha }^{\mu }D_{l}l^{\alpha }  \notag \\
& -\left( i-1\right)\left( m_{0}^{2}\hat{\mathcal{I}}_{i-2,j+1}-
\hat{\mathcal{I}}_{i,j+1}+\hat{\mathcal{I}}_{i-2,j+3}\right) l_{\alpha }
\tilde{\nabla}^{\mu }u^{\alpha }  \notag \\
& -j\left( m_{0}^{2}\hat{\mathcal{I}}_{i,j-1}-\hat{\mathcal{I}}_{i+2,j-1}
+\hat{\mathcal{I}}_{i,j+1}\right) l_{\alpha }\tilde{\nabla}^{\mu}u^{\alpha }\;.  \label{General_aniso_vector}
\end{align}
Here, $D=u^{\mu }\partial _{\mu }$ denotes the comoving derivative and $D_{l}=-l^{\mu }\partial _{\mu }$ 
is the derivative in the direction of the
anisotropy. The spatial gradient in the directions orthogonal to both 
$u^{\mu }$ and $l^{\mu }$ is $\tilde{\nabla}_{\mu }=\Xi _{\mu \nu }\partial^{\nu }$, 
while the expansion scalars are defined as $\tilde{\theta}=\tilde{\nabla}_{\mu }u^{\mu }$ 
and $\tilde{\theta}_{l}=\tilde{\nabla}_{\mu }l^{\mu} $.

The particle-number conservation equation follows from Eq.\ (\ref{General_aniso_scalar}) 
by choosing $i=1$ and $j=0$. Using Eqs.\ (\ref{n_hat})--(\ref{P_t_hat}) we obtain
\begin{eqnarray}
0 =\partial _{\mu }\hat{N}^{\mu }& \equiv &D\hat{n}-D_{l}\hat{n}_{l}+\hat{n}
\tilde{\theta}+\hat{n}_{l}\tilde{\theta}_{l}  \notag \\
&&+\hat{n}\, l_{\mu }D_{l}u^{\mu } -\hat{n}_{l}\, l_{\mu}Du^{\mu } \;,  \label{conservation_eq_particle}
\end{eqnarray}
where due to particle-number conservation $\hat{\mathcal{C}}_{00}=0$.

The energy-conservation equation follows from Eq.\ (\ref{General_aniso_scalar})
by choosing $i=2$ and $j=0$,
\begin{eqnarray}
0 =u_{\nu }\partial _{\mu }\hat{T}^{\mu \nu }&\equiv &D\hat{e}-D_{l}\hat{M}
+\left( \hat{e}+\hat{P}_{\perp }\right) \tilde{\theta}+\hat{M}\tilde{\theta}_{l}  \notag \\
&&+\left( \hat{e}+\hat{P}_{l}\right) l_{\mu }D_{l}u^{\mu } -2
\hat{M}\, l_{\mu }Du^{\mu } \,,  \label{conservation_eq_energy}
\end{eqnarray}
while the conservation equation for the momentum in the $l^{\mu }-$direction can be
obtained for $i=1$ and $j=1$, 
\begin{eqnarray}
0 =l_{\nu }\partial _{\mu }\hat{T}^{\mu \nu }&\equiv& -D\hat{M}+D_{l}\hat{P}_{l}
-\hat{M}\tilde{\theta}+\left( \hat{P}_{\perp }-\hat{P}_{l}\right) \tilde{\theta}_{l}  \notag \\
&&-2\hat{M}\, l_{\mu }D_{l}u^{\mu } +\left( \hat{e}+\hat{P}_{l}\right)  l_{\mu }Du^{\mu }\; ,
\label{conservation_eq_momenta_l}
\end{eqnarray}
where $\hat{\mathcal{C}}_{10}=0$ and $\hat{\mathcal{C}}_{01}=0$ vanish due
to energy and momentum conservation, respectively.

The conservation equation for the momentum transverse to $l^{\mu }$
can be obtained from Eq.\ (\ref{General_aniso_vector}) for $i=1$ and $j=0$.
Using $\hat{\mathcal{C}}_{10}^{\left\{ \mu \right\} }=0$ we obtain, 
\begin{align}
0& =\Xi _{\nu }^{\alpha }\partial _{\mu }\hat{T}^{\mu \nu }\equiv \left( 
\hat{e}+\hat{P}_{\perp }\right) \left( Du^{\alpha }+l^{\alpha } l_{\nu}Du^{\nu }\right)  \notag \\
& -\tilde{\nabla}^{\alpha }\hat{P}_{\perp }+\left( \hat{P}_{\perp }-\hat{P}_{l}\right) 
\left( D_{l}l^{\alpha }+u^{\alpha } l_{\nu }D_{l}u^{\nu}\right)   \notag \\
& +\hat{M}\left( Dl^{\alpha }+u^{\alpha } l_{\nu }Du^{\nu }\right) 
-\hat{M}\left( D_{l}u^{\alpha }+l^{\alpha } l_{\nu}D_{l}u^{\nu }\right) \;.  \label{conservation_eq_momenta_t}
\end{align}

In order to close the five conservation equations in terms of 
fluid-dynamical quantities  we
need to supply Eqs.\ (\ref{conservation_eq_particle})--(\ref{conservation_eq_momenta_t}) 
with an additional equation of motion. To this
end, it is natural to select the equation of motion for $\hat{n}_{l}$ or $\hat{M}$ 
(depending on the choice of the LR frame), or $\hat{P}_{l}$.
However, as we have already discussed in the Introduction, alternatively we may use any
higher moment of the Boltzmann equation to close the
conservation equations. 
The choice of closure is the main question that we will further investigate in the following sections.

\section{Applications}
\label{applications}

\subsection{The Romatschke-Strickland distribution function and properties}

As a simple, and at the same time relevant, example we take the anisotropic
distribution function introduced by Romatschke and Strickland (RS) \cite{Romatschke:2003ms} 
\begin{equation}
\hat{f}_{RS}\equiv \left[ \exp \left( -\alpha _{RS}+\beta _{RS}
\sqrt{k^{\mu}k^{\nu }\Omega _{\mu \nu }}\right) +a\right] ^{-1}\;,  \label{f_RS}
\end{equation}
where 
\begin{equation}
\Omega ^{\mu \nu }=u^{\mu }u^{\nu }+\xi \ l^{\mu }l^{\nu }\;.
\end{equation}
Here, $\xi$ denotes the so-called anisotropy
parameter. For $\xi < 0$, $\hat{f}_{RS}$ is a prolate spheroid 
and for $\xi >0$ it is an oblate spheroid with respect to the $z-$axis 
in momentum space and in the LR frame.

Comparing to Eq.\ (\ref{hat_f->f_0}) we identify $\hat{\alpha}\equiv \alpha _{RS}$, 
$\hat{\beta}_{u}\equiv \beta _{RS}$, and $\hat{\beta}_{l}\equiv \beta_{RS} \sqrt{\xi} $.
Furthermore, in order to calculate the fluid-dynamical quantities using the RS distribution function,
we shall introduce a new set of thermodynamic integrals, $\hat{I}_{nrq}^{RS}\left( \alpha _{RS},\beta _{RS},\xi \right) $,
which correspond to the replacement $\hat{f}_{0\mathbf{k}}\rightarrow \hat{f}_{RS}$ in Eq.\ (\ref{I_nrq}), 
\begin{equation}
\hat{I}_{nrq}^{RS}=\frac{\left( -1\right) ^{q}}{\left( 2q\right) !!}\int
dK\, E_{\mathbf{k}u}^{n-r-2q}E_{\mathbf{k}l}^{r}\left( \Xi ^{\mu \nu }k_{\mu}k_{\nu }\right)^{q}\ \hat{f}_{RS}\;.  \label{I_nrq_RS}
\end{equation}
These integrals are most easily evaluated for a massless Boltzmann
gas, i.e., $m_{0}=0$ and $a=0$. As shown in Ref.\ \cite{Martinez:2009ry} 
assuming $m_0=0$ leads to factorization of the $\xi-$dependent part, 
\begin{equation}
\hat{I}_{nrq}^{RS}\left( \alpha _{RS},\beta _{RS},\xi \right) =
I_{nq}\left( \alpha_{RS},\beta _{RS}\right) \, R_{nrq}\left( \xi \right) \;,  \label{R_nrq_RS}
\end{equation}
where the standard thermodynamic integrals are 
\begin{equation}
I_{nq}\left( \alpha _{0},\beta _{0}\right) =\frac{\left( -1\right) ^{q}}{
\left( 2q+1\right) !!}\left\langle E_{\mathbf{k}u}^{n-2q}
\left( \Delta^{\alpha \beta }k_{\alpha }k_{\beta }\right) ^{q}\right\rangle _{0}\;.
\label{I_nq}
\end{equation}
Considering the tensor decomposition of the first and the second moment of the equilibrium distribution function,
$N_{0}^{\mu }\equiv I_{10}u^{\mu }=I_{100}u^{\mu }$ and
$T_{0}^{\mu \nu }\equiv I_{20}u^{\mu }u^{\nu }-I_{21}\Delta ^{\mu\nu }
=I_{200}u^{\mu }u^{\nu }+I_{220}l^{\mu }l^{\nu }-I_{201}\Xi ^{\mu \nu }$, respectively,
the particle density is $n_{0}\equiv N_{0}^{\mu }u_{\mu }=I_{100}\equiv I_{10}$, the energy
density is $e_{0}\equiv T_{0}^{\mu \nu }u_{\mu }u_{\nu }=I_{200}\equiv I_{20}$, and the
thermodynamic pressure is $P_{0}\equiv -\frac{1}{3}T_{0}^{\mu \nu }\Delta _{\mu \nu}$.
The latter is necessarily isotropic, such that $P_{0l}\equiv T_{0}^{\mu \nu}l_{\mu }l_{\nu }=I_{220}$ 
and $P_{0\perp }\equiv -\frac{1}{2}T_{0}^{\mu \nu}\Xi _{\mu \nu }=I_{201}$ are identical, 
$P_{0}=P_{0l}\equiv P_{0\perp }$.
Furthermore, in equilibrium $n_{0l}\equiv -N_{0}^{\mu }l_{\mu }=I_{110}=0$
and $M_{0}\equiv T_{0}^{\mu \nu }u_{\mu }l_{\nu }=I_{210}=0$.

The first and second moments of the RS distribution function are 
\begin{eqnarray}
\hat{N}_{RS}^{\mu } &=&\hat{n}u^{\mu }\;,  \label{N_mu_RS} \\
\hat{T}_{RS}^{\mu \nu } &=&\hat{e}u^{\mu }u^{\nu }+\hat{P}_{l}l^{\mu }l^{\nu}-\hat{P}_{\perp }\Xi ^{\mu \nu }\;,  \label{T_munu_RS}
\end{eqnarray}
where the quantities defined in Eqs.\ (\ref{n_hat})--(\ref{P_t_hat}) can be written with the help of Eq.\ 
(\ref{R_nrq_RS}) as
\begin{eqnarray}
\hat{n} &\equiv &\hat{I}_{100}^{RS}=n_{0}\left( \alpha _{RS},\beta
_{RS}\right) R_{100}\left( \xi \right) \; ,\   \label{n_hat_RS} \\
\hat{e} &\equiv &\hat{I}_{200}^{RS}=e_{0}\left( \alpha _{RS},\beta
_{RS}\right) R_{200}\left( \xi \right) \; ,  \label{e_hat_RS} \\
\hat{P}_{l} &\equiv &\hat{I}_{220}^{RS}=e_{0}\left( \alpha _{RS},\beta
_{RS}\right) R_{220}\left( \xi \right) \; ,  \label{Pl_hat_RS} \\
\hat{P}_{\perp } &\equiv &\hat{I}_{201}^{RS}=P_{0}\left( \alpha _{RS},\beta
_{RS}\right) R_{201}\left( \xi \right) \; .  \label{Pt_hat_RS}
\end{eqnarray}
The isotropic pressure, Eq.\ (\ref{P_iso_relation}), leads to the well-known 
massless ideal gas relation, 
\begin{eqnarray}
\hat{P}\left( \alpha _{RS},\beta _{RS},\xi \right) &\equiv &P_{0}\left(
\alpha _{RS},\beta _{RS}\right) R_{200}\left( \xi \right)  \notag \\
&=& \frac{\hat{e}\left( \alpha _{RS},\beta _{RS},\xi \right)}{3}\;.
\end{eqnarray}
This is similar to
\begin{equation}
P_{0}\left( \alpha _{RS},\beta _{RS}\right) \equiv \frac{n_{0}\left(\alpha _{RS},\beta _{RS}\right)}{\beta _{RS}}
= \frac{e_{0}\left( \alpha _{RS},\beta _{RS}\right)}{3}\;,
\end{equation}
which is obtained from Eq.\ (\ref{I_nq}).
All thermodynamic integrals and ratios $R_{nrq}$ are listed in Appendix \ref{appendix_aniso_integrals}. 
Note that, for the RS distribution function, $\hat{n}_{l}\equiv \hat{I}_{110}^{RS}=0$ 
and $\hat{M}\equiv \hat{I}_{210}^{RS}=0$.
This means that the fluid-dynamical flow velocity does not depend on
our choice of LR frame.

Since chemical potential and temperature are quantities defined exclusively
in thermodynamical equilibrium, the parameters of the anisotropic distribution function, 
$\alpha_{RS} $, $\beta _{RS}$, and $\xi $, have no real physical meaning. However, one can relate them
to chemical potential and temperature, or equivalently $\alpha _{0}$ and $\beta _{0}$, of a ``fictitious''
equilibrium state by imposing the so-called Landau matching conditions $(\hat{N}^{\mu}_{RS} -
N^{\mu}_0)u_{\mu} = 0$ and $(\hat{T}^{\mu\nu}_{RS} - T^{\mu \nu}_0)u_{\mu}
u_{\nu} = 0$, or
\begin{eqnarray}
\hat{n}\left( \alpha _{RS},\beta _{RS},\xi \right) &=&n_{0}\left( \alpha
_{0},\beta _{0}\right)\; ,  \label{matching_n} \\
\hat{e}\left( \alpha _{RS},\beta _{RS},\xi \right) &=&e_{0}\left( \alpha
_{0},\beta _{0}\right)\; .  \label{matching_e}
\end{eqnarray}
Now, using Eqs.\ (\ref{n_hat_RS}) and (\ref{e_hat_RS}) together with Eqs.\ (\ref{I_nq_massless}) and
(\ref{I_nrq_RS_massless}) we obtain
\begin{eqnarray}
\beta _{0} &=&\beta _{RS}\frac{R_{100}\left( \xi \right) }{R_{200}\left( \xi
\right) }\;,  \label{Landau_matching_beta0} \\
\lambda _{0} &=&\lambda _{RS}\frac{\left[ R_{100}\left( \xi \right) \right]^{4}}{
\left[ R_{200}\left( \xi \right) \right]^{3}}\;,
\label{Landau_matching_lambda0}
\end{eqnarray}
where $\lambda _{0}\equiv \exp  \alpha _{0} =\exp \left( \mu \beta _{0}\right) $ 
and $\lambda _{RS}\equiv \exp \alpha_{RS} =\exp \left( \mu _{RS}\beta _{RS}\right) $ denote the
corresponding fugacities. Thus, using Eq.\ (\ref{R_nrq_RS}) together with
these results we obtain the following general relation between the
thermodynamic integrals 
\begin{equation}
\hat{I}_{nrq}^{RS}\left( \alpha _{RS},\beta _{RS},\xi \right) =I_{nq}\left(
\alpha _{0},\beta _{0}\right) R_{nrq}\left( \xi \right) \frac{\left[
R_{200}\left( \xi \right) \right] ^{1-n}}{\left[ R_{100}\left( \xi \right) \right] ^{2-n}}\;.  \label{Matched_I_nrq_RS}
\end{equation}

However, in case the particle number is not conserved, i.e., $\hat{\alpha} = \alpha_{RS} = 0$, 
the inverse temperature inferred from the Landau
matching condition (\ref{matching_e}) is
\begin{equation}
\beta _{0}=\frac{\beta _{RS}}{\left[ R_{200}\left( \xi \right) \right] ^{1/4}}\;.  \label{temp_matching_RS}
\end{equation}
Thus, similarly to Eq.\ (\ref{Matched_I_nrq_RS}) we now obtain 
\begin{equation}
\hat{I}_{nrq}^{RS}\left( \beta _{RS},\xi \right) =I_{nq}\left( \beta_{0}\right) 
\frac{R_{nrq}\left( \xi \right) }{\left[ R_{200}\left( \xi \right) \right] ^{(n+2)/4}}\;.  \label{Matched_I_nrq_RS_beta0}
\end{equation}
This result was also obtained e.g.\ in Refs.\ 
\cite{Florkowski:2010cf,Ryblewski:2010bs,Ryblewski:2011aq,Ryblewski:2012rr} and
Refs.\ \cite{Martinez:2009ry,Martinez:2010sc,Martinez:2010sd}. 

After applying the matching conditions we can calculate the
equilibrium pressure, $P_{0}=P_{0}\left( \alpha _{0},\beta _{0}\right) $,
and hence define the bulk viscous pressure 
\begin{equation}
\hat{\Pi}\equiv -\frac{1}{3}\hat{T}^{\mu \nu }\Delta _{\mu \nu }=\hat{P}
\left( \alpha _{RS},\beta _{RS},\xi \right) -P_{0}\left( \alpha _{0},\beta_{0}\right) \;,
\end{equation}
which vanishes for a massless ideal gas, $\lim_{m_{0}\rightarrow 0}\hat{\Pi}=0$.

\subsection{0+1 dimensional boost-invariant expansion}
\label{Results_BJ}

We now investigate how the solution of the fluid-dynamical equations of motion is
influenced by the choice of moment to close them.
We study this for a very simple case only, the 0+1 dimensional boost-invariant
expansion of matter, known as Bjorken flow \cite{Bjorken:1982qr}.
To this end, it is advantageous to transform the usual space-time coordinates $(t,z)$ to proper
time $\tau =\sqrt{t^{2}-z^{2}}$ and space-time rapidity 
$\eta_s =\frac{1}{2}\ln \frac{t+z}{t-z}$. The inverse transformation then reads
$t=\tau \cosh \eta_s $ and $z=\tau \sinh\eta_s $. 
In Bjorken flow, the velocity of matter is given by $v_{z}\equiv z/t=\tanh \eta_s $, such that
\begin{eqnarray}  \label{u_BJ}
u^{\mu } &\equiv&\left( \frac{t}{\tau} ,0,0,\frac{z}{\tau}\right) 
= \left(\cosh \eta_s ,0,0,\sinh \eta_s \right) \;, \\
l^{\mu } &\equiv&\left( \frac{z}{\tau} ,0,0,\frac{t}{\tau} \right) 
= \left(\sinh \eta_s ,0,0,\cosh \eta_s \right) \; .  \label{l_BJ}
\end{eqnarray}
Then, $D\equiv u^{\mu }\partial _{\mu }
=\frac{\partial }{\partial \tau }$, $D_{l}\equiv -l^{\mu }\partial _{\mu }
=-\frac{\partial }{\tau \partial \eta_s }$, and $Du^{\mu }=Dl^\mu = 0$, $D_{l}u^{\mu }=-\frac{1}{\tau }
l^{\mu }$, $D_{l}l^{\mu }=-\frac{1}{\tau }u^{\mu }$, while 
$\tilde{\theta}=\tilde{\theta}_l=0$. Inserting Eq.\ (\ref{I_ij_tens_expanded}) into Eq.\ (\ref{General_aniso_scalar}),
and using the fact that in Bjorken flow all thermodynamic quantities are independent of $\eta_s$, we obtain
\begin{equation}
\frac{\partial \hat{I}_{i+j,j,0}}{\partial \tau }+\frac{1}{\tau }\left[
\left( j+1\right) \hat{I}_{i+j,j,0}+\left( i-1\right) \hat{I}_{i+j,j+2,0}
\right] =\hat{\mathcal{C}}_{i-1,j} \; . \label{Main_eq_motion}
\end{equation}
We also assume that the collision term is given by
the relaxation-time approximation (RTA) \cite{Bhatnagar:1954zz}, i.e., 
$C[ \hat{f}] =-\frac{E_{\mathbf{k}u}}{\tau _{eq}}
\left( \hat{f}_{0\mathbf{k}}-f_{0\mathbf{k}}\right)$. 
This means that the anisotropic distribution function $\hat{f}_{0\mathbf{k}}$ 
is assumed to approach
the equilibrium distribution $f_{0\mathbf{k}}$ on a timescale set by $\tau_{eq}$. 
Thus, in RTA the r.h.s.\ of Eq.\ (\ref{Main_eq_motion}) reads
\begin{equation}
\hat{\mathcal{C}}_{i-1,j}=-\frac{1}{\tau _{eq}} 
\left( \hat{I}_{i+j,j,0}-I_{i+j,j,0}\right) \;.  \label{Coll_Int_RTA}
\end{equation}
For $\tau_{eq}$ we will either use a constant value 
or parametrize it using the relation between $\tau_{eq}$ and shear viscosity \cite{Florkowski:2013lza,Florkowski:2013lya}
\begin{equation} \label{tau_eq}
\tau_{eq}(\tau) = 5\beta_0(\tau)\frac{\eta}{s} \; ,
\end{equation}
where $\eta/s$ denotes the ratio of shear viscosity to entropy density, which we assume to
take a constant value.

In order to obtain the fluid-dynamical equations of motion for the
RS distribution function we substitute $\hat{I}_{nrq}\rightarrow \hat{I}_{nrq}^{RS}$ 
in Eqs.\ (\ref{Main_eq_motion}) and (\ref{Coll_Int_RTA}).
Furthermore, using the matching conditions Eqs.\ (\ref{matching_n}) and 
(\ref{matching_e}), the conservation equations for particle number 
(\ref{conservation_eq_particle}) and energy
(\ref{conservation_eq_energy}) read 
\begin{equation}
\frac{\partial n_{0}\left( \alpha _{0},\beta _{0}\right) }{\partial \tau }+
\frac{1}{\tau }n_{0}\left( \alpha _{0},\beta _{0}\right) =0 \;,
\label{BJ_n_cons}
\end{equation}
and
\begin{equation}
\frac{\partial e_{0}\left( \alpha _{0},\beta _{0}\right) }{\partial \tau }
+\frac{1}{\tau }\left[ e_{0}\left( \alpha _{0},\beta _{0}\right) +\hat{P}_{l}
\left( \alpha _{RS},\beta _{RS},\xi \right) \right] =0 \; .
\label{BJ_e_cons}
\end{equation}
The conservation of momentum in the direction of the anisotropy, Eq.\ (\ref{conservation_eq_momenta_l}), 
leads to $D\hat{M}=-2\hat{M}/\tau$, but since $\hat{M} =0$ for the RS distribution function, this equation does
not provide any additional information. Likewise,
the conservation of momentum in the direction transverse to the anisotropy, 
Eq.\ (\ref{conservation_eq_momenta_t}), gives $\tilde{\nabla}^{\alpha }\hat{P}_{\perp }=0$, which
also contains no additional information since the system is homogeneous in the transverse direction.

Now we discuss various choices to close the conservation equations (\ref{BJ_n_cons}) and (\ref{BJ_e_cons}).
As mentioned above, in principle there are infinitely many equations that can be 
selected from the hierarchy of balance equations (\ref{Hierarchy_moments}) to serve this purpose. 
Here we restrict ourselves to a few representative examples. These follow from Eqs.\ (\ref{Main_eq_motion}) 
and (\ref{Coll_Int_RTA}) by choosing particular values for the indices $i$ and $j$.

(i) $i=0$, $j=2$: This choice gives the time-evolution equation for the
longitudinal pressure $\hat{P}_{l}\left( \alpha _{RS},\beta _{RS},\xi\right) $,
\begin{equation}
\frac{\partial \hat{P}_{l}}{\partial \tau }+\frac{1}{\tau }\left( 3\hat{P}_{l}-\hat{I}_{240}^{RS}\right) 
=-\frac{1}{\tau _{eq}}\left( \hat{P}_{l}-P_{0}\right) \; .  \label{BJ_Pl_relax}
\end{equation}
Here we can explicitly express $\hat{P}_{l}$ and $\hat{I}_{240}^{RS}$ via
Eq.\ (\ref{Matched_I_nrq_RS}),
\begin{eqnarray}
\hat{P}_{l}\left( \alpha _{RS},\beta _{RS},\xi \right) &=& 
e_{0}\left( \alpha_{0},\beta _{0}\right) \frac{R_{220}\left(\xi \right)}{R_{200}\left( \xi\right) }\;,  
\label{Matched_Pl_RS} \\
\hat{I}_{240}^{RS}\left( \alpha _{RS},\beta _{RS},\xi \right) &=& 
e_{0}\left(\alpha _{0},\beta _{0}\right) \frac{R_{240}\left(\xi \right)}{R_{200}\left( \xi \right) }\;,  
\label{Matched_I240_RS}
\end{eqnarray}
where the $R_{nrq}$ are listed in 
Eqs.\ (\ref{R_200}), (\ref{R_220}), and (\ref{R_240}). 
Note that Eqs.\ (\ref{Matched_Pl_RS}) and 
(\ref{Matched_I240_RS}) are formally unchanged when expressing $\hat{P}_{l}$ 
and $\hat{I}_{240}^{RS}$ through Eq.\ (\ref{Matched_I_nrq_RS_beta0}). 
This means that the dynamical equation for $\hat{P}_{l}$ is a very special choice for closure 
of the conservation equations, since Eq.\ (\ref{BJ_Pl_relax}) is formally independent 
of whether we conserve particle number or not.
Furthermore, as it should be, in the limit of small deviations from local thermodynamical equilibrium, 
i.e., $\xi \ll 1$, Eqs.\ (\ref{BJ_e_cons}) and (\ref{BJ_Pl_relax}) lead precisely to
the equations of motion of second-order fluid dynamics as obtained in Refs.~\cite{Jaiswal:2013npa,Jaiswal:2014isa,Denicol:2012es,Denicol:2012cn}, 
for more details see App.\ \ref{2nd_order_hydro}.
  
(ii) $i=1$, $j=0$: This choice was made e.g.\ in Refs.\ \cite{Martinez:2009ry,Martinez:2010sc,Martinez:2010sd}.
It is possible only if particle number is not conserved
(such that the chemical potential is always zero),
\begin{equation}  \label{BJ_n_relax}
\frac{\partial \hat{n}}{\partial \tau } + \frac{1}{\tau }\hat{n}
=-\frac{1}{\tau _{eq}}\left( \hat{n}-n_{0}\right)\;.
\end{equation}

(iii) $i=3$, $j=0$: This choice is analogous to the one of Israel and Stewart \cite{Israel:1979wp}, which use the 
second moment of the Boltzmann equation to close the conservation equations,
\begin{equation}
\frac{\partial \hat{I}_{300}^{RS}}{\partial \tau }+\frac{1}{\tau }\left( 
\hat{I}_{300}^{RS}+2\hat{I}_{320}^{RS}\right) 
=-\frac{1}{\tau _{eq}}\left( \hat{I}_{300}^{RS}-I_{300}\right) \;.  \label{BJ_I300_relax}
\end{equation}

(iv) $i=1$, $j=2$: This choice is analogous to the previous one, in the sense that it also results from the second
moment of the Boltzmann equation,
\begin{equation}
\frac{\partial \hat{I}_{320}^{RS}}{\partial \tau }+\frac{3}{\tau }\hat{I}_{320}^{RS}
=-\frac{1}{\tau _{eq}}\left( \hat{I}_{320}^{RS}-I_{320}
\right) \;.  \label{BJ_I320_relax}
\end{equation}

Note that according to Eq.\ (\ref{Hierarchy_projections}) the cases (iii) and (iv) 
follow from different projections of $\partial _{\lambda }
\hat{\mathcal{I}}_{00}^{\mu \nu \lambda }=\hat{\mathcal{C}}_{00}^{\mu \nu }$. 
Also note that the choice $i=2$, $j=1$ is trivial since $\hat{I}_{310}^{RS}=0$.

Finally, we also tested the following choices:

(v) $i=0$, $j=0$:
\begin{equation}
\frac{\partial \hat{I}_{000}^{RS}}{\partial \tau }+\frac{1}{\tau }
\left( \hat{I}_{000}^{RS}-\hat{I}_{020}^{RS}\right) =-\frac{1}{\tau _{eq}}
\left( \hat{I}_{000}^{RS}-I_{000}\right) \;.  \label{BJ_I000_relax}
\end{equation}

(vi) $i=0$, $j=4$:
\begin{equation}
\frac{\partial \hat{I}_{440}^{RS}}{\partial \tau }+\frac{1}{\tau }
\left( 5\hat{I}_{440}^{RS}-\hat{I}_{460}^{RS}\right) 
= -\frac{1}{\tau _{eq}}\left( \hat{I}_{440}^{RS}-I_{440}\right) \; ,  \label{BJ_I440_relax}
\end{equation}

(vii) $i=1$, $j=4$:
\begin{equation}
\frac{\partial \hat{I}_{540}^{RS}}{\partial \tau }+\frac{5}{\tau }\hat{I}
_{540}^{RS}=-\frac{1}{\tau _{eq}}\left( \hat{I}_{540}^{RS}-I_{540}\right)\; .  \label{BJ_I540_relax}
\end{equation}

\section{Results and discussions}
\label{results}

In this section, we solve the conservation equations (\ref{BJ_n_cons}) and (\ref{BJ_e_cons})
and study the impact of different ways to close them, i.e., choosing one of the moment equations 
(\ref{BJ_Pl_relax}), (\ref{BJ_n_relax}), (\ref{BJ_I300_relax}), 
\ref{BJ_I320_relax}), (\ref{BJ_I000_relax}),  (\ref{BJ_I440_relax}), or (\ref{BJ_I540_relax}). We will also compare the 
fluid-dynamical solutions to the solution of the Boltzmann equation, in order to identify which one of the 
moment equations gives the best agreement with the latter.

We always initialize the system with temperature $T_0=300$ MeV at initial time $\tau_0 = 1.0$ fm,
for three choices of the initial anisotropy, $\xi(\tau_{0})\equiv\xi_{0} = \left\{ 0,10,100\right\}$. 
We investigate separately the cases with and without particle-number conservation. 
In the case with particle-number conservation, we take an initial fugacity $\lambda_0 = 1$.
The initial value of the temperature and the anisotropy parameter 
are shown in the headlines of the following figures. If the particle number is conserved, the initial
fugacity is also shown. We use either a constant relaxation time $\tau_{eq} = 1$ fm, 
or the temperature-dependent one from Eq.~(\ref{tau_eq}) with $\eta/s = \left\{1/4\pi, 10/4\pi, 100/4\pi \right\}$.

For the comparison of the choice of moment in Sec.\ \ref{results_choice}, we also solve the conservation
equations for an ideal fluid, $\frac{\partial e_{0}}{\partial \tau }+\frac{1}{\tau }\left( e_{0} + P_0\right) =0$, 
where $P_0 \equiv n_0 T_0 = e_0/3$,
together with $\frac{\partial n_{0}}{\partial \tau }+ \frac{1}{\tau }n_{0}=0$. 
Note that in this case the two conservation equations are independent from
each other, hence if the system was initially in chemical
equilibrium it will stay in chemical equilibrium.
Furthermore, in the case of an ideal fluid $\xi(\tau)=0$, the time evolution of the 
fugacity is simply given by $\lambda(\tau) = 1$, while the pressure is necessarily 
isotropic, hence  $\hat{P}_l(\tau)/\hat{P}_\perp(\tau) = 1$. 
These constant horizontal lines are redundant and will not be shown in the respective figures.

\subsection{The choice of moment}
\label{results_choice} 

%%%%%%%%%%%%%%%%%%%%% FIGURE %%%%%%%%%%%%%%%%%%%%%%%%%%%%%%%%
\begin{figure*}[htb!]
\centering
\includegraphics[width=7.8cm]{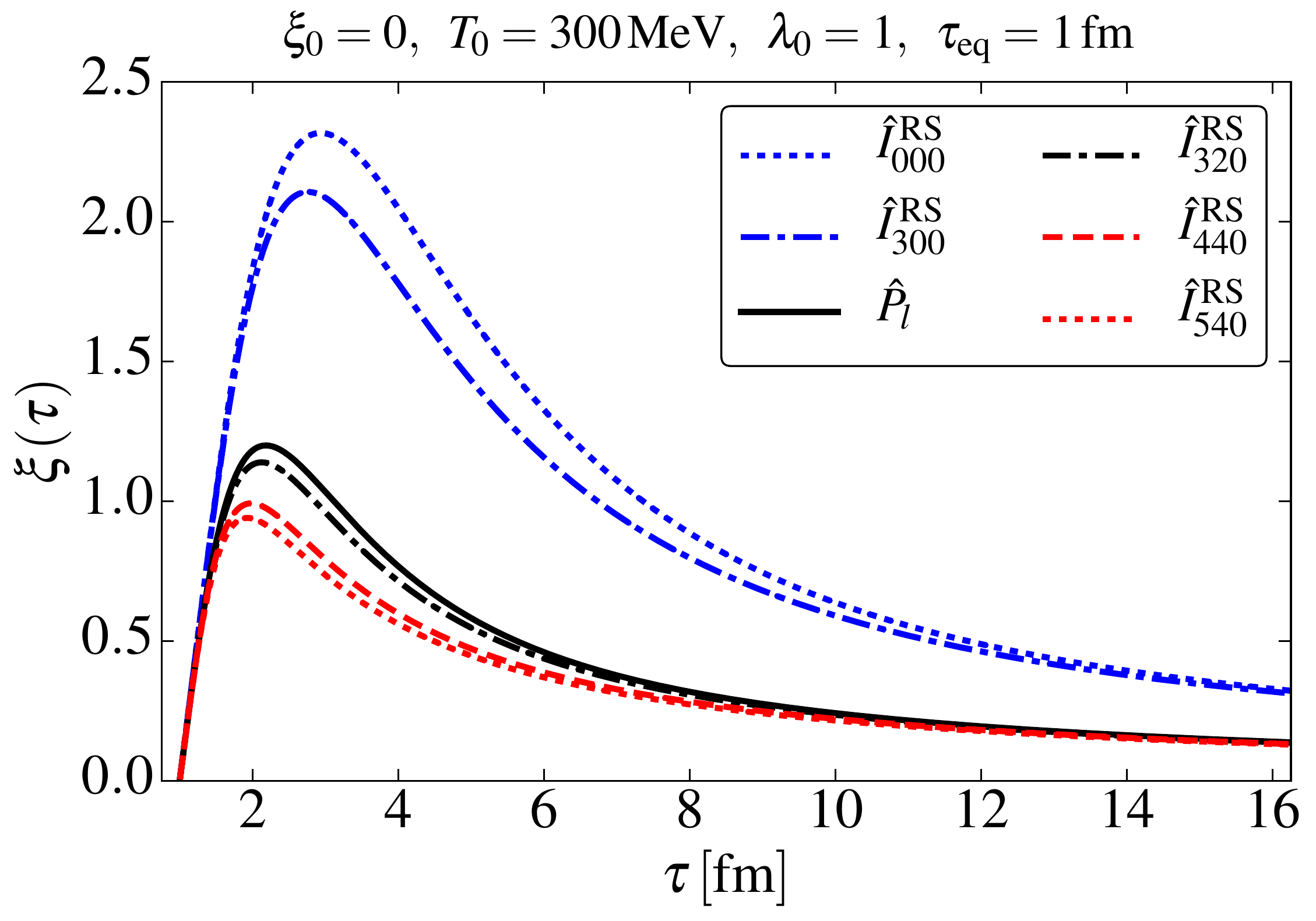} 
\includegraphics[width=7.8cm]{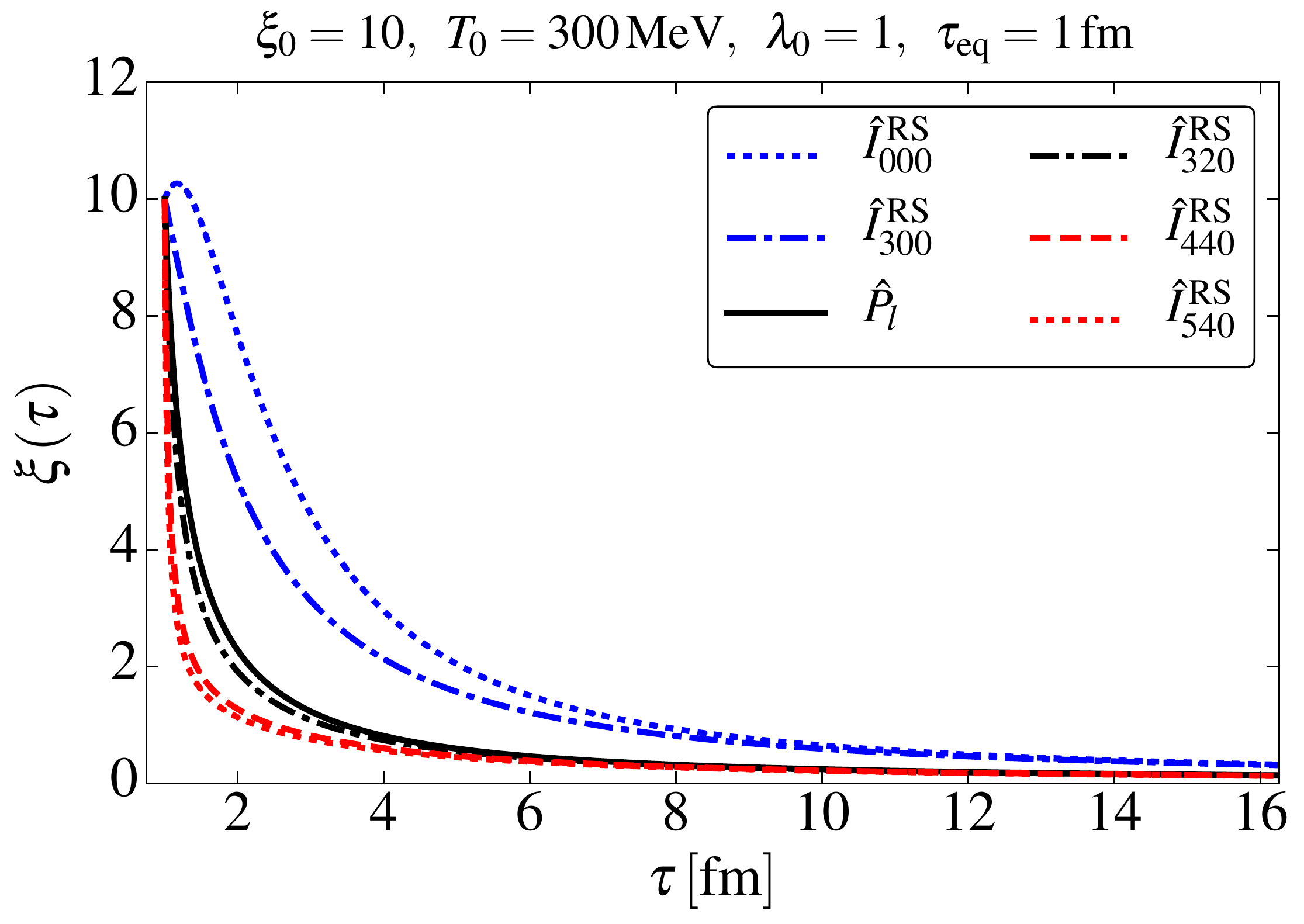}   
\includegraphics[width=7.8cm]{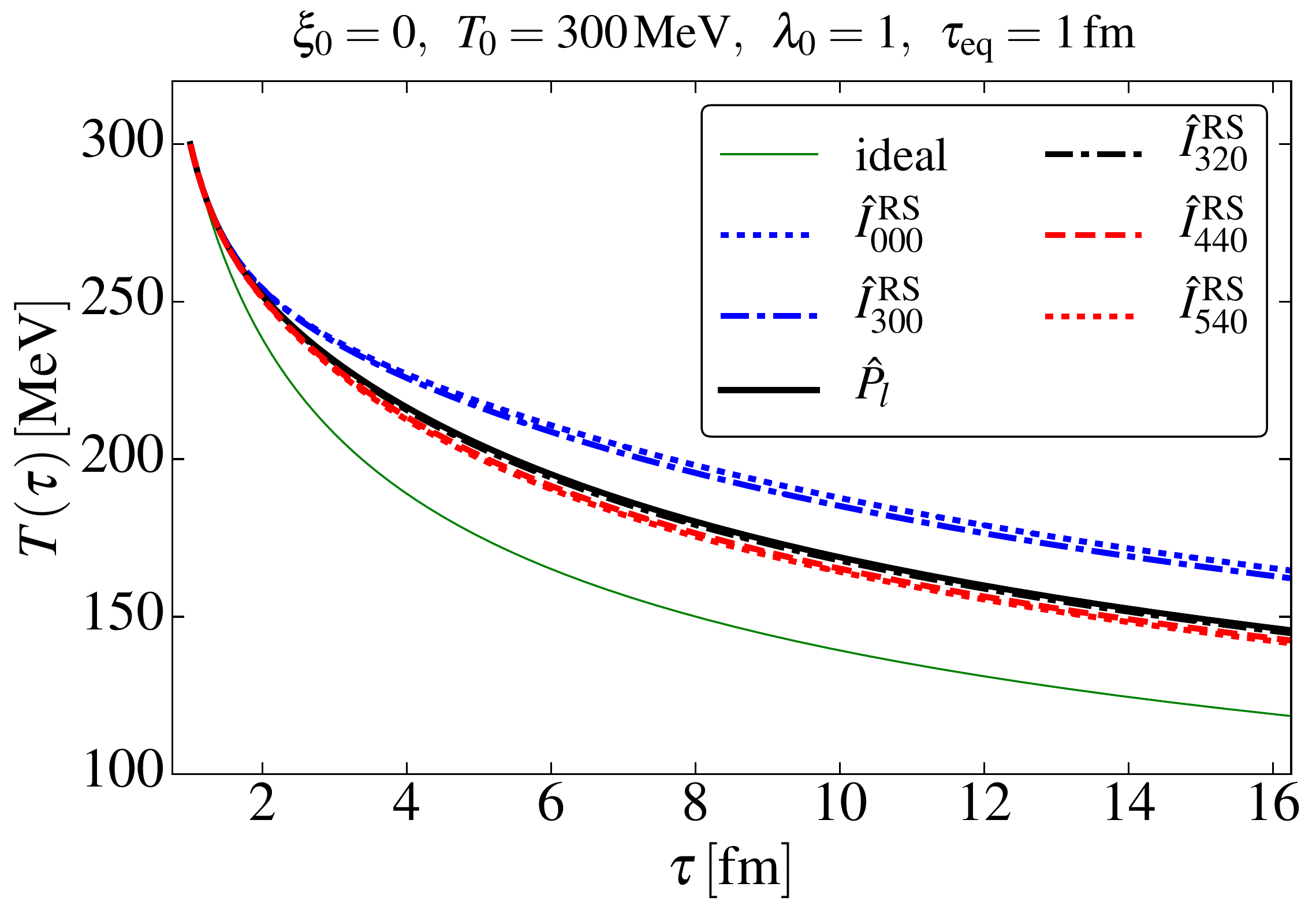}
\includegraphics[width=7.8cm]{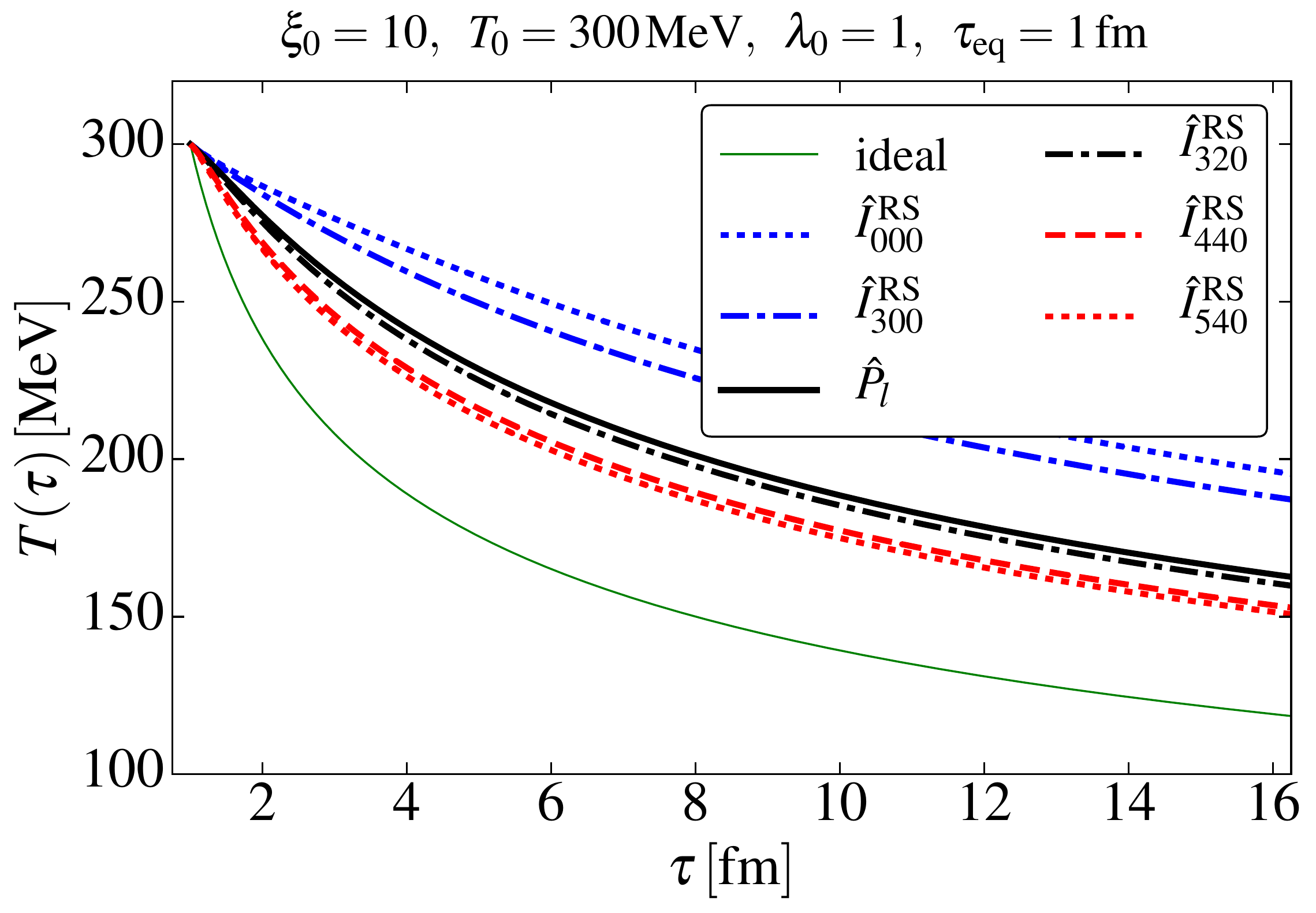}
\includegraphics[width=7.8cm]{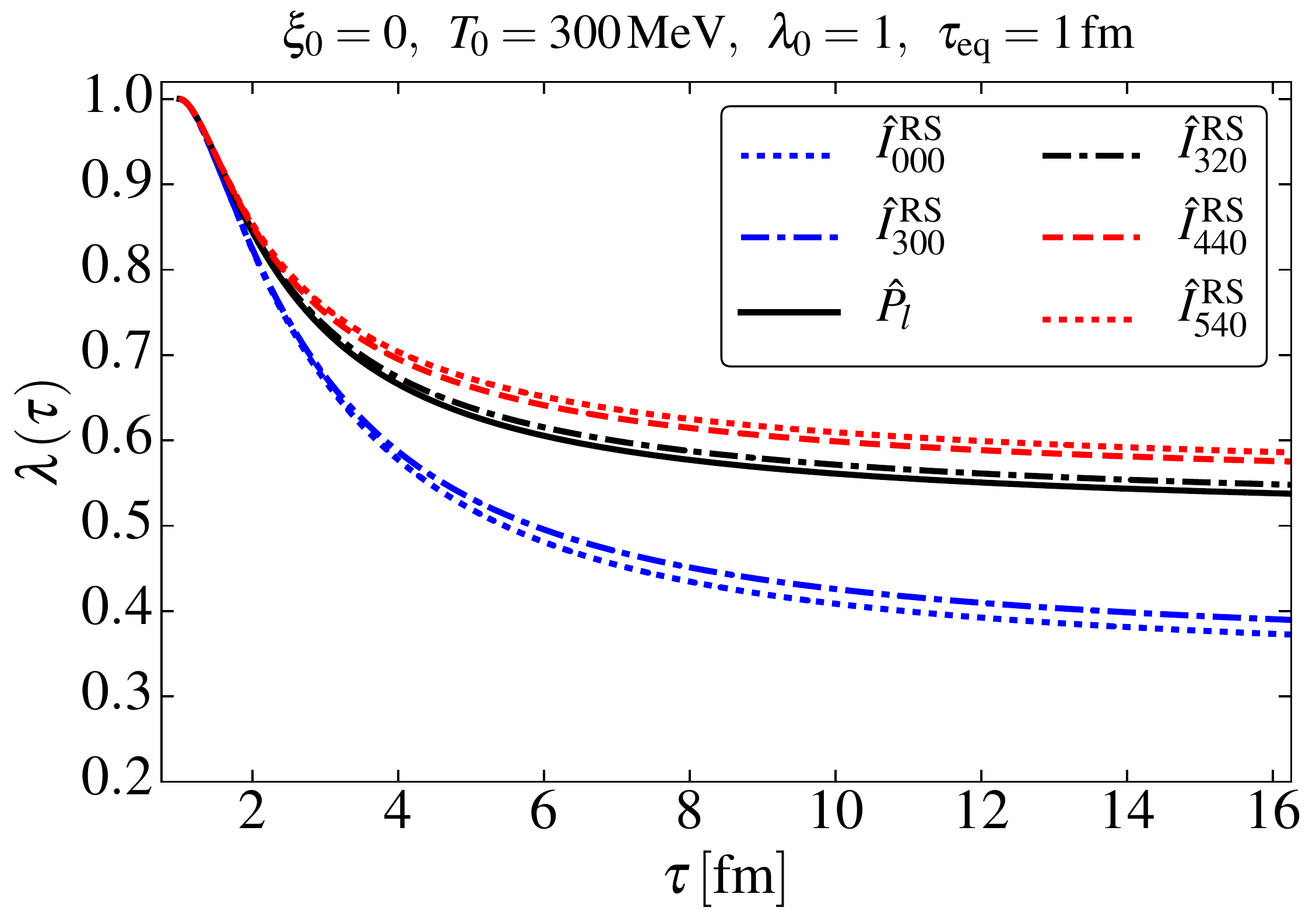} 
\includegraphics[width=7.8cm]{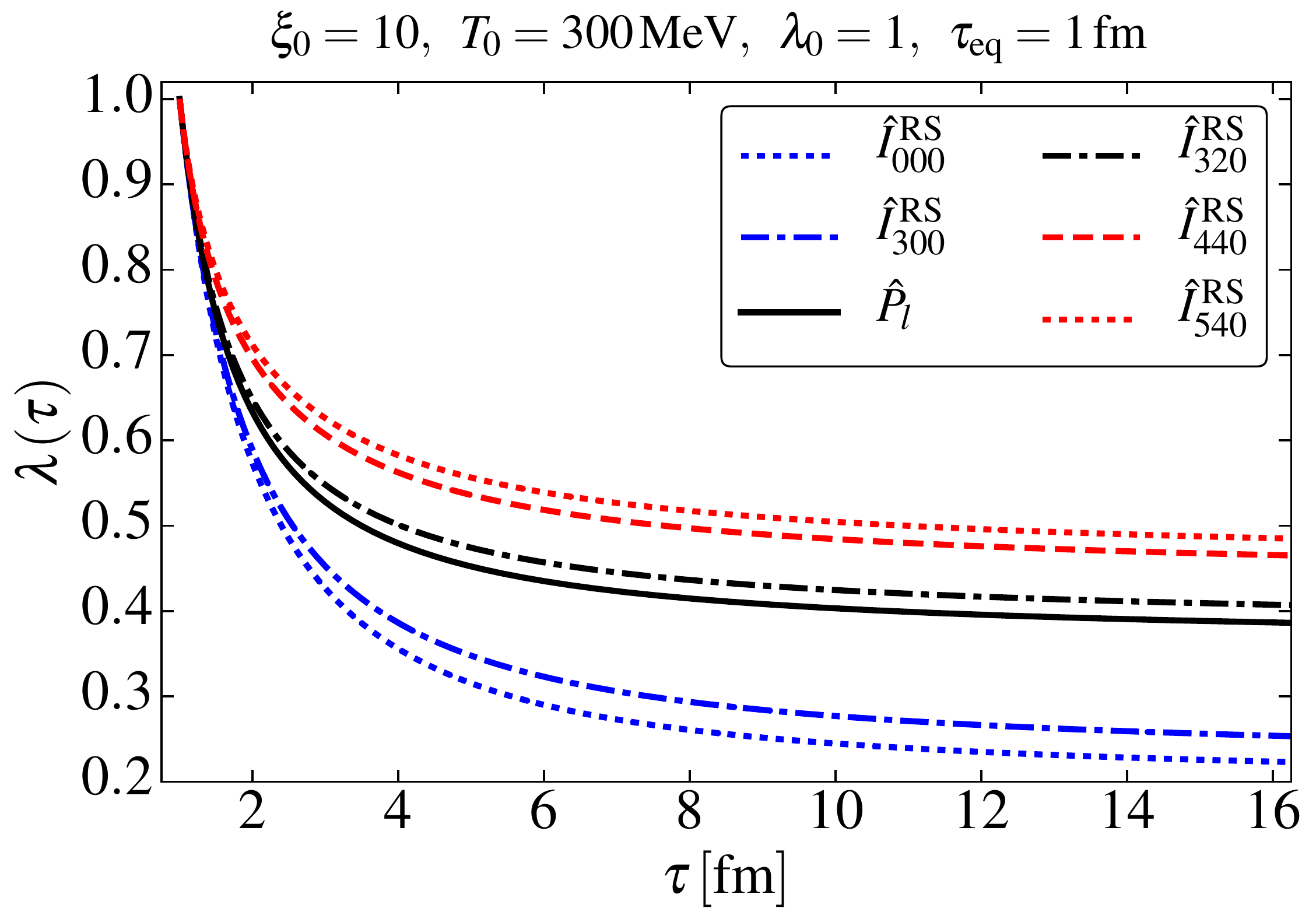} 
\includegraphics[width=7.8cm]{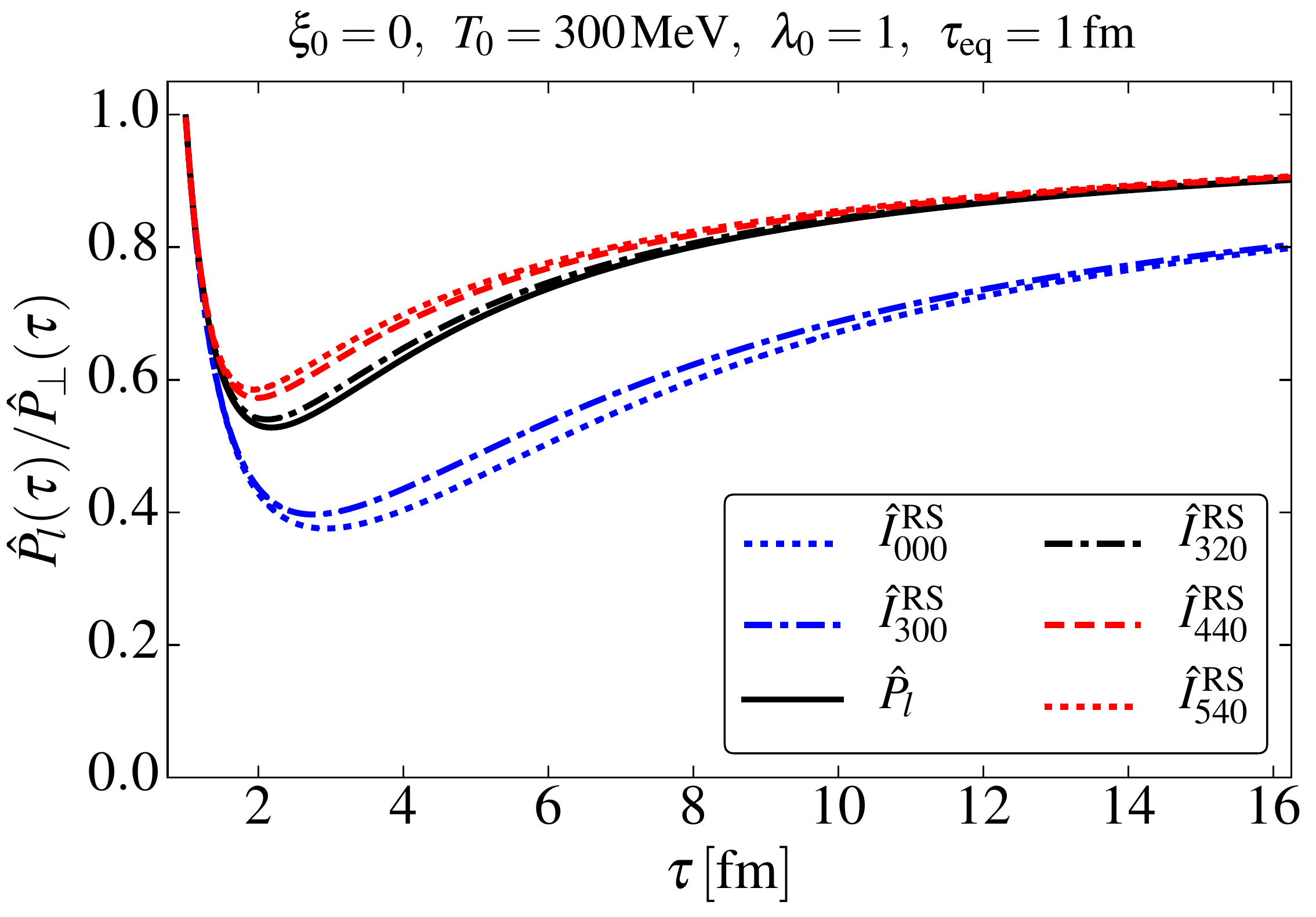} 
\includegraphics[width=7.8cm]{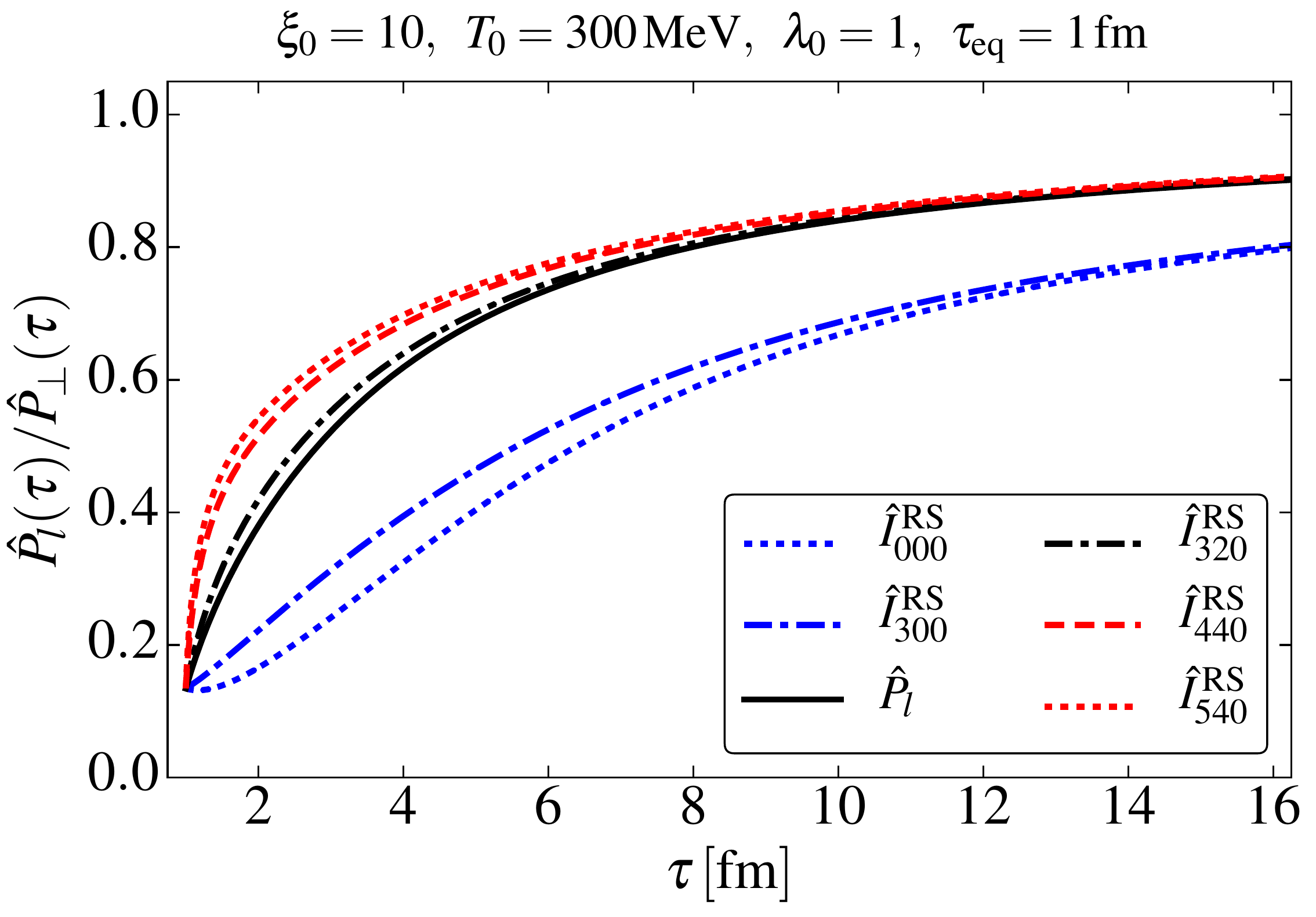}
\vspace{-0.2cm} 
\caption{(Color online) From top to bottom: the evolution of the anisotropy
parameter $\protect\xi$, temperature $T$, fugacity $\protect\lambda$, and
the ratio of longitudinal and transverse pressure components $\hat{P}_l/\hat{P}_{\perp}$  
as a function of proper time $\tau$. The thin green line in the figures in the second row
represents the temperature evolution for an ideal fluid. 
The other lines are the solution of the conservation equations (\protect\ref{BJ_n_cons}), (\protect\ref{BJ_e_cons}) 
closed by different moment equations: the dotted blue line ($\hat{I}^{RS}_{000}$) 
corresponds to Eq.\ (\protect\ref{BJ_I000_relax}), the dash-dotted blue
line ($\hat{I}^{RS}_{300}$) to Eq.\ (\protect\ref{BJ_I300_relax}),
the full black line ($\hat{P}_l$) to Eq.\ (\ref{BJ_Pl_relax}), the dash-dotted black line 
($\hat{I}^{RS}_{320}$) to Eq.\ (\ref{BJ_I320_relax}), the dashed  red ($\hat{I}^{RS}_{440}$)
to Eq.\ (\ref{BJ_I440_relax}), and the dotted red line ($\hat{I}^{RS}_{540}$) to Eq.\ 
(\ref{BJ_I540_relax}), respectively.}
\label{fig:1set}
\end{figure*}
%%%%%%%%%%%%%%%%%%%%% FIGURE %%%%%%%%%%%%%%%%%%%%%%%%%%%%%%%% 

%%%%%%%%%%%%%%%%%%%%% FIGURE %%%%%%%%%%%%%%%%%%%%%%%%%%%%%%%%
\begin{figure*}[htb!]
\centering
\includegraphics[width=7.8cm]{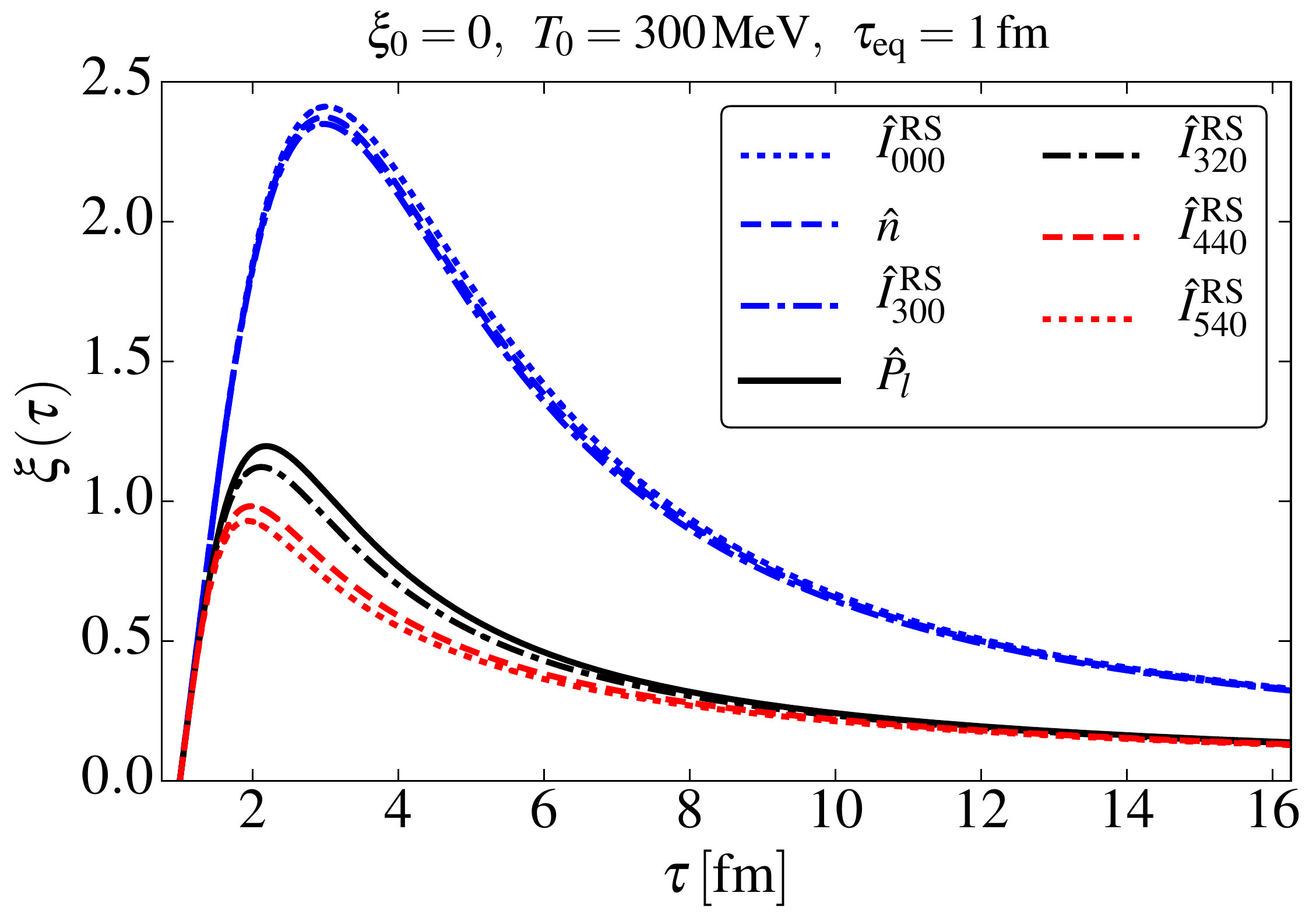} 
\includegraphics[width=7.8cm]{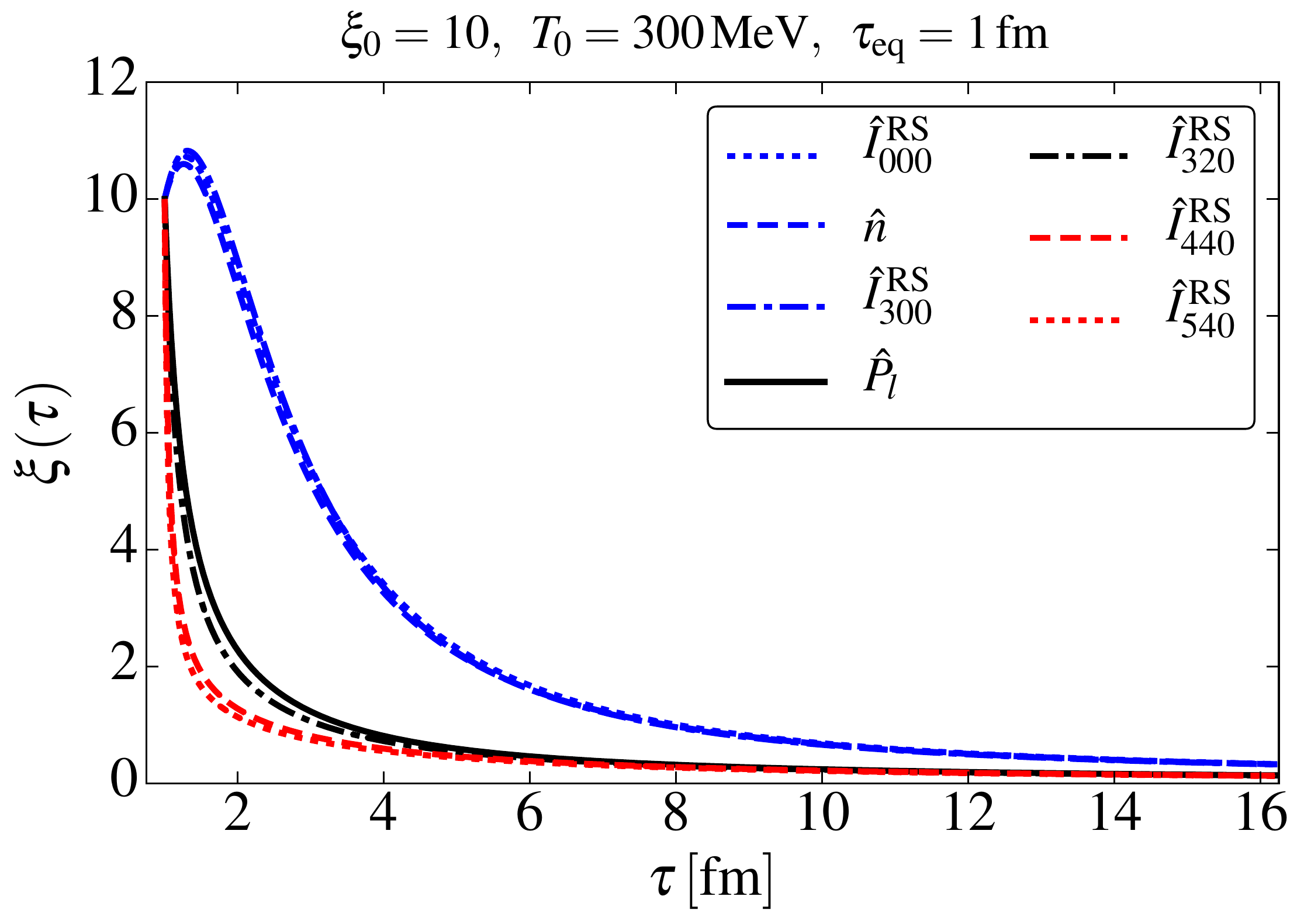}  
\includegraphics[width=7.8cm]{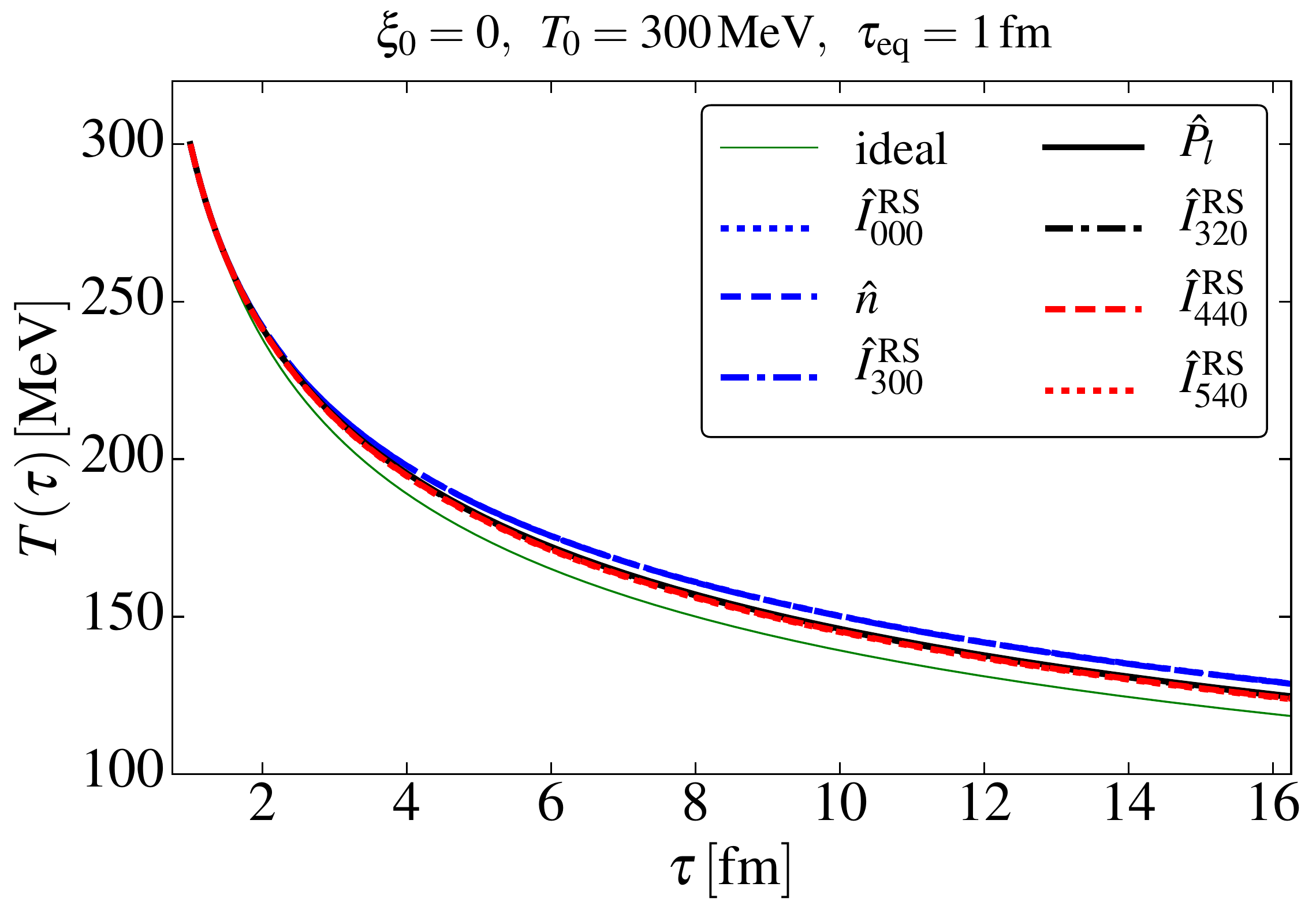} 
\includegraphics[width=7.8cm]{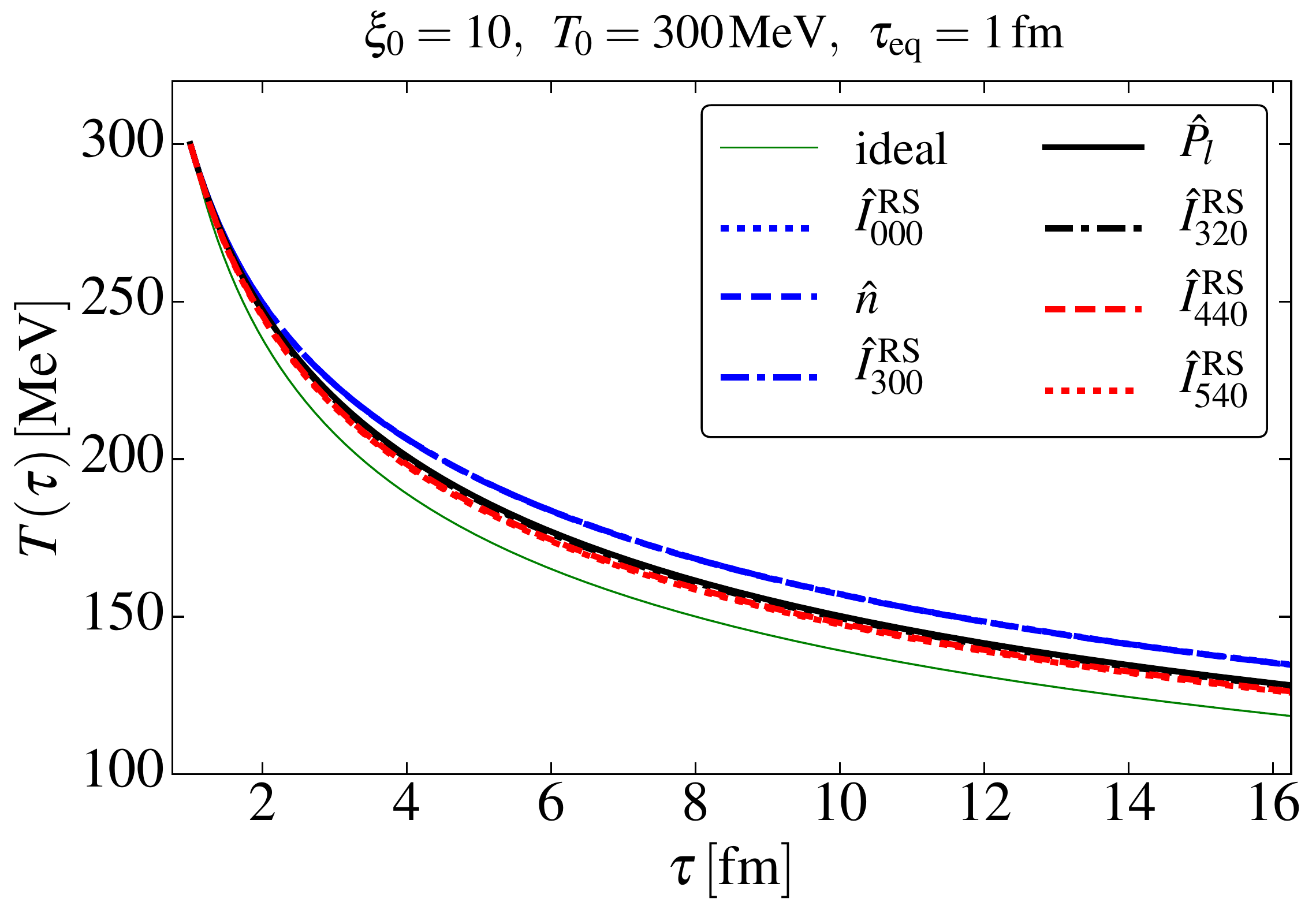} 
\includegraphics[width=7.8cm]{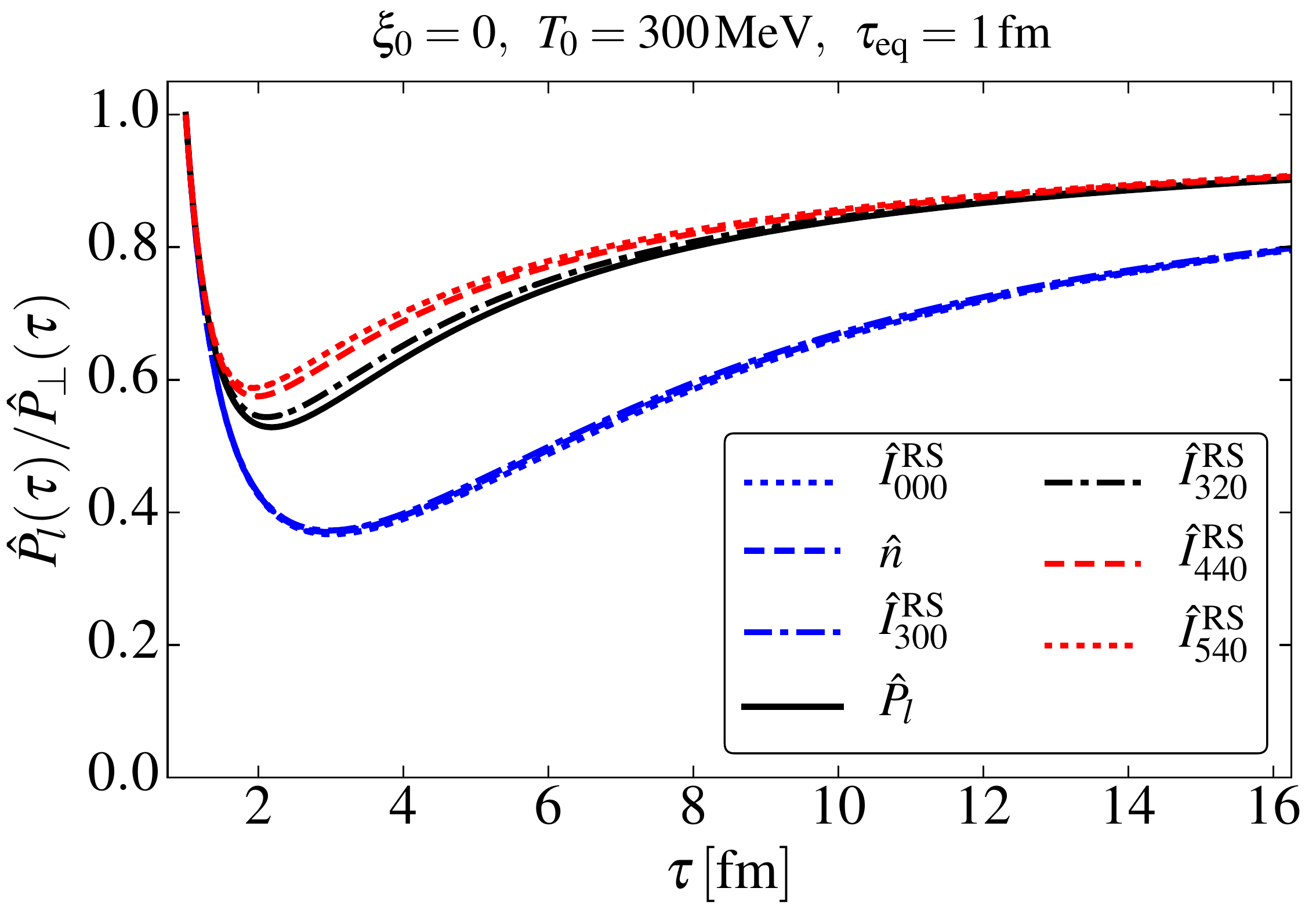} 
\includegraphics[width=7.8cm]{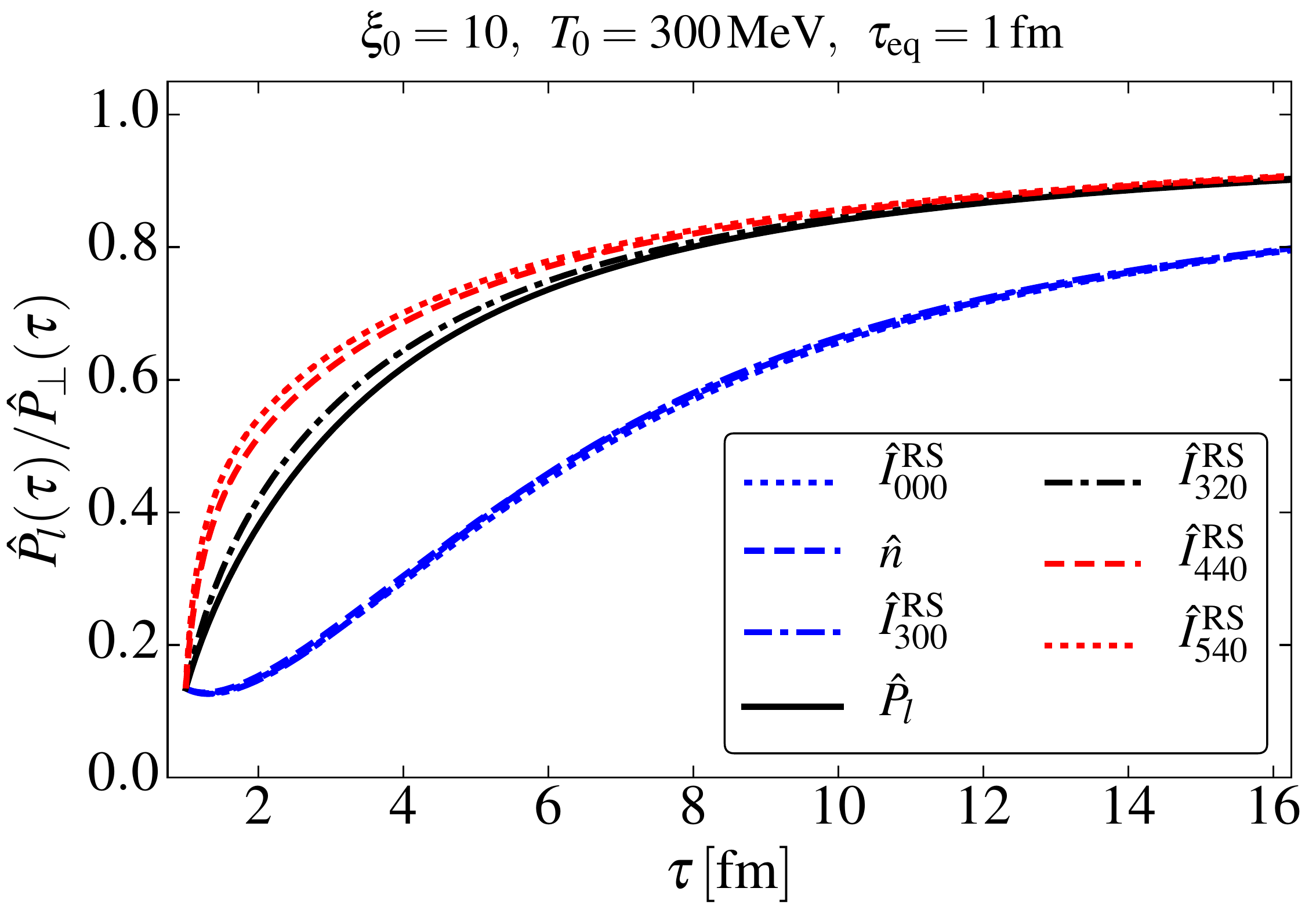}
\vspace{-0.2cm} 
\caption{(Color online) Similar to Fig.\ \protect\ref{fig:1set}, but for the case without
particle-number conservation, such that always $\lambda (\tau)=1$ (and thus not explicitly shown). 
The only difference to Fig.\ \protect\ref{fig:1set} is that now Eq.\ (\ref{BJ_n_relax}) is available
to close the energy-conservation equation. The respective dashed blue line is labelled $\hat{n}$.}
\label{fig:2set}
\end{figure*}
%%%%%%%%%%%%%%%%%%%%% FIGURE %%%%%%%%%%%%%%%%%%%%%%%%%%%%%%%% 

%%%%%%%%%%%%%%%%%%%%% FIGURE %%%%%%%%%%%%%%%%%%%%%%%%%%%%%%%%
\begin{figure*}[htb!]
\centering
\includegraphics[width=7.8cm]{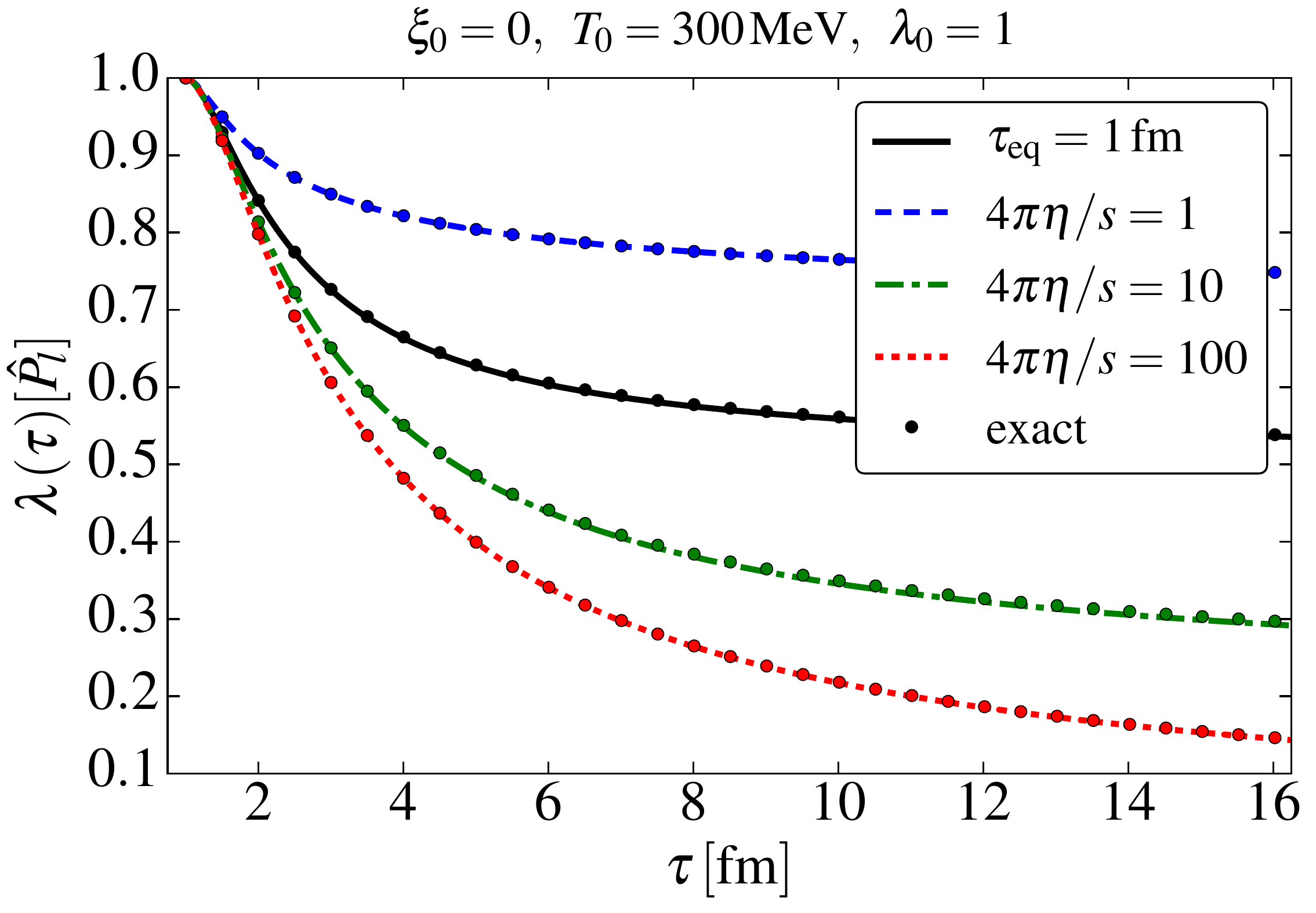} 
\includegraphics[width=7.8cm]{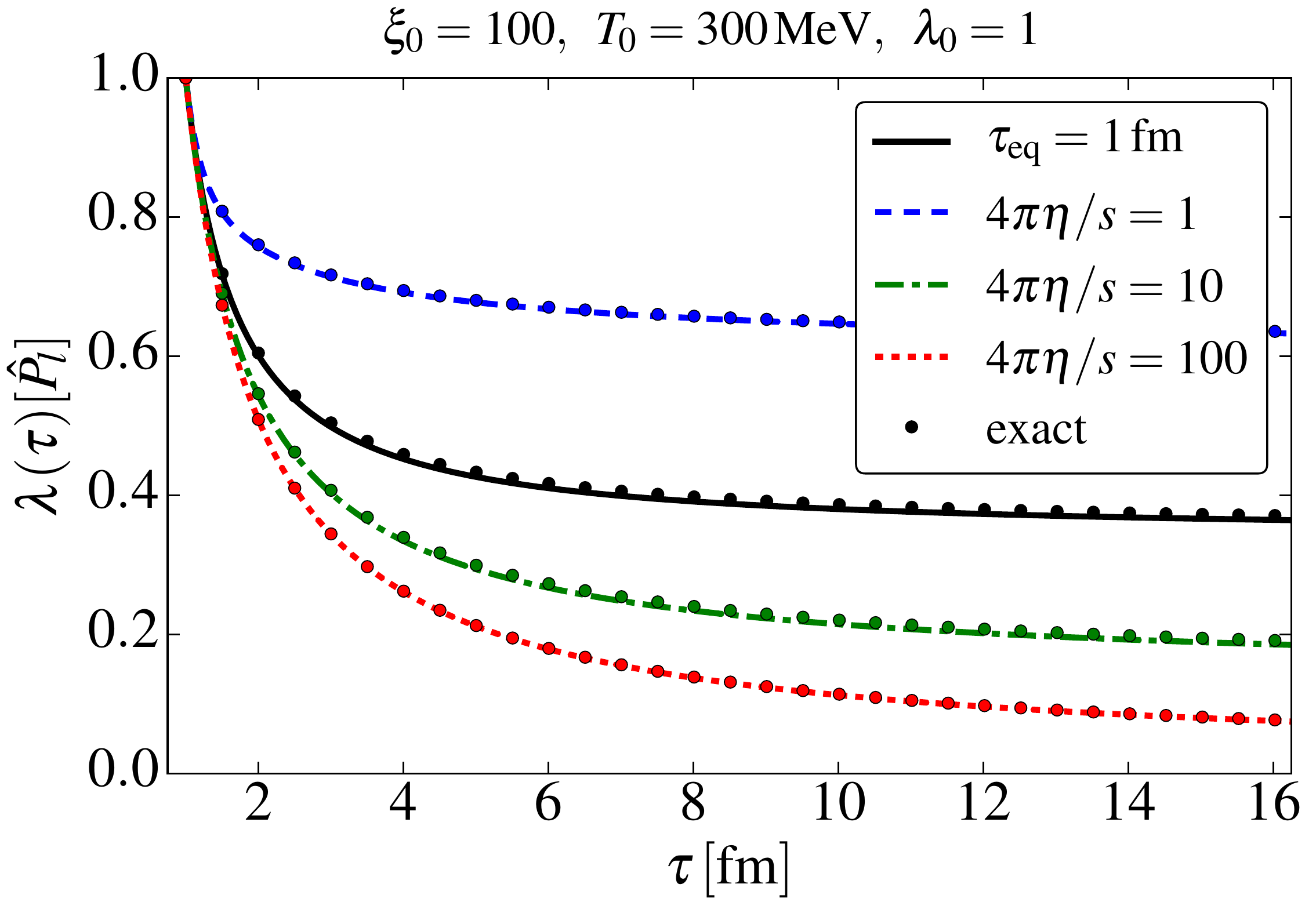}  
\includegraphics[width=7.8cm]{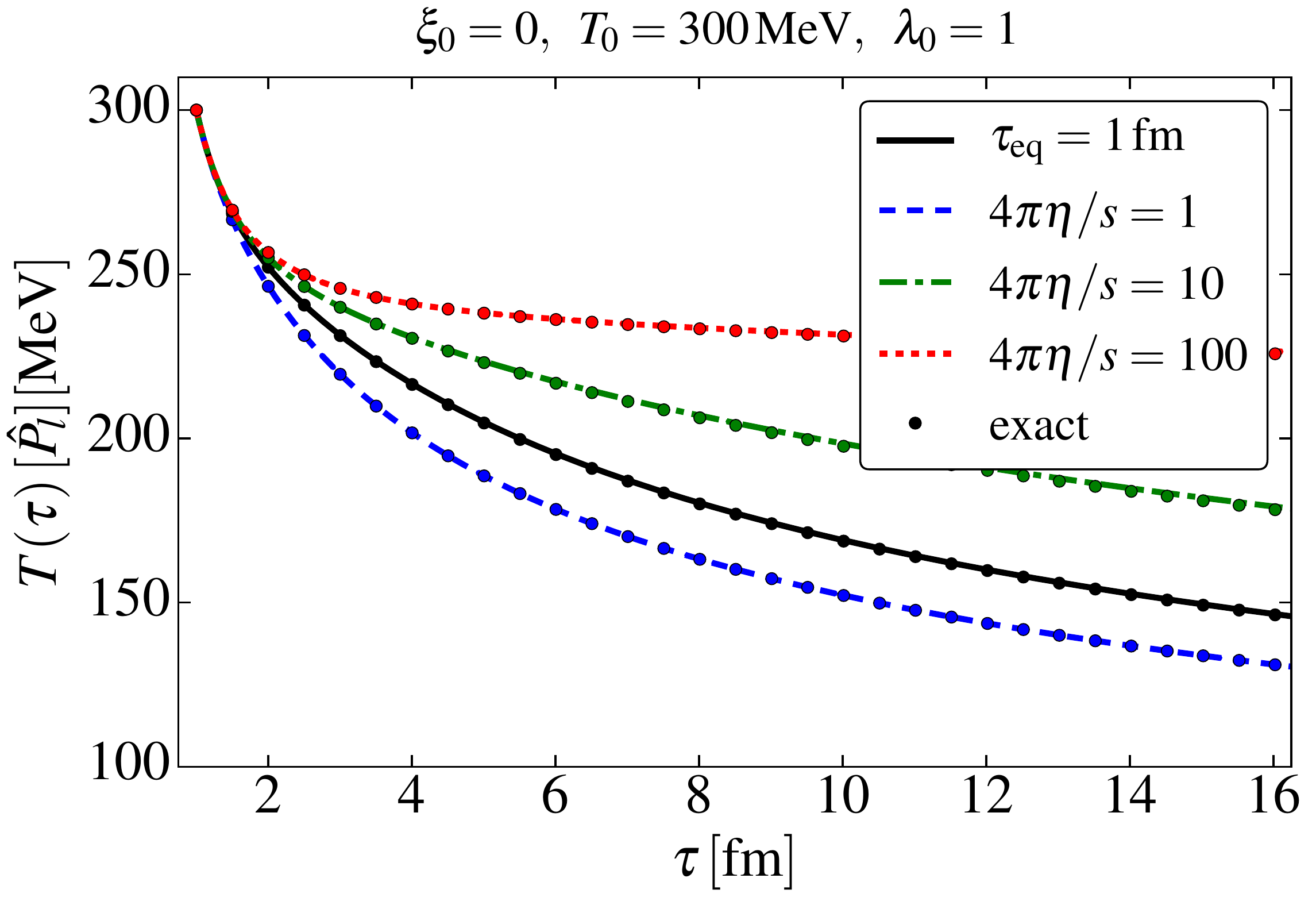} 
\includegraphics[width=7.8cm]{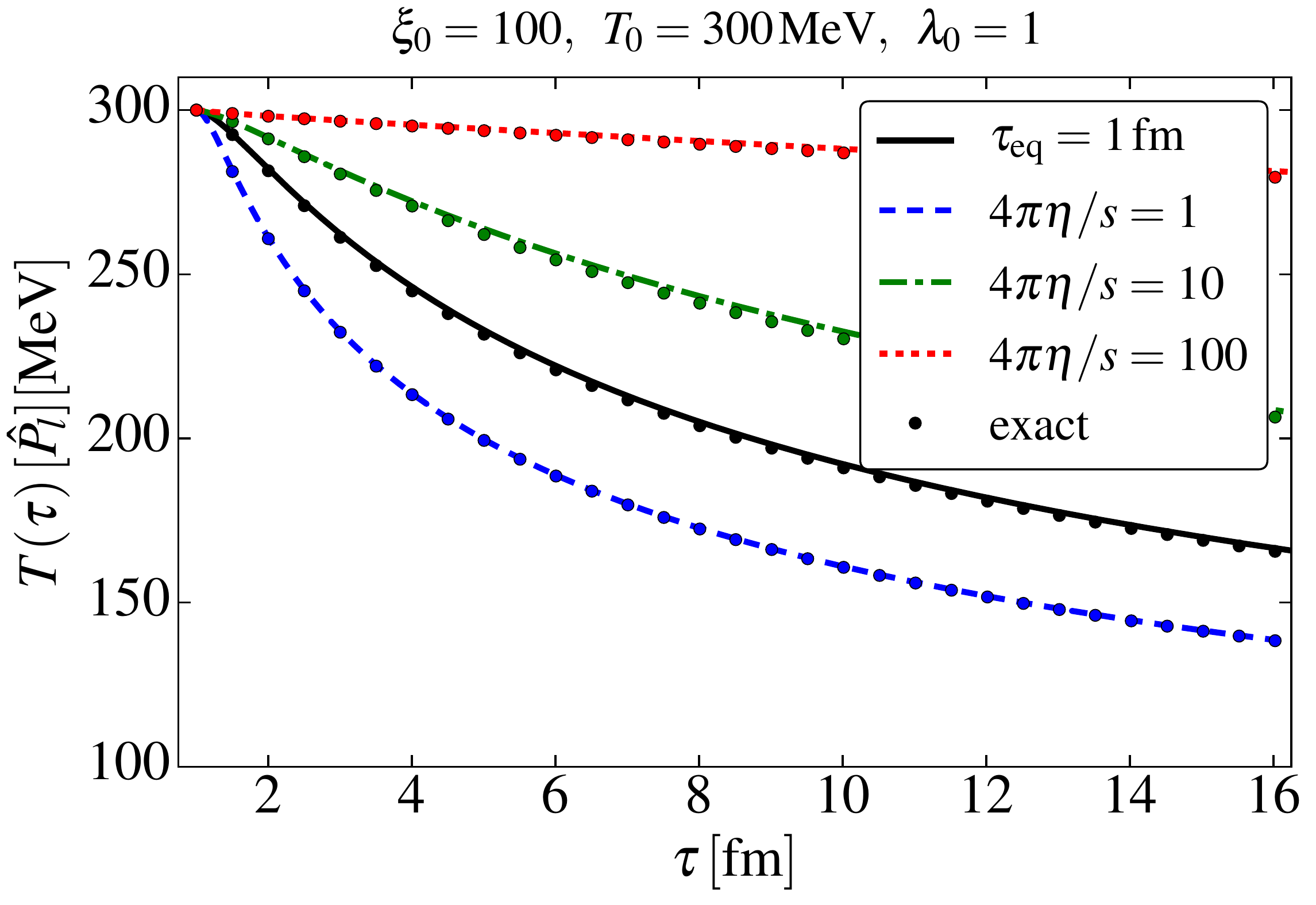}  
\includegraphics[width=7.8cm]{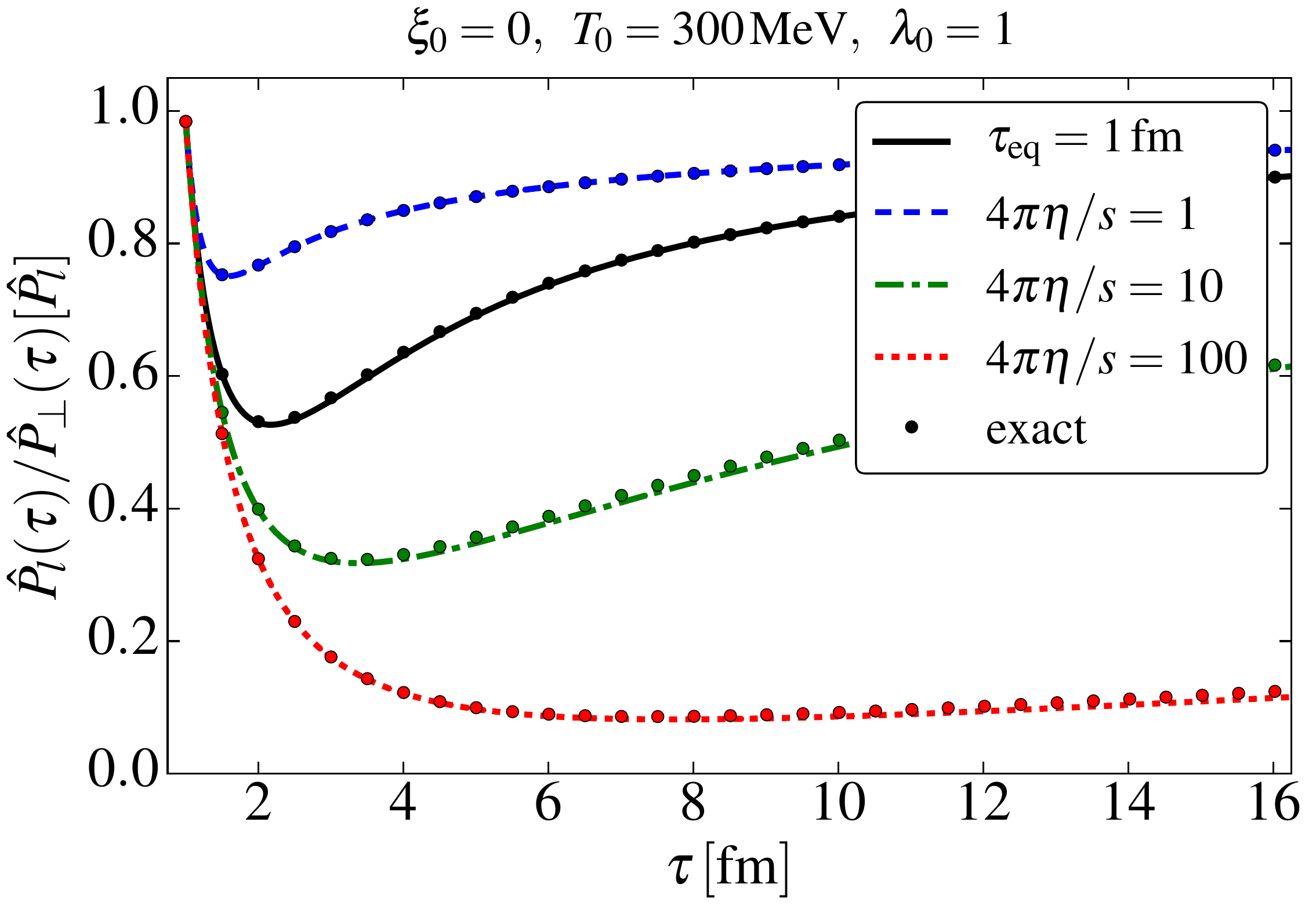} 
\includegraphics[width=7.8cm]{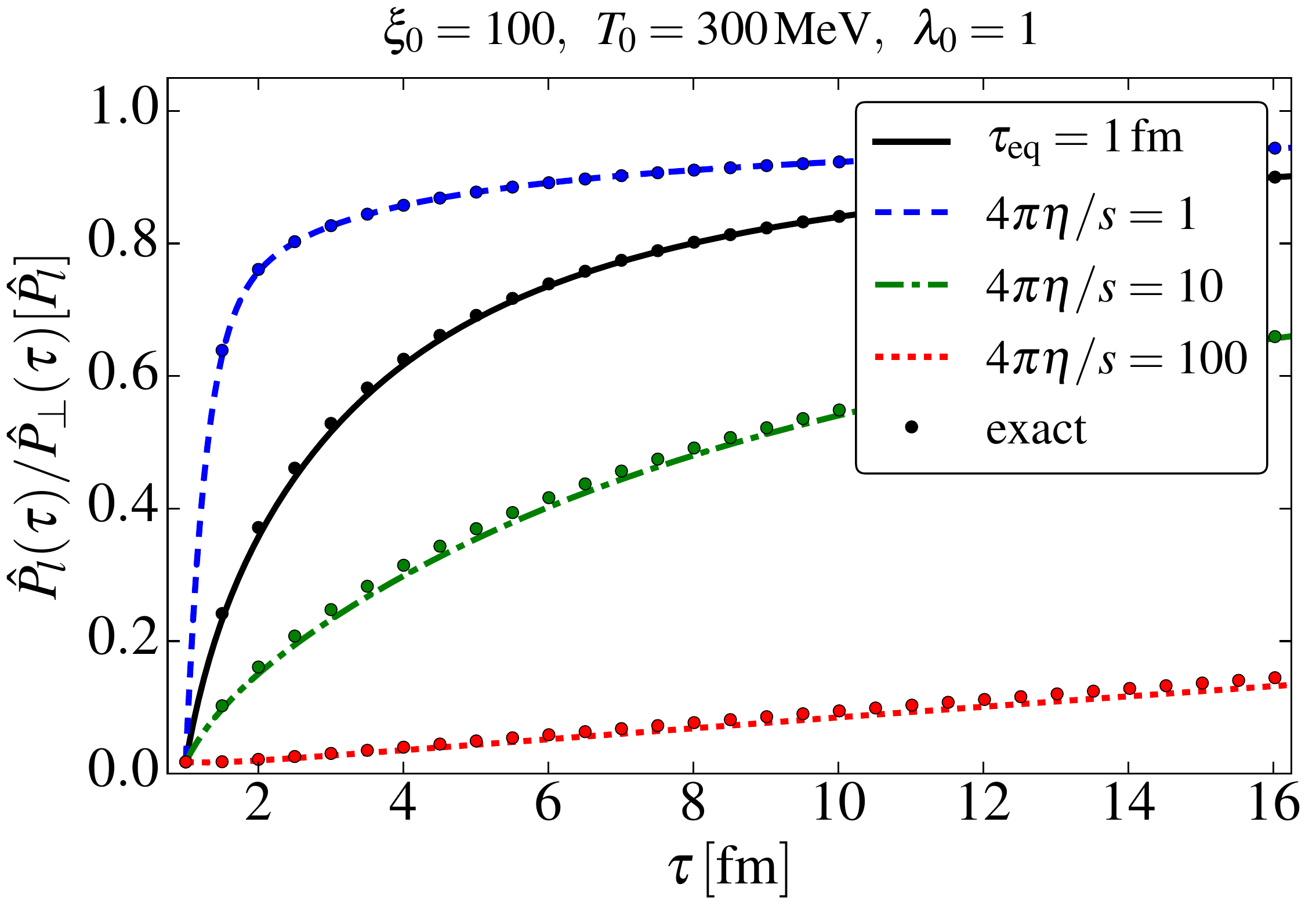}   
\vspace{-0.2cm} 
\caption{(Color online) The evolution of temperature $T$, fugacity $\lambda$, 
and the ratio of longitudinal and transverse pressure components $\hat{P}_l/\hat{P}_{\perp}$
as a function of proper time $\tau$.
The full black, the dashed blue, the dashed-dotted green and the dotted red lines are 
the solution of the conservation equations closed by the relaxation equation for 
$\hat{P}_l$ for $\tau_{eq}=1$ fm and for $\tau_{eq}$ from Eq.\ (\ref{tau_eq}) with the three different choices for $\eta/s$.
The large dots show the corresponding solution of the Boltzmann equation.}
\label{fig:3set}
\end{figure*}
%%%%%%%%%%%%%%%%%%%%% FIGURE %%%%%%%%%%%%%%%%%%%%%%%%%%%%%%%% 

%%%%%%%%%%%%%%%%%%%%% FIGURE %%%%%%%%%%%%%%%%%%%%%%%%%%%%%%%%
\begin{figure*}[htb!]
\centering
\includegraphics[width=7.8cm]{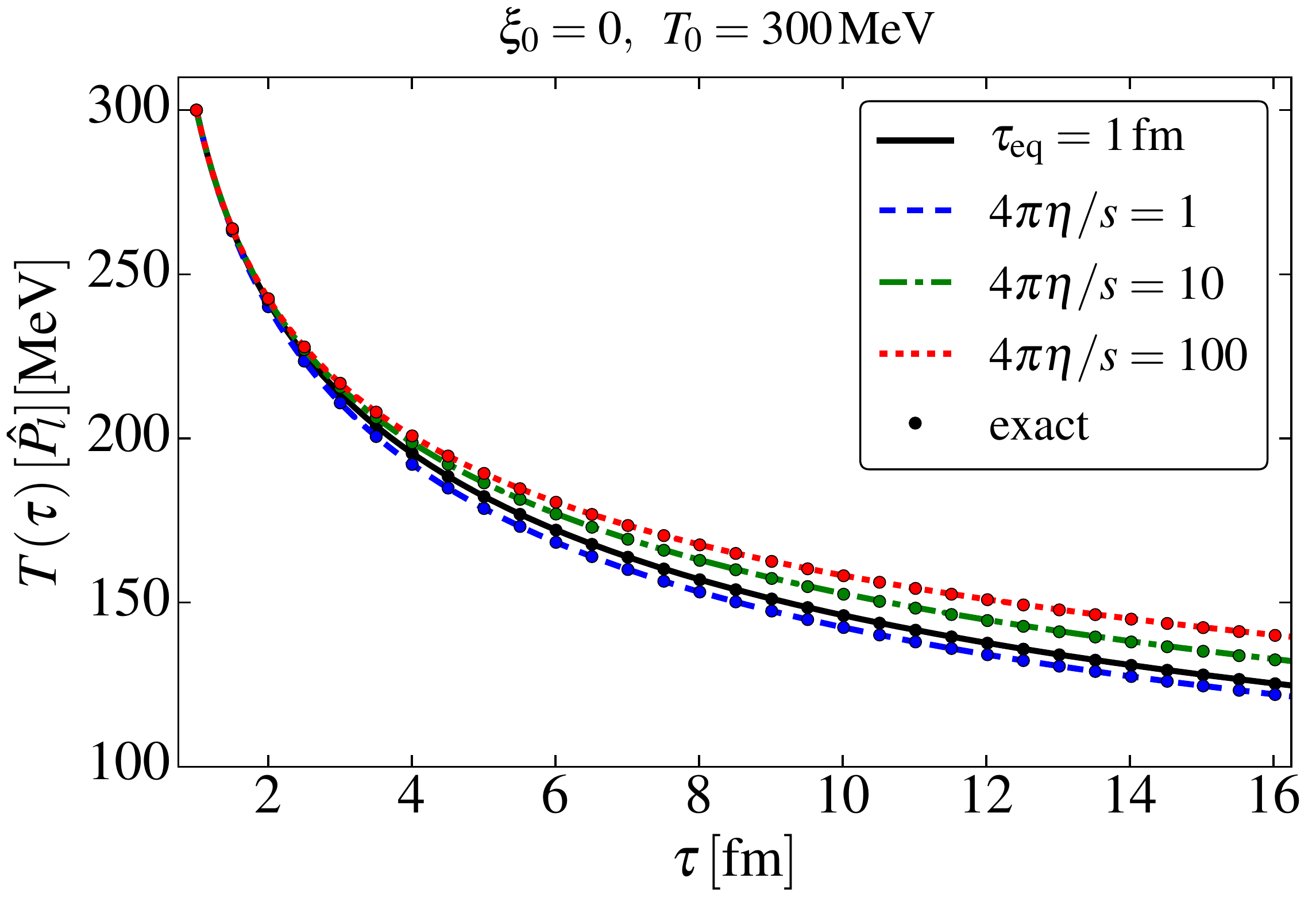} 
\includegraphics[width=7.8cm]{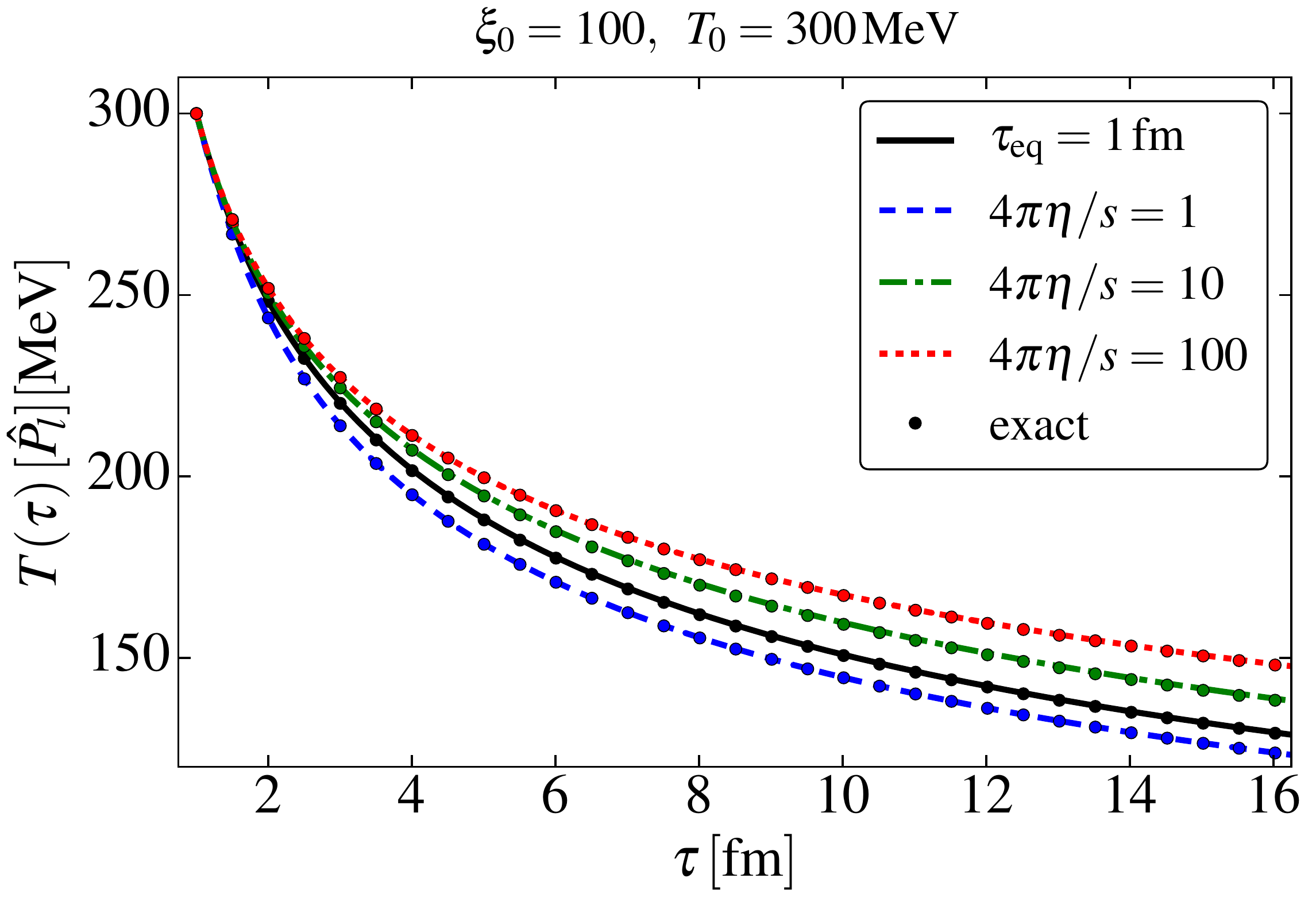}  
\includegraphics[width=7.8cm]{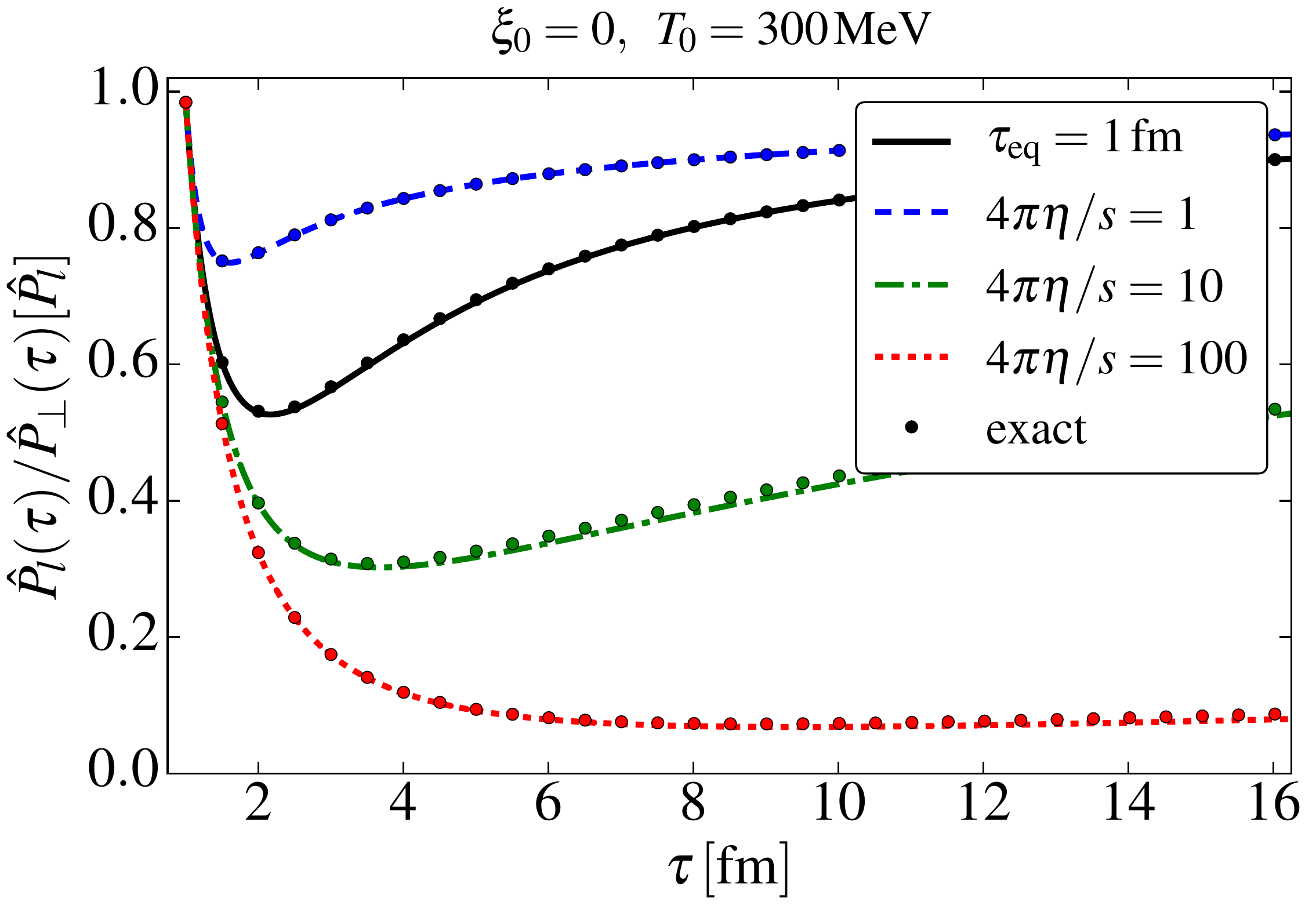} 
\includegraphics[width=7.8cm]{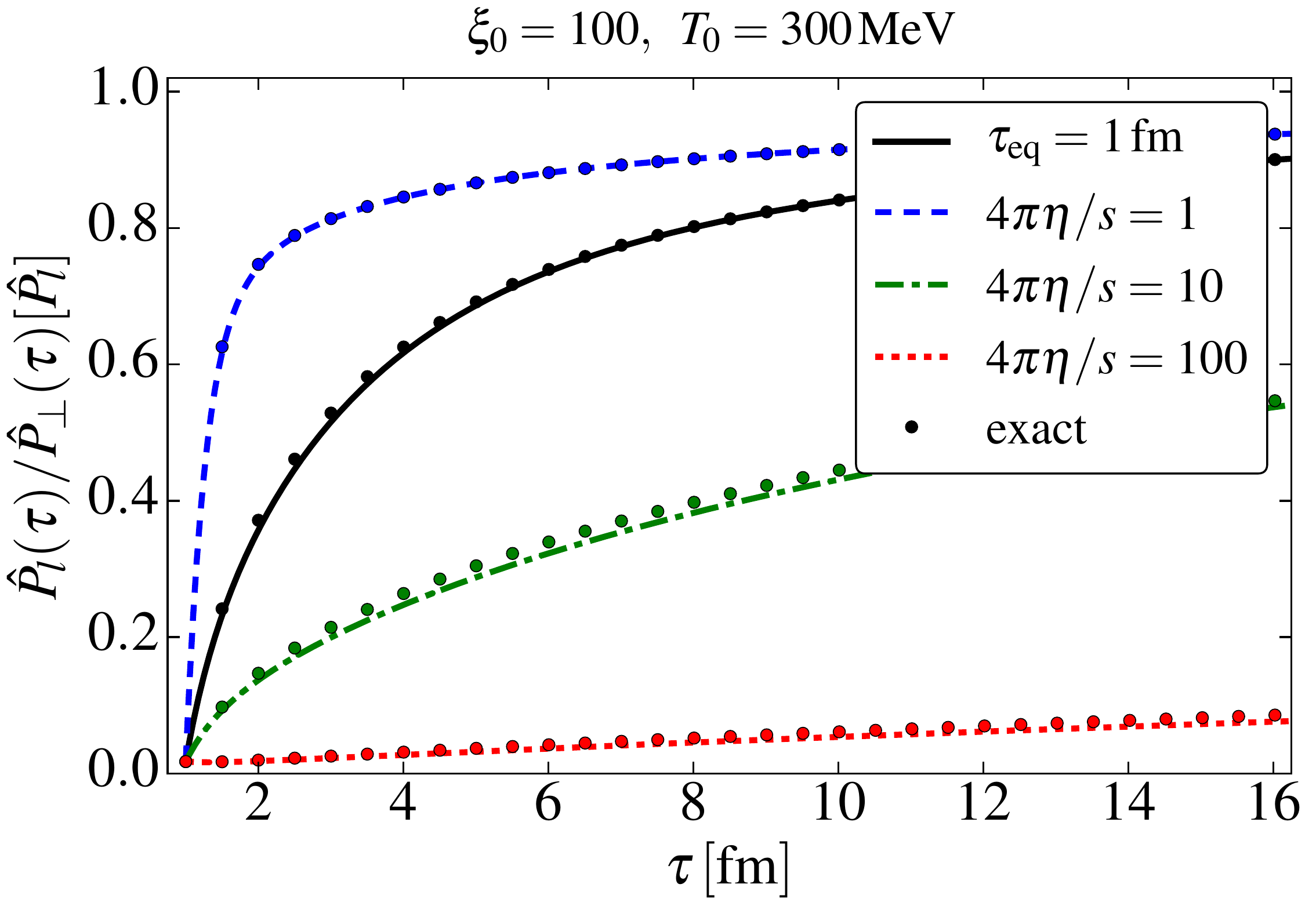}  
\vspace{-0.2cm} 
\caption{(Color online) Similar to  Fig.\ \ref{fig:3set}, but for the case without particle-number conservation
(such that $\lambda(\tau) = 1$ and thus not shown explicitly).}
\label{fig:4set}
\end{figure*}
%%%%%%%%%%%%%%%%%%%%% FIGURE %%%%%%%%%%%%%%%%%%%%%%%%%%%%%%%% 

%%%%%%%%%%%%%%%%%%%%% FIGURE %%%%%%%%%%%%%%%%%%%%%%%%%%%%%%%%
\begin{figure*}[htb!]
\centering
\includegraphics[width=7.8cm]{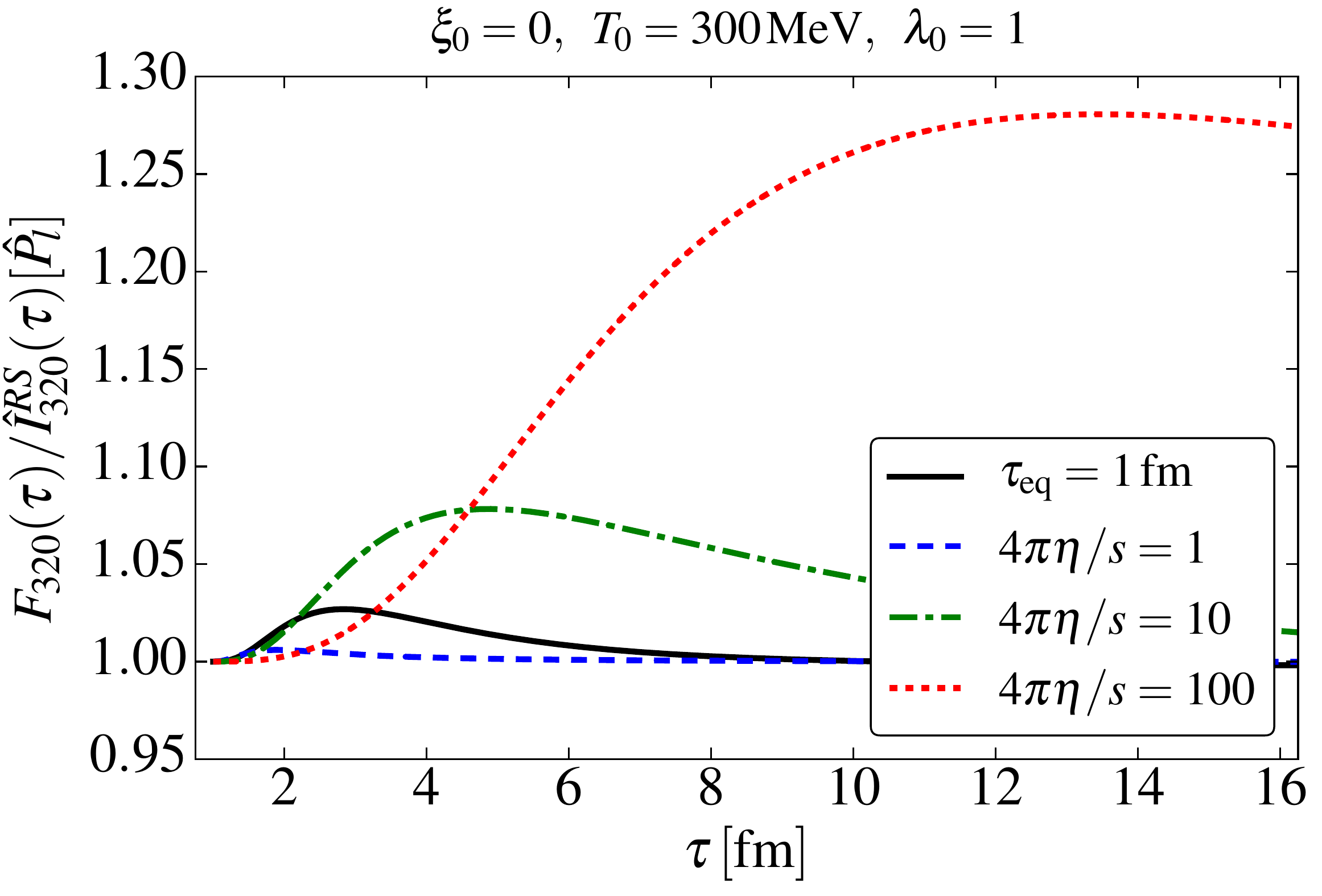} 
\includegraphics[width=7.8cm]{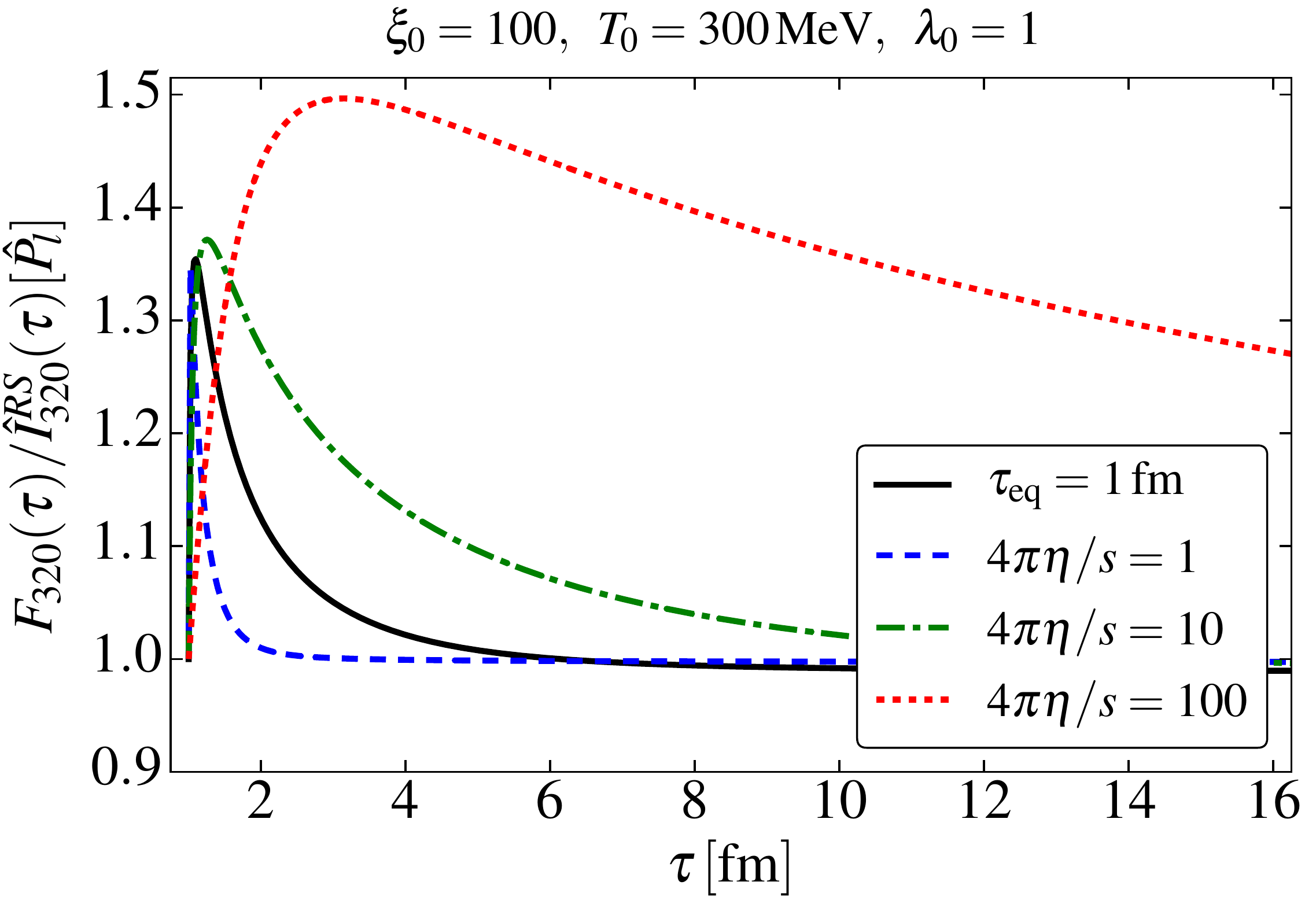} 
\vspace{-0.2cm} 
\caption{(Color online) 
The ratio of the exact solution $F_{320}(\tau)$ to the corresponding fluid-dynamical solution $\hat{I}^{RS}_{320}(\tau)$.
obtained from Eqs.~(\ref{BJ_n_cons}), (\ref{BJ_e_cons}), and (\ref{BJ_Pl_relax}). 
The choice of moment to close the equations is indicated in brackets, $[\hat{P}_l]$, behind the label on the ordinate.
The various lines are similar to Figs.\ \ref{fig:3set} and \ref{fig:4set}. }
\label{fig:5set}
\end{figure*}
%%%%%%%%%%%%%%%%%%%%% FIGURE %%%%%%%%%%%%%%%%%%%%%%%%%%%%%%%% 

%%%%%%%%%%%%%%%%%%%%% FIGURE %%%%%%%%%%%%%%%%%%%%%%%%%%%%%%%%
\begin{figure*}[htb!]
\centering
\includegraphics[width=7.8cm]{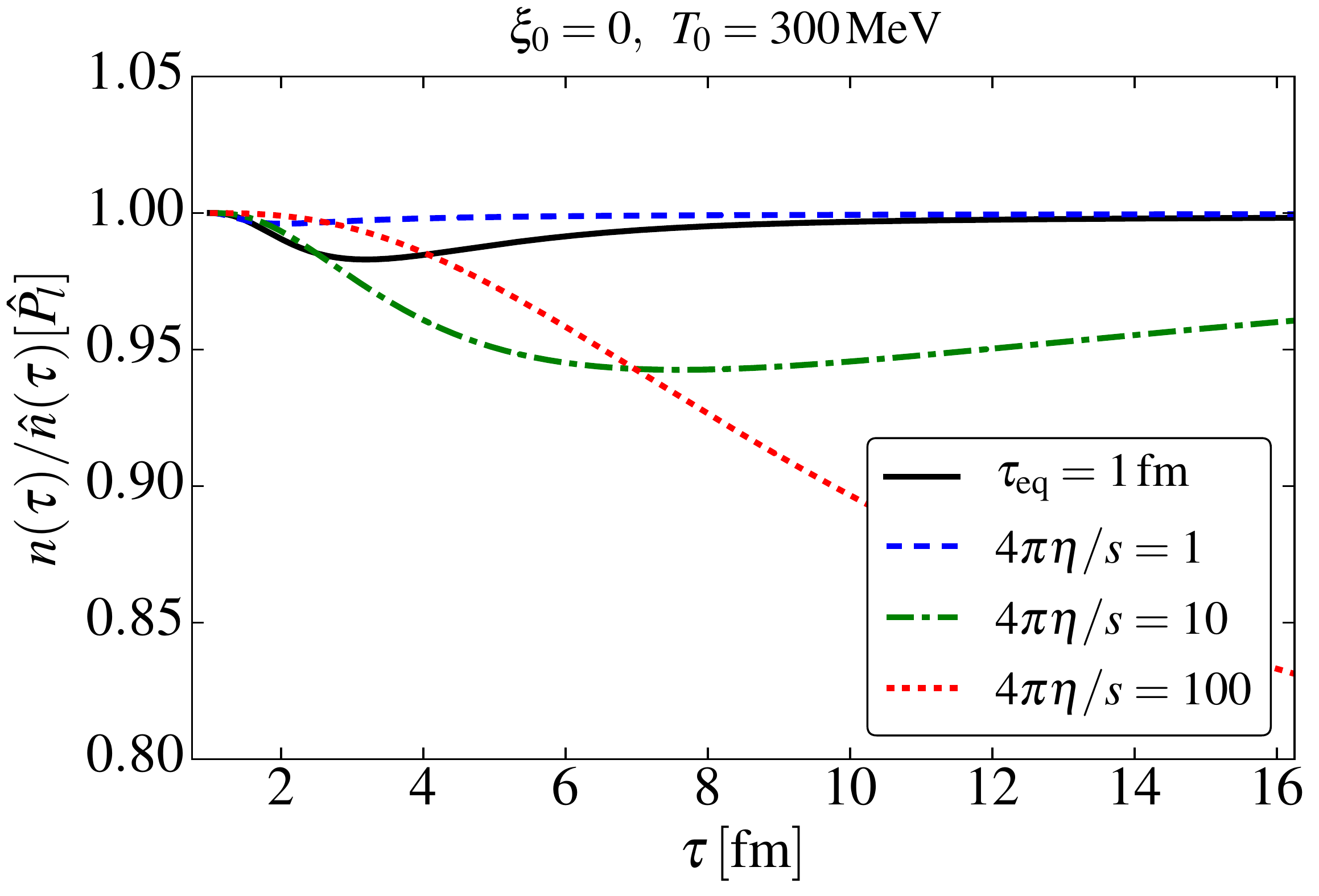} 
\includegraphics[width=7.8cm]{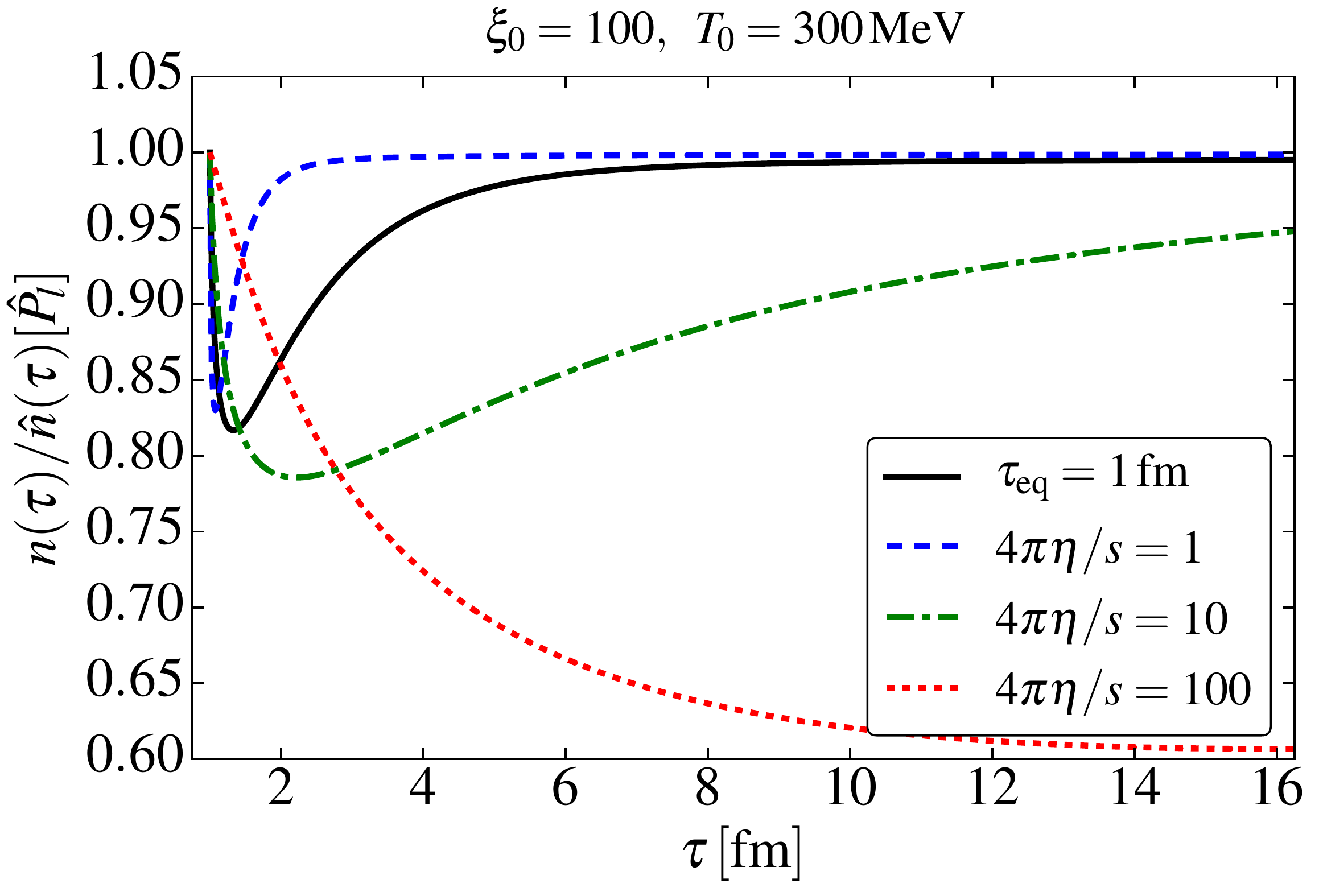} 
\vspace{-0.2cm} 
\caption{(Color online) Similar to Fig.\ \ref{fig:5set}, but without particle-number conservation.
Here $n(\tau) \equiv F_{100}(\tau)$ represents the solution of the Boltzmann equation while 
$\hat{n}(\tau) \equiv \hat{I}^{RS}_{100}(\tau)$ is computed from the solution of 
Eqs.\ (\ref{BJ_e_cons}) and (\ref{BJ_Pl_relax}).
The choice of moment to close the equations is indicated in brackets, $[\hat{P}_l]$, behind the label on the ordinate.
}
\label{fig:6set}
\end{figure*}
%%%%%%%%%%%%%%%%%%%%% FIGURE %%%%%%%%%%%%%%%%%%%%%%%%%%%%%%%% 

%%%%%%%%%%%%%%%%%%%% FIGURE %%%%%%%%%%%%%%%%%%%%%%%%%%%%%%%%
\begin{figure*}[htb!]
\centering
\includegraphics[width=7.8cm]{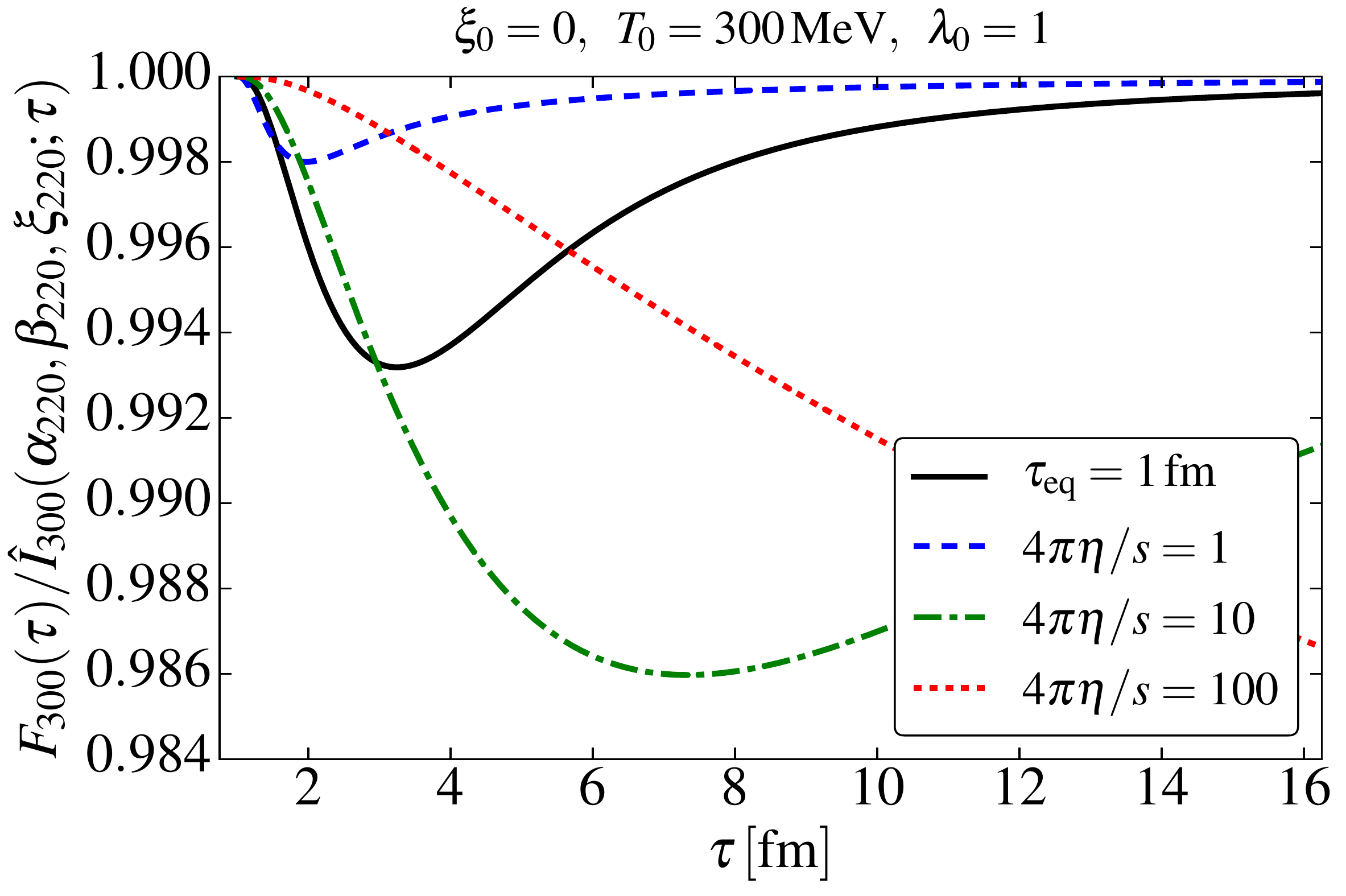} 
\includegraphics[width=7.8cm]{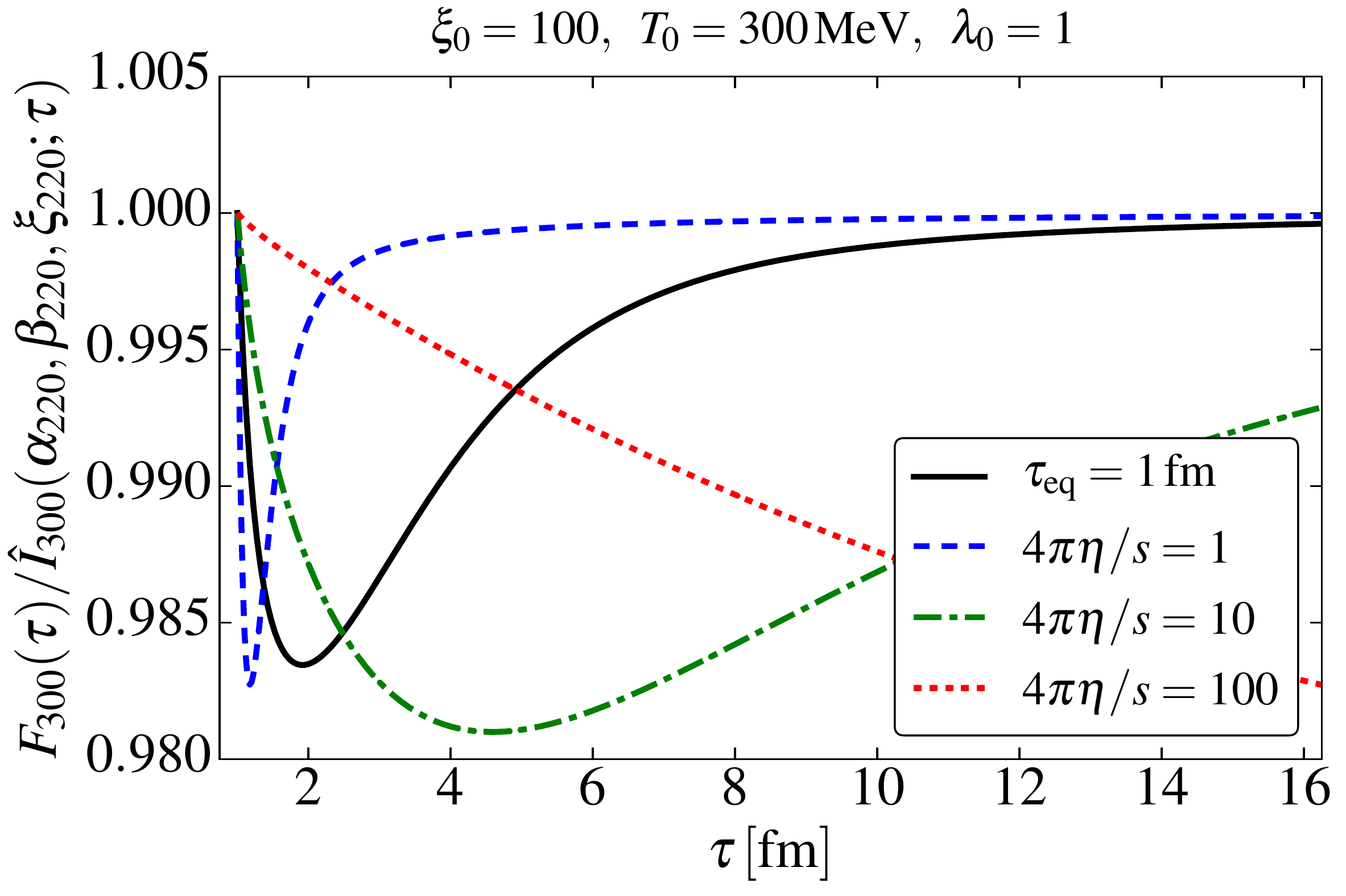}  
\vspace{-0.2cm} 
\caption{(Color online) 
The ratio of the exact solution $F_{300}(\tau)$ to $\hat{I}^{RS}_{300}(\alpha_{220}, \beta_{220}, \xi_{220};\tau)$,
where $\alpha_{220}$, $\beta_{220}$, and $\xi_{220}$ were obtained by matching to $F_{220}(\tau)$.
See text for more details. }
\label{fig:matching_pl}
\end{figure*}
%%%%%%%%%%%%%%%%%%%%% FIGURE %%%%%%%%%%%%%%%%%%%%%%%%%%%%%%%% 

%%%%%%%%%%%%%%%%%%%% FIGURE %%%%%%%%%%%%%%%%%%%%%%%%%%%%%%%%
\begin{figure*}[htb!]
\centering
\includegraphics[width=7.8cm]{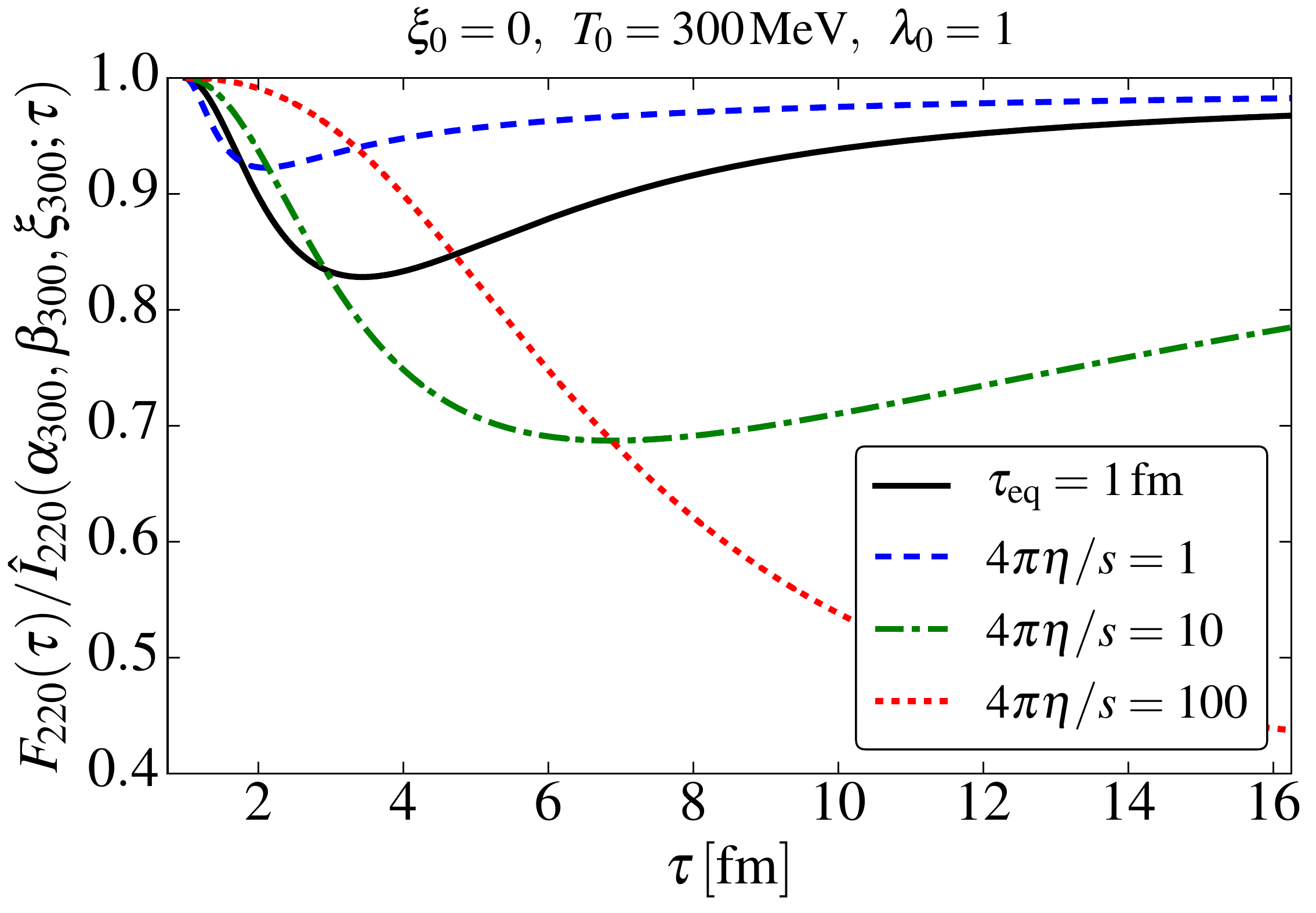} 
\includegraphics[width=7.8cm]{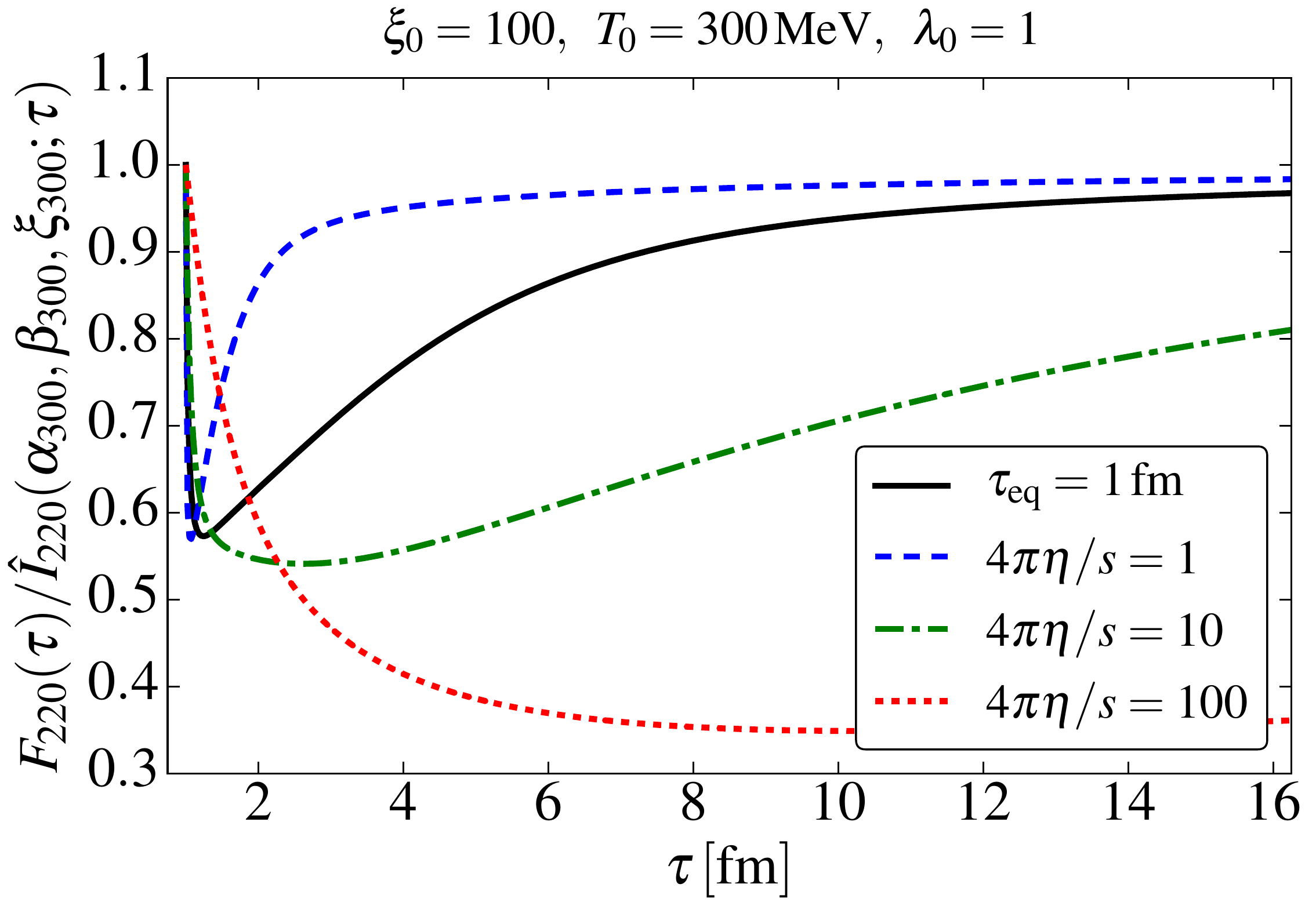} 
\vspace{-0.2cm} 
\caption{(Color online) Similar to Fig.~\ref{fig:matching_pl}. The ratio of the exact solution $F_{220}(\tau)$ to 
$\hat{I}^{RS}_{220}(\alpha_{300}, \beta_{300}, \xi_{300};\tau)$,
where $\alpha_{300}$, $\beta_{300}$, and $\xi_{300}$ were obtained by matching to $F_{300}(\tau)$.
See text for more details.}
\label{fig:matching_i300}
\end{figure*}
%%%%%%%%%%%%%%%%%%%%% FIGURE %%%%%%%%%%%%%%%%%%%%%%%%%%%%%%%% 

The results shown in Fig.\ \ref{fig:1set} were obtained by solving both the
particle-number conservation equation (\ref{BJ_n_cons}) and the energy-conservation equation 
(\ref{BJ_e_cons}), closed by one of the moment equations 
(\ref{BJ_Pl_relax}),  (\ref{BJ_I300_relax}), \ref{BJ_I320_relax}), 
(\ref{BJ_I000_relax}), (\ref{BJ_I440_relax}), or (\ref{BJ_I540_relax}). 
Correspondingly, the results in Fig.\ \ref{fig:2set} were obtained without 
particle-number conservation, i.e., we only solved the energy-conservation 
equation (\ref{BJ_e_cons}) coupled to a particular moment equation. In this case,
the first moment of the Boltzmann equation, Eq.\ (\ref{BJ_n_relax}), can also 
be used to provide closure (in addition to the previously listed relaxation equations).

Figure \ref{fig:1set} shows the evolution of the anisotropy parameter 
$\xi$, temperature $T$, fugacity $\lambda$, and the ratio of longitudinal 
and transverse pressure components $\hat{P}_l/\hat{P}_{\perp}$, as a function of proper time $\tau$.
All figures in the left column are for $\xi_0 = 0$, while those in the right column are for $\xi_0 = 10$. 
In Fig.\ \ref{fig:2set}, the same is presented for the case without particle-number conservation.

Focusing on the evolution of the anisotropy parameter $\xi$ we observe that in
case the system was initially isotropic ($\xi_0 = 0$, left column), the longitudinal expansion drives
the system out of equilibrium. This lasts for about $1-2$ fm, after which the system 
starts to approach the isotropic state again, $\xi \rightarrow 0$.
The approach to equilibrium 
becomes much faster for a nonzero initial anisotropy ($\xi_0 = 10$, right column). The late-time 
behavior is quite similar in both cases, as both 
reach a similar value for the anisotropy around $\tau \sim 6-7$ fm. 
The behaviour of the longitudinal to 
transverse pressure ratio, $\hat{P}_l /\hat{P}_\perp$, is
quite similar to that of the anisotropy isotropy parameter.
Both the evolution of $\xi$ and $\hat{P}_l / \hat{P}_\perp$ are
similar in the cases with and without particle-number conservation, cf.\
top and bottom rows of Figs.\ \ref{fig:1set} and \ref{fig:2set}. This can be explained by
the fact that $\hat{P}_l / \hat{P}_\perp$ is mainly determined by the momentum anisotropy $\xi$.

The temperature, second row of Figs.\ \ref{fig:1set} and \ref{fig:2set},
decreases as the system expands, but the decrease is slower for a
nonzero initial anisotropy. This is due to the fact $\hat{P}_l$
decreases with increasing anisotropy, hence the driving force to expand
(and cool) the system is smaller for a
larger initial anisotropy. Note that for the case without particle-number conservation,
Fig.\ \ref{fig:2set},
the evolution of the temperature is much closer to the one for an ideal fluid than
for the case with particle-number conservation. This holds for all choices of closure
of the conservation equations. Vice versa, the spread in the curves is much larger
for the case with particle-number conservation, cf.\ Fig.\ \ref{fig:1set}. The reason is
that the deviation from chemical equilibrium parametrized by the fugacity, third row
of Fig.\ \ref{fig:1set}, has to be compensated by an increase in temperature. Thus,
the smaller the $\lambda(\tau)$, the larger $T(\tau)$ has to be. This also explains why
all curves lie above the case for an ideal fluid (green lines).

These observations are generally valid for all choices of closure for the
conservation equations. However, there are striking
differences between the various choices.
The first observation is that there is a grouping according to 
the power $j$ of longitudinal momentum $E_{\mathbf{k}l}$
appearing in the particular moment $\hat{I}^{RS}_{i+j,j,0}$, cf.\ 
$\hat{I}^{RS}_{000}$ and $\hat{I}^{RS}_{300}$ (blue lines), 
$\hat{P}_l \equiv \hat{I}^{RS}_{220}$ and $\hat{I}^{RS}_{320}$ (black lines),
as well as $\hat{I}^{RS}_{440}$ and $\hat{I}^{RS}_{540}$ (red lines). 
Apparently, the larger the $j$, the
faster the approach to isotropization. This behavior is universal and can
be observed in both Figs.\ \ref{fig:1set} and \ref{fig:2set}.

We also remark that the solutions provided by $\hat{I}^{RS}_{440}$ and $\hat{I}^{RS}_{540}$ 
stay closer to the solution provided by $\hat{P}_l$ than the ones provided by $\hat{I}^{RS}_{000}$, 
$\hat{I}^{RS}_{300}$, and, in the case without particle-number conservation, $\hat{n}$. 
In particular for $\xi$ and $\hat{P}_l/\hat{P}_\perp$, the latter ones sometimes
deviate by more than a factor of two from the solution provided by $\hat{P}_l$. 
As we shall see in the next subsection, it turns out that the solution provided by $\hat{P}_l$ is
closest to the one of the Boltzmann equation. Note that, in order to improve the agreement of the
fluid-dynamical solution given by $\hat{n}$ with the solution of the Boltzmann equation,
in some earlier works \cite{Florkowski:2013lza,Florkowski:2013lya} a rescaled 
relaxation time $\tau^{AH}_{eq} = \tau_{eq}/2$ was used.
We checked that also Eqs.\ (\ref{BJ_I000_relax}) and
(\ref{BJ_I300_relax}) with $\tau^{AH}_{eq}$ instead of $\tau_{eq}$ lead to results that resemble the 
ones for $\hat{P}_l$. We remark, however, that one is actually not free to adjust the relaxation time 
in the various moment equations, since in the RTA, cf.\ Eq.\ (\ref{Coll_Int_RTA}), it is the same as the one
appearing in the Boltzmann equation.

\subsection{Comparisons to the exact solution}
\label{results_comparisions} 

In this section we compare the solution of the conservation equations 
closed by Eq.\ (\ref{BJ_Pl_relax}) for $\hat{P}_l$ to the solution of the Boltzmann equation in the RTA.
The numerical method to solve the Boltzmann equation is discussed in detail 
in Refs.\ \cite{Florkowski:2013lza,Florkowski:2013lya} as well as in App.\ \ref{Numeric_Boltzmann}.
In analogy to Eq.\ (\ref{I_nrq}) we introduce the moments of the solution $f_{\mathbf{k}}$ of the Boltzmann equation,
\begin{equation}
F_{nrq} =\frac{\left( -1\right)^{q}}{\left( 2q\right) !!}\int dKE_{\mathbf{k}u}^{n-r-2q}E_{\mathbf{k}l}^{r}\left(
\Xi ^{\mu \nu }k_{\mu }k_{\nu }\right) ^{q}f_{\mathbf{k}} \;. \label{F_nrq}
\end{equation}
On the other hand, the moments $\hat{I}^{RS}_{nrq}$, cf.\ Eq.\ (\ref{I_nrq_RS}), can be computed from
solving the fluid-dynamical equations, which provide $\alpha_0$, $\beta_0$, and $\xi$ 
required to compute these moments according to Eqs.\ (\ref{Matched_I_nrq_RS}) or (\ref{Matched_I_nrq_RS_beta0}).
We then compare $F_{nrq}$ to $\hat{I}^{RS}_{nrq}$ in order to estimate 
how much the anisotropic distribution function $\hat{f}_{RS}$
deviates from the full solution of the Boltzmann equation.

In order to compare with the results of 
Refs.\ \cite{Florkowski:2013lza,Florkowski:2013lya} we have also used 
the temperature-dependent relaxation time from Eq.\ (\ref{tau_eq}). 
The solution of the Boltzmann equation is obtained choosing the RS distribution function 
as the initial condition at proper time 
$\tau_0=1.0$ fm, i.e., $f_{\mathbf{k}} (\tau_0) = f_{RS} (\tau_0)$.

In Figs.~\ref{fig:3set} and \ref{fig:4set} the fluid-dynamical solution for a
constant relaxation time is shown by the black lines. 
For $\xi_0=0$ (left columns of these figures) these are identical to 
the black lines in the left columns of Figs.\ \ref{fig:1set} and \ref{fig:2set}. 
The other curves in Figs.~\ref{fig:3set} and \ref{fig:4set} correspond to relaxation times
chosen according to Eq.\ (\ref{tau_eq}).

For all quantities shown, the fluid-dynamical solution agrees very well with the 
exact solution, even for very large $\eta/s = 100/4\pi$, and 
very large initial anisotropy $\xi_0 = 100$. This is a strong indication that the conservation 
equations closed by the relaxation equation for $\hat{P}_l$ provides the best match
to the Boltzmann equation, at least for quantities which appear in the energy-momentum tensor.

Now that we have identified the apparent best match for closure we will investigate how well the other moments of the 
Boltzmann equation are reproduced using this particular choice for closure.
This comparison is shown in Figs.\ \ref{fig:5set} and \ref{fig:6set}. As can be seen, the very good agreement 
is not necessarily inherited by other moments. In Fig.~\ref{fig:5set} we show the ratio of the exact moment
$F_{320}$ to the fluid-dynamical solution $\hat{I}^{RS}_{320}$ in the case of
conserved particle number, and in Fig.~\ref{fig:6set} the ratio of the exact number density $n \equiv F_{100}$
to $\hat{n}\equiv \hat{I}^{RS}_{100}$ in the case without particle-number conservation. As can be seen, the 
deviations between the fluid-dynamical and exact solutions can be as large as 50 \%, even
if the agreement between the primary fluid-dynamical quantities, i.e., the quantities that
appear in the energy-momentum tensor itself, is almost perfect. cf.\ Figs.\  \ref{fig:3set} and \ref{fig:4set}.

\subsection{Matching to the solution of the Boltzmann equation}

In general, a given functional form of the anisotropic distribution function, 
e.g.\ $\hat{f}_{RS}$, does not agree exactly with the solution $f_{\mathbf{k}}$ of
the Boltzmann equation. Thus, there is also no reason to expect that all moments $\hat{I}^{RS}_{nrq}$ of 
$\hat{f}_{RS}$ agree with all moments $F_{nrq}$ of $f_{\mathbf{k}}$. In other words, computing the parameters
$\alpha_0$, $\beta_0$, and $\xi$ that determine $\hat{f}_{RS}$ from matching a certain \emph{subset} of the moments
$\hat{I}^{RS}_{nrq}$ to the moments $F_{nrq}$ of the exact solution, does not necessarily lead to a 
good agreement for \emph{all} other moments. In this subsection, we provide evidence for this observation
through an explicit calculation.

First, however, let us make a few remarks  (for the sake of simplicity we discuss only the case with 
particle-number conservation):
\begin{itemize}
\item The anisotropic distribution function is characterized by three parameters, $\alpha_0$, $\beta_0$, and $\xi$.
Correspondingly, three matching conditions are required to determine these parameters.

\item The matching conditions can in general be chosen from Eq.~(\ref{Matched_I_nrq_RS}), for
any values of $n$, $r$, and $q$. 

\item In the RTA the usual Landau matching conditions (\ref{matching_n}) and (\ref{matching_e}) 
are not only convenient, but also necessary to ensure the conservation of 
energy, momentum, and particle number. These
correspond to matching with $(n,r,q) = (2,0,0)$ and $(n,r,q) = (1,0,0)$. 
The third matching condition can be provided by any other choice for $(n,r,q)$.

\item When we choose a particular moment equation to close the conservation equations, at the same time 
we also choose a specific matching condition. For example, closing the system by using Eq.~(\ref{BJ_Pl_relax})
also implies that $\alpha_0$, $\beta_0$, and $\xi$ are matched to $\hat{P}_l \equiv \hat{I}^{RS}_{220}$. 
Alternatively, choosing Eq.~(\ref{BJ_I300_relax}) implies 
that they are matched to $\hat{I}^{RS}_{300}$. Note that in general different choices lead to different values of 
$\alpha_0$, $\beta_0$, and $\xi$.

\item Once the matching conditions are fixed, any other moment can be calculated according to the r.h.s.\ of 
Eq.~(\ref{Matched_I_nrq_RS}).
\end{itemize}

To further investigate the importance of the matching conditions, we match 
$\alpha_0(\tau)$, $\beta_0(\tau)$, and $\xi(\tau)$
to a particular moment of the solution of the Boltzmann equation, instead of its 
fluid-dynamical approximation (\ref{Main_eq_motion}).

In Fig.~\ref{fig:matching_pl} we use $P_l \equiv F_{220} = \hat{I}^{RS}_{220}(\alpha_{220}, \beta_{220}, \xi_{220}) 
\equiv \hat{P}_l$ as a matching
conditon, and plot the ratio of the exact solution $F_{300}$ to the moment
$\hat{I}^{RS}_{300}(\alpha_{220}, \beta_{220}, \xi_{220})$ obtained 
through such a choice of matching [the r.h.s.\ of Eq.~(\ref{Matched_I_nrq_RS})].

In Fig.~\ref{fig:matching_i300} we show the opposite scenario,
i.e., the ratio of $F_{220}$ to $\hat{I}^{RS}_{220}(\alpha_{300}, \beta_{300},\xi_{300})$ obtained by matching 
$\alpha_{300}$, $\beta_{300}$, and $\xi_{300}$
to always reproduce the exact moment $F_{300}$. As can be seen from the figures, 
if we choose the $\hat{P}_l$ matching,
then $\hat{I}^{RS}_{300}(\alpha_{220}, \beta_{220},\xi_{220})$ is a good approximation to $F_{300}$, but 
the opposite is not true: matching to $F_{300}$
does not give the correct $P_l = F_{220} \neq \hat{P}_l(\alpha_{300}, \beta_{300},\xi_{300})$. 

We note that when we solve anisotropic fluid dynamics by choosing Eq.~(\ref{BJ_I300_relax}), we 
implicitly use the latter matching. In other words, 
the values for $\alpha_0,\, \beta_0$, and $\xi$ are obtained by matching to 
$\hat{I}^{RS}_{300}$, $e_0$, and $n_0$. However, as can be seen from Fig.\ \ref{fig:matching_i300},
the values for $\hat{P}_l$ obtained in this way can deviate by more 
than 50 \% from the exact solution. Since $\hat{P}_l$ appears explicitly in the energy-conservation equation, 
this deviation in $\hat{P}_l$ will also lead to deviations from the exact solution.

On the other hand, closing the conservation equations with Eq.~(\ref{BJ_Pl_relax}), which corresponds to 
matching with $\hat{P}_l$, gives an overall good agreement with the exact solution. The comparison
of Figs.~\ref{fig:matching_pl} and \ref{fig:matching_i300} indicates why this is the case:
the matching to $\hat{P}_l$ directly leads to the correct driving force in the energy-conservation equation.
It is then obvious that this choice gives the best agreement with the Boltzmann equation, as well as smaller 
deviations from the exact solution for all other moments.

We note that in for the (0+1)--dimensional expansion this choice corresponds to the one proposed in 
Ref.~\cite{Tinti:2015xwa},
where the anisotropy parameters were matched to the components of $T^{\mu\nu}$, and the equations
of motion were closed by using the exact equations of motion for the dissipative quantities, e.g.\ for the shear-stress
tensor $\pi^{\mu\nu}$. This approach was originally proposed
in Ref.~\cite{Denicol:2010xn} for conventional fluid dynamics.

\section{Conclusions}
\label{Conclusions}

Starting from the relativistic Boltzmann equation, we have derived the equations of motion for a fluid 
which has a certain anisotropic single-particle distribution function $\hat{f}_{0\mathbf{k}}$ 
in momentum space in the LR frame. Choosing as an example $\hat{f}_{0\mathbf{k}} = \hat{f}_{RS}$ 
we have solved these equations in a simple 0+1 dimensional
boost-invariant expansion scenario. We have pointed out the importance of the
choice of moment equation to close the conservation equations.
The solution of the Boltzmann equation was most accurately 
reproduced by the equations of anisotropic fluid dynamics when the latter are closed
using the relaxation equation for the longitudinal pressure $\hat{P}_l$, i.e., a quantity which also appears
in the energy-momentum tensor. Other choices for the moments to close the conservation equations lead to
a less good agreement with the solution of the Boltzmann equation. 
In the future, one should extend the present study to more realistic geometries
(with non-trivial transverse and longitudinal dynamics) and include corrections to $\hat{f}_{0\mathbf{k}}$, using
the framework developed in Ref.\ \cite{Molnar:2016vvu}.

\begin{acknowledgments}
  The authors thank W.\ Florkowski, U.\ Heinz, P.\ Huovinen, and M.\ Strickland for comments and suggestions, 
  and M.\ Strickland for sharing the source code of the Boltzmann equation solver in the RTA.
  The work of E.M.\ was partially supported
  by the Alumni Program of the Alexander von Humboldt Foundation and by BMBF grant no.\
  05P15RFCA1.
  H.N.\ has received funding from the European Union's Horizon 2020 research and innovation
  programme under the Marie Sklodowska-Curie grant agreement no.\ 655285.
  E.M.\ and H.N.\ were partially supported by the Helmholtz International Center for
  FAIR within the framework of the LOEWE program launched by the State
  of Hesse. 
\end{acknowledgments}

%%%
\appendix

\section{Thermodynamic integrals in the massless Boltzmann limit}
\label{appendix_aniso_integrals}

Here we evaluate the thermodynamic integrals assuming
Boltzmann statistics in the massless limit. In the LR frame, where $u_{LR}^{\mu
}=\left(1,0,0,0\right) $, Eq.\ (\ref{Equilibrium_dist}) reads 
\begin{equation}
f_{0\mathbf{k}}\equiv \exp \left( \alpha _{0}-\beta _{0}E_{\mathbf{k}u}\right) 
=\lambda _{0}\exp \left( -\beta _{0}\sqrt{m_{0}^{2}+k^{2}}\right) \; ,
\end{equation}
where the fugacity is $\lambda _{0}=\exp  \alpha _{0}$ and 
$E_{\mathbf{k}u,LR}=k_{0}$. Similarly, the RS distribution function (\ref{f_RS}) 
reads in the LR frame
\begin{eqnarray}
\hat{f}_{RS} &\equiv &\exp \left( \alpha _{RS}-\beta _{RS}\sqrt{E_{\mathbf{k}u}^{2}
+\xi E_{\mathbf{k}l}^{2}}\right)  \notag \\
&=&\lambda _{RS}\exp \left( -\beta _{RS}\sqrt{m_{0}^{2}+k^{2}+\xi k_{z}^{2}}\right) \; ,
\end{eqnarray}
where the fugacity is $\lambda _{RS}=\exp \alpha _{RS}$ and $E_{\mathbf{k}l,LR}=k_{z}$.

Using these distribution functions the different thermodynamic integrals appearing in Eqs.\ (\ref{I_nrq_RS}) 
and (\ref{I_nq}) are evaluated in spherical coordinates, 
where $k=\left\vert \mathbf{k}\right\vert$, $k_{x}=k\cos \varphi \sin \theta $, $k_{y}=k\sin \varphi \sin \theta $,
$k_{z}=k\cos \theta $. Therefore, $dK\equiv A_{0}\frac{d^{3}\mathbf{k}}{k^{0}}
=A_{0}\frac{k^{2}}{k^{0}}dk\sin \theta d\theta d\varphi $, where $\theta \in \left[ 0,\pi \right] $, $
\varphi \in \left[ 0,2\pi \right) $, $k\in \left[ 0,\infty \right) $, and 
$A_{0}=g/(2\pi)^{3}$.
Furthermore, we will also need the LR frame values of $-\left( \Delta
^{\alpha \beta }k_{\alpha }k_{\beta }\right) _{LR}=k^{2}$ and $-\left( \Xi
^{\mu \nu }k_{\mu }k_{\nu }\right) _{LR}\equiv k_{\perp
}^{2}=k_{x}^{2}+k_{y}^{2}=k^{2}\sin ^{2}\theta $.

In the massless limit, i.e., $\lim_{m_{0}\rightarrow 0}\sqrt{m_{0}^{2}+k^{2}}=|k|$,
we obtain the following result for Eq.\ (\ref{I_nq}), 
\begin{eqnarray}
\lim_{m_{0}\rightarrow 0}I_{nq} &\equiv & \lambda _{0}\frac{\left( -1\right)
^{2q}4\pi A_{0}}{\left( 2q+1\right) !!}\int_{0}^{\infty }dkk^{n+1}\exp
\left( -\beta _{0}k\right)  \notag \\
&=&\lambda _{0}\frac{4\pi A_{0}\left( n+1\right) !}{\beta _{0}^{n+2}\left(
2q+1\right) !!} \;.  \label{I_nq_massless}
\end{eqnarray}
Here we list some of these integrals explicitly, 
\begin{eqnarray}
I_{00}\left( \alpha _{0},\beta _{0}\right) &\equiv &I_{000}=\lambda _{0}
\frac{4\pi A_{0}}{\beta _{0}^{2}} \;,\   \label{app_I_00} \\
I_{10}\left( \alpha _{0},\beta _{0}\right) &\equiv &I_{100}=\lambda _{0}
\frac{8\pi A_{0}}{\beta _{0}^{3}}=n_{0} \;,\   \label{app_I_10} \\
I_{20}\left( \alpha _{0},\beta _{0}\right) &\equiv &I_{200}=\lambda _{0}
\frac{24\pi A_{0}}{\beta _{0}^{4}}=e_{0},\   \label{app_I_20} \\
I_{21}\left( \alpha _{0},\beta _{0}\right) &\equiv &I_{201}=I_{220}=P_{0} \;,
\label{app_I_21}
\end{eqnarray}
where $P_{0}\equiv n_{0}/\beta _{0}=e_{0}/3$, and 
\begin{eqnarray}
I_{30}\left( \alpha _{0},\beta _{0}\right) &\equiv &I_{300}=\lambda _{0}
\frac{96\pi A_{0}}{\beta _{0}^{5}} \; ,  \label{app_I_30} \\
I_{31}\left( \alpha _{0},\beta _{0}\right) &\equiv &I_{301}=I_{320}
=\frac{I_{30}\left( \alpha _{0},\beta _{0}\right) }{3} \;.  \label{app_I_31}
\end{eqnarray}

The RS distribution function leads to the following thermodynamical
integral in the massless limit,
\begin{align}
\lim_{m_{0}\rightarrow 0}\hat{I}_{nrq}^{RS}& \equiv \lambda _{RS}
\frac{\left( -1\right) ^{2q}2\pi A_{0}}{\left( 2q\right) !!}\int_{0}^{\pi }d\theta
\cos ^{r}\theta \sin ^{2q+1}\theta  \notag \\
& \times \int_{0}^{\infty }dkk^{n+1}\exp \left[ -\beta _{RS}k 
\sqrt{1+\xi \cos ^{2}\theta }\right]  \notag \\
& =\lambda _{RS}\frac{2\pi A_{0}\left( n+1\right) !}{\beta _{RS}^{n+2}\left(
2q\right) !!}\int_{0}^{\pi }d\theta \frac{\cos ^{r}\theta \sin ^{2q+1}\theta 
}{\left( 1+\xi \cos ^{2}\theta \right) ^{\frac{n+2}{2}}} \;.
\label{I_nrq_RS_massless}
\end{align}
Therefore, the ratio between the RS and equilibrium thermodynamical integrals
defined in Eq.\ (\ref{R_nrq_RS}) reads 
\begin{equation}
R_{nrq}=\frac{\left( 2q+1\right) !!}{2\left( 2q\right) !!} 
\int_{0}^{\pi}d\theta \frac{\cos ^{r}\theta \sin ^{2q+1}\theta }
{\left( 1+\xi \cos^{2}\theta \right) ^{\frac{n+2}{2}}} \;,
\end{equation}
hence the values that correspond to Eqs.\ (\ref{n_hat_RS}) -- (\ref{Pt_hat_RS}) are 
\begin{eqnarray}
R_{100}\left( \xi \right) &=&\frac{1}{\sqrt{1+\xi }} \;,  \label{R_100} \\
R_{200}\left( \xi \right) &=&\frac{1}{2}\left( \frac{1}{1+\xi }
+\frac{\arctan \sqrt{\xi }}{\sqrt{\xi }}\right) \; ,  \label{R_200} \\
R_{201}\left( \xi \right) &=&\frac{3}{2\xi }\left[ \frac{1}{1+\xi }-\left(
1-\xi \right) R_{200}\left( \xi \right) \right] \; ,  \label{R_201} \\
R_{220}\left( \xi \right) &=&-\frac{1}{\xi }\left[ \frac{1}{1+\xi }
-R_{200}\left( \xi \right) \right] \; .  \label{R_220}
\end{eqnarray}
Note that these results were obtained previously by Martinez and Strickland,
see for example Ref.\ \cite{Martinez:2010sc}, such that $R_{100}=\mathcal{R}_{0}$
and $R_{200}=\mathcal{R}$, $R_{201}=\mathcal{R}_{T}$, while the last term
differs from the results of Ref.\ \cite{Martinez:2010sc} by a factor of $
I_{20}/I_{21}=3$ since they calculated $\hat{I}_{220}^{RS}/I_{21}=3R_{220}
\left( \xi \right) $, i.e., $R_{220}=\mathcal{R}_{L}/3$.

Furthermore, for the other moment equations (\ref{BJ_Pl_relax}), (\ref{BJ_I300_relax}),  
(\ref{BJ_I320_relax}), (\ref{BJ_I000_relax}), (\ref{BJ_I440_relax}), and (\ref{BJ_I540_relax})
we also need the following $R_{nrq}(\xi)$ ratios 
\begin{eqnarray}
R_{000}\left( \xi \right) &=&\frac{\arctan \sqrt{\xi }}{\sqrt{\xi }} \; ,
\label{R_000} 
\;\;\;\;R_{020}\left( \xi \right) = \frac{1 - R_{000} \left( \xi \right)}{\xi } \;,  \label{R_020} \\
R_{240}\left( \xi \right) &=&\frac{1}{\xi ^{2}}\left[ \frac{3+\xi }{1+\xi }
-3R_{200}\left( \xi \right) \right] \; ,  \label{R_240} \\
R_{300}\left( \xi \right) &=&\frac{ 3+2\xi  }{3\left( 1+\xi\right) ^{3/2}} \; ,  
\label{R_300} \;\;\;\;
R_{301}\left( \xi \right) =  R_{100}\left( \xi \right) \; ,  \label{R_301} \\
R_{320}\left( \xi \right) &=& \frac{1}{3\left( 1+\xi \right) ^{3/2}} \; ,\\
\label{R_320}
R_{440}\left( \xi \right) &=&-\frac{1}{8\xi^{2}}\left( \frac{3+5\xi }{(1+\xi)^2 }
-3\frac{\arctan \sqrt{\xi }}{\sqrt{\xi }} \right) \; ,  \label{R_440} \\
R_{460}\left( \xi \right) &=&\frac{1}{8\xi^{3}}\left( \frac{15+25\xi+8\xi^2 }{(1+\xi)^2 } 
-15\frac{\arctan \sqrt{\xi }}{\sqrt{\xi }} \right) \; ,  \;\;\;\;\;\;\;\;   \label{R_460} \\
R_{540}\left( \xi \right) &=&\frac{1}{5\left( 1+\xi \right) ^{5/2}} \; . \label{R_540}
\end{eqnarray}
Note that, for any odd $r$, $\hat{I}_{nrq}^{RS}=0$.

\section{Second-order fluid dynamics in the limit of small anisotropy}
\label{2nd_order_hydro}

Here we recall the equations of second-order fluid dynamics describing the 0+1 dimensional 
boost-invariant expansion. Neglecting bulk viscosity we have
\begin{eqnarray}
\frac{\partial e_{0}}{\partial \tau } &=&
-\frac{1}{\tau }\left( e_{0} + P_0 - \pi\right)  \; , \\  \label{BJ_e_cons_2nd}
\tau_\pi \frac{\partial\pi}{\partial \tau }  &=&
\frac{4}{3}\frac{\eta}{\tau } - \pi - \left(\frac{1}{3}\tau_{\pi \pi} + \delta_{\pi \pi} \right) \frac{\pi}{\tau}  \; ,
\label{BJ_pi_relax}
\end{eqnarray}
where the equation for particle-number conservation is given in Eq.\ (\ref{BJ_n_cons}).
Here $\pi = \pi^{00} - \pi^{zz}$ enters the shear-stress tensor 
$\pi^{\mu \nu} \equiv T^{\alpha \beta} \Delta^{\mu \nu}_{\alpha \beta} = \textrm{diag}(0, \pi/2,\pi/2,-\pi)$,
where the corresponding symmetric, orthogonal, and traceless projection
operator is $\Delta _{\alpha \beta }^{\mu \nu } = \frac{1}{2}\left( \Delta _{\alpha}^{\mu }\Delta _{\beta }^{\nu }
+\Delta _{\beta }^{\mu }\Delta _{\alpha}^{\nu }\right) -\frac{1}{3}\Delta ^{\mu \nu }\Delta _{\alpha \beta }$.
Furthermore, the coefficients in the massless limit are, see for example Refs.\ \cite{Jaiswal:2013npa, Jaiswal:2014isa, 
Denicol:2012cn, Denicol:2012es, Denicol:2014vaa, Denicol:2014mca},
\begin{eqnarray}
\tau_{\pi \pi} &=& \frac{10}{7} \tau_\pi \; , \;\;\;\;
\delta_{\pi \pi} = \frac{4}{3} \tau_\pi \; .
\end{eqnarray}
These equations and coefficients can be derived by using the method of moments~\cite{Denicol:2012cn}, 
where $\pi^{\mu \nu} = \int dK k^{\left \langle \mu \right.} k^{\left. \nu \right\rangle} \delta f_{\mathbf{k}}$ and 
$\delta f_{\mathbf{k}} = f_{\mathbf{k}} - f_{0 \mathbf{k}}$ is the deviation from the equilibrium distribution function.

In the present case a similar approximation leads to $\hat{f}_{\mathbf{k}} = f_{0 \mathbf{k}} + \delta f_{\mathbf{k}}(\xi)$ 
which corresponds to a series expansion of the fluid-dynamical quantities for small $\xi$. 
Expanding Eqs.\ (\ref{Matched_Pl_RS}) and (\ref{Matched_I240_RS}) and neglecting corrections of order 
$\mathcal{O}(\xi^{2})$ we obtain 
\begin{eqnarray}
\hat{P}_{l} &\equiv& e_0 \left( \frac{1}{3} - \frac{8}{45}\xi \right) 
= P_0 - \pi\; , \\
\hat{I}_{240}^{RS} &\equiv& e_0 \left(\frac{1}{5} - \frac{16}{105}\xi \right) 
= \frac{3}{5}P_0 - \frac{6}{7} \pi \; .
\end{eqnarray}
Applying these results to Eqs.\ (\ref{BJ_e_cons}) and (\ref{BJ_Pl_relax}) together with Eq.\ (\ref{tau_eq}) we get
\begin{equation}
\tau_{eq}\frac{\partial\pi}{\partial \tau }  =
\frac{4}{3}\frac{\eta}{\tau } - \pi - \frac{38}{21} \frac{\pi}{\tau} \tau_{eq} \, .
\label{eq:hydrolimit}
\end{equation}
which, after noting that in RTA $\tau_{eq} = \tau_\pi$, leads precisely to Eq.\ (\ref{BJ_e_cons_2nd}).
In the massive case it was shown \cite{Tinti:2015xwa} that this closure leads to the fluid-dynamical limit 
calculated in Refs.~\cite{Jaiswal:2013npa,Jaiswal:2014isa}, which in the massless case 
reduces to Eq.~(\ref{eq:hydrolimit}).

\section{Numerical solution of the Boltzmann equation in the RTA}
\label{Numeric_Boltzmann}

For the sake of completeness we will repeat the discussion related to 
the numerical solution of the Boltzmann equation based on the derivation 
of Refs.\ \cite{Florkowski:2013lza,Florkowski:2013lya}. 
Let us first introduce the following Lorentz-invariant variables, 
\begin{eqnarray}
v &\equiv &\tau E_{\mathbf{k}u}=k_{0}t-k_{z}z=\sqrt{w^{2}+(k_{\perp
}^{2}+m_{0}^{2})\tau ^{2}}\; , \\
w &\equiv &\tau E_{\mathbf{k}l}=k_{z}t-k_{0}z=\sqrt{v^{2}-(k_{\perp
}^{2}+m_{0}^{2})\tau ^{2}}\; .
\end{eqnarray}
The inverse transformation reads
\begin{eqnarray}
k_{0} &\equiv &\frac{vt+wz}{\tau ^{2}}=\frac{\cosh \eta }{\tau }(v+w\tanh\eta ) \; , \\
k_{z} &\equiv &\frac{wt+vz}{\tau ^{2}}=\frac{\cosh \eta }{\tau }(w+v\tanh\eta )\; .
\end{eqnarray}
Using these new boost-invariant variables the Boltzmann equation in RTA
becomes a first-order linear differential equation 
\begin{equation}
\frac{\partial f_{\mathbf{k}}}{\partial \tau }=\frac{f_{\mathbf{k}0}-f_{\mathbf{k}}}{\tau _{eq}}\;,
\end{equation}
with the formal solution 
\begin{align}
f_{\mathbf{k}}(\tau) =D(\tau ,\tau _{0})f_{\mathbf{k}}(\tau _{0}) 
 +\int_{\tau _{0}}^{\tau }\frac{d\tau ^{\prime }}{\tau _{eq}(\tau ^{\prime
})}D(\tau ,\tau ^{\prime })f_{\mathbf{k}0}(\tau ^{\prime })\;,
\label{BTE_RTA_solution}
\end{align}
where $D(\tau _{2},\tau _{1})$ is a so-called damping function 
\begin{equation}
D(\tau _{2},\tau _{1})=\exp \left( -\int_{\tau _{1}}^{\tau _{2}}\frac{d\tau
^{\prime }}{\tau _{eq}(\tau ^{\prime })}\right) \;.
\end{equation}
In order to obtain the relevant quantities we will need to calculate the moments 
of the solution (\ref{BTE_RTA_solution}). Hence, similarly to Eq.\ (\ref{I_ij_tens_expanded}), 
we calculate the moments of the equilibrium and anisotropic
distribution functions. The equilibrium thermodynamic integrals 
$I_{nrq}\left( \tau ,\tau^{\prime },\alpha _{0}\left( \tau ^{\prime }\right) ,\beta _{0}
\left( \tau^{\prime }\right) \right) = I_{nrq}$ read 
\begin{align}
I_{nrq}& \equiv \frac{\left( -1\right) ^{q}}{\left( 2q\right) !!}
\int dKE_{\mathbf{k}u}^{n-r-2q}E_{\mathbf{k}l}^{r}\left( \Xi ^{\mu \nu }k_{\mu }
k_{\nu}\right)^{q}f_{\mathbf{k}0}\left( \tau ^{\prime }\right)  \notag \\
& =\frac{2\pi A_{0}\left( n+1\right) !}{\left(2q\right) !!}
\frac{\lambda _{0}\left( \tau ^{\prime }\right) }{\beta _{0}^{n+2}
\left(\tau ^{\prime }\right) }H_{nrq}\left( \tau ,\tau ^{\prime }\right) \;,  \label{I_nrq_tau}
\end{align}
where we introduced the following integral: 
\begin{equation}
H_{nrq}\left( \tau ,\tau ^{\prime }\right) =\int_{0}^{\pi }d\theta 
\frac{\cos ^{r}\theta \sin ^{2q+1}\theta }{\left[ \left( \frac{\tau }{\tau
^{\prime }}\right) ^{2}\cos ^{2}\theta +\sin ^{2}\theta \right]^{\frac{n+2}{2}}}\;.  \label{H_nrq}
\end{equation}
In some cases of interest these integrals were already calculated in Ref.\ \cite{Florkowski:2013lya}, 
hence $H_{200}\left( \tau,\tau ^{\prime }\right) =\mathcal{H}\left( \tau ^{\prime }/\tau \right) $, 
$H_{220}\left( \tau ,\tau ^{\prime }\right) =\mathcal{H}_{L}\left( \tau
^{\prime }/\tau \right) $ and $H_{201}\left( \tau ,\tau ^{\prime }\right) =
\mathcal{H}_{T}\left( \tau ^{\prime }/\tau \right) $, see Eq.\ (A1) of Ref.\
\cite{Florkowski:2013lya}. Also note that these integrals were calculated
using boost-invariant variables and additionally applying a second variable
change, $p\cos \theta =\beta \left( \tau' \right) w/\tau' $ and $p\sin \theta
=\beta \left( \tau' \right) k_{\perp }$.

Similarly, for the RS distribution function we have 
$\hat{I}_{nrq}^{RS}\left(\tau ,\tau ^{\prime },\alpha _{RS}(\tau ^{\prime}) ,\beta _{RS}( \tau ^{\prime}),
\xi( \tau ^{\prime}) \right) =\hat{I}_{nrq}^{RS}$, 
\begin{align}
\hat{I}_{nrq}^{RS}& \equiv \frac{\left( -1\right) ^{q}}{\left( 2q\right) !!}
\int dKE_{\mathbf{k}u}^{n-r-2q}E_{\mathbf{k}l}^{r}\left( \Xi ^{\mu \nu
}k_{\mu }k_{\nu }\right) ^{q}f_{RS}\left( \tau ^{\prime }\right)  \notag \\
& =\frac{2\pi A_{0}\left( n+1\right) !}{\left( 2q\right) !!}
\frac{\lambda _{RS}\left( \tau ^{\prime }\right) }{\beta_{RS}^{n+2}\left( \tau ^{\prime }\right) }
H_{nrq}\left( \tau ,\frac{\tau ^{\prime }}{\sqrt{1+\xi
(\tau ^{\prime }) }}\right) \;,  \label{I_nrq_RS_tau}
\end{align}
where the argument of the $H_{nrq}$ integral is scaled by a factor of $1/\sqrt{1+\xi
\left( \tau ^{\prime }\right) }$ compared to Eq.\ (\ref{I_nrq_tau}).

Now, applying the definition of the moments on both sides of Eq.\ (\ref%
{BTE_RTA_solution}) together with the formal integral from Eq.\ (\ref{F_nrq}),
\begin{equation}
F_{nrq}(\tau ,\tau ^{\prime })\! = \!\frac{\left( -1\right) ^{q}}{%
\left( 2q\right) !!} \!\int \! dK E_{\mathbf{k}u}^{n-r-2q}E_{\mathbf{k}l}^{r}\left(
\Xi ^{\mu \nu }k_{\mu }k_{\nu }\right) ^{q}f_{\mathbf{k}}\left( \tau
^{\prime }\right) \;,
\end{equation}%
we obtain an integral equation that can be solved for various initial conditions
\begin{align}\nonumber
F_{nrq}(\tau ,\tau) & =D(\tau ,\tau _{0})F_{nrq}\left( \tau,\tau _{0}\right)  \\ 
& +\int_{\tau _{0}}^{\tau }\frac{d\tau ^{\prime }}{\tau _{eq}(\tau ^{\prime})}
D(\tau ,\tau ^{\prime }) I_{nrq}\left( \tau ,\tau ^{\prime }\right) \;. \label{F_nrq_general}
\end{align}%
Assuming that the initial distribution function is of the RS form, i.e., 
$f_{\mathbf{k}}(\tau _{0},k)=f_{RS}(\tau _{0},k)$, leads to the following equation for 
the energy density, $e(\tau) = F_{200}(\tau ,\tau)$,
\begin{align}
e(\tau) & =D(\tau ,\tau _{0})\hat{I}_{200}\left( \tau
,\tau _{0}\right)  \notag \\
& +\int_{\tau _{0}}^{\tau }\frac{d\tau ^{\prime }}{\tau _{eq}(\tau ^{\prime
})}D(\tau ,\tau ^{\prime })I_{200}\left( \tau ,\tau ^{\prime }\right) \;.
\end{align}

However, since $\hat{I}_{nrq}$ depends on a different set of parameters than $I_{nrq}$ we also need to obtain 
$\alpha_{RS}(\tau) $ and $\beta _{RS}(\tau) $ in terms of the equilibrium quantities 
$\alpha_{0}(\tau) $  and $\beta _{0}(\tau) $ at $\tau=\tau_0$. 
This is done via the Landau matching conditions as shown in Sec.\ \ref{applications}, hence in 
case that particle number is not conserved, i.e., $\hat{I}_{200}(\tau_0 ,\tau_{0}) =I_{200}(\tau_0 ,\tau _{0})$,
Eqs.\ (\ref{I_nrq_tau}) and (\ref{I_nrq_RS_tau}) lead to 
\begin{equation}
\beta _{RS}\left( \tau _{0}\right) =\beta _0\left( \tau_0\right) \left[ 
\frac{H_{200}( \tau_0 ,\tau _0/\sqrt{1+\xi_0})}{H_{200}(\tau_0 ,\tau _0)}\right] ^{1/4},
\end{equation}%
where $\xi_0 = \xi(\tau_0)$ while $H_{200}\left( \tau_0 ,\tau _0\right) = 2$ and so this expression is obviously 
equivalent to Eq.\ (\ref{temp_matching_RS}).

Finally, using the Landau matching condition for the l.h.s.\ of the integral equation 
$e(\tau) \equiv F_{200}\left( \tau ,\tau\right) = I_{200}\left( \tau ,\tau\right)$, we obtain the following 
equation for the evolution of the temperature 
\begin{align}
T^{4}(\tau) & = T^{4}( \tau_0) 
D(\tau,\tau _{0})\frac{H_{200}( \tau ,\tau _0/\sqrt{1+\xi_0 }) }
{H_{200}(\tau_0 ,\tau _0 /\sqrt{1+\xi_0 })} \notag \\
& +\int_{\tau _{0}}^{\tau }\frac{d\tau ^{\prime }}{\tau _{eq}(\tau ^{\prime
})} T^{4}\left( \tau ^{\prime }\right) D(\tau ,\tau ^{\prime })
\frac{H_{200}(\tau ,\tau ^{\prime })}{H_{200}(\tau ,\tau)} \;.
\end{align}%
In order to obtain the temperature as a function of proper time we use a combination of iteration 
and interpolation technique designed to obtain numerical solutions of integral equations 
\cite{StanRichardson}.

Once the temperature is obtained we can calculate any moment of the distribution function 
from Eq.\ (\ref{F_nrq_general}) as
\begin{align}
F_{nrq}(\tau) & = \frac{2\pi A_0 (n+1)!}{(2q)!!} \times \notag \\
& \left[ \frac{T^{n+2}(\tau_0) D(\tau ,\tau _{0})}{H^{-(n+2)/4}_{200}(\tau_0 ,\tau _0)}
\frac{H_{nrq}(\tau ,\tau _0/\sqrt{1+\xi_0 })}
{H^{(n+2)/4}_{200}(\tau_0 ,\tau _0 /\sqrt{1+\xi_0 }) } \right. \notag \\
& \left. +\int_{\tau _{0}}^{\tau }\frac{d\tau ^{\prime }}{\tau _{eq}(\tau ^{\prime
})}T^{n+2}(\tau') D(\tau ,\tau ^{\prime })  H_{nrq}( \tau ,\tau ^{\prime }) \right] \;.
\end{align}
The method presented here can be extended to the case where particle number is conserved.

%%%

\end{document}